\documentclass[12pt]{article}

\usepackage{graphicx}
\usepackage{amsmath}

\oddsidemargin  - 5 mm
\evensidemargin - 5 mm
\newtheorem{theorem}{Theorem}

\newtheorem{proposition}[theorem]{Proposition}

\newenvironment{proof}[1][Proof]{\textbf{#1.} }{\ \rule{0.5em}{0.5em}}
\input{tcilatex}

\begin{document}

\date{}
\title{Cubic Algebraic Equations in Gravity Theory, Parametrization with the
Weierstrass Function and Non-Arithmetic Theory of Algebraic Equations}

\author{Bogdan G. Dimitrov\\
\thanks{
Electronic mail: bogdan@thsun1.jinr.ru}
Bogoliubov Laboratory for Theoretical Physics\\
Joint Institute for Nuclear Research \\
Dubna 141980, Russia}
\maketitle

\begin{abstract}
A cubic algebraic equation for the effective parametrizations of the
standard gravitational Lagrangian has been obtained without applying any
variational principle. It was suggested that such an equation may find
application in gravity theory, brane, string and Rundall-Sundrum theories.
The obtained algebraic equation was brought by means of a linear-fractional
transformation to a parametrizable form, expressed through the elliptic
Weierstrass function, which was proved to satisfy the standard
parametrizable form, but with $g_{2}$ and $g_{3}$ functions \ of a complex
variable instead of the definite complex numbers (known from the usual
(arithmetic) theory of elliptic functions and curves). The generally
divergent (two) infinite sums of the inverse first and second powers of the
poles in the complex plane were shown to be convergent in the investigated
particular case, and the case of the infinite point of the linear-fractional
transformation was investigated. Some relations were found,which ensure the
parametrization of the cubic equation in its general form with the
Weierstrass function.

\end{abstract}




\newpage

\section*{\protect\bigskip I.{\protect\Huge \ }INTRODUCTION}

The synthesis of algebraic geometry and physics has been known for a long
time, beginning from chiral Potts model, algebraic Bethe ansatz (for a
review of these aspects see $^{1}$) and ending up with orbifold models of
string compactification.$^{2}$ In the context of string theories, the
application of algebraic curves, related to Fermat's theorem, has also been
pointed out.$^{3}$

Concerning gravitational physics, which is an inherent constituent of any
string, brane or ADS - theories, any applications of the theory of algebraic
curves are almost absent. In this aspect perhaps one of the most serious
attempts was undertaken in the recent paper of Kraniotis and Whitehouse.$%
^{4} $ Based on a suitably chosen metric of an inhomogeneous cosmological
model and introducing a pair of complex vaiables, the authors have succeeded
to obtain a \textit{nonlinear partial differential equation }\ for a
function, entering the trial solution of the field equations. The most
peculiar and important feature of the obtained equation is that it can be
parametrized by the well-known \textit{Weierstrass function }(for a
classical introduction in the theory of elliptic and Weierstrass functions
see Ref.5-7 ). This convenient representation enabled the authors to express
important physical quantities such as the Hubble constant and the scale
factor through the Weierstrass and the Jacobi theta functions. In fact, an
analogy has been used with examples on the motion of a body in the field of
an central force, depending on the inverse powers of the radial distance $r$%
. The cases of certain inverse powers of $r$, when the solution of the
trajectory equation is expressed in terms of elliptic and Weierstrass
functions, have been classified in details.$^{8}$

Three important conclusions immediately follow from the paper of Kraniotis
and Whitehouse,$^{4}$ and they provide an impetus towards further
investigations. The first two conclusions are correctly noted by the authors
themselves: 1. Other cases may exist, when solutions of nonlinear equations
of General Relativity might be expressed in terms of Weierstrass or theta
functions$^{9}$ , associated with Riemann surfaces. 2. The differential
equations of General Relativity in a much broader context might be related
to the mathematical theory of \textit{elliptic curves and modular forms }
(for an introduction, see Refs. 10-12 ) and even to the famous
Taniyama-Shimura conjecture, stating that every elliptic curve over the
field of rational numbers is a modular one. For a short review of some of
the recent developments in the \textit{arithmetic theory of elliptic curves,
}the interested reader may consult also the monograph$^{13}$ . In fact, the
eventual connection of General Relativity Theory with Number and Elliptic
Functions Theory was formulated even in the form of a conjecture that ''%
\textit{all nonlinear exact solutions of General Relativity with a non-zero
cosmological constant }$\Lambda $\textit{\ can be given in terms of the
Weierstrass Jacobi Modular Form}''.$^{4}$ Of course, such a conjecture is
expressed for the first time, and yet there are no other solutions derived
in terms of elliptic functions, not to speak about any classification of the
solutions on that bases.

The present paper will not have the purpose to present any new solutions of
the Einstein's equations by applying elliptic functions, nor will give any
new physical interpretation, which in principle should be grounded on
previously developed mathematical techniques. Rather than that, in this
paper an \ essential algebraic ''feature'' of the gravitational Lagrangian
will be proved, which is inherent in its structure, mostly in its partial
derivatives. This '\textit{algebraically inherent structure}'' represents
the third conclusion, which in a sense may be related to the problems,
discussed in Ref. 4.

However, this algebraic feature will become evident under some special
assumptions. While in standard gravitational theory it is usually assumed
that the metric tensor has an inverse one, in the so called \textit{theories
of spaces with covariant and contravariant metrics (and affine connections)}$%
^{14}$ instead of an inverse metric tensor one may have another
contravariant tensor $\overline{g}^{jk}$, satisfying the relation $g_{ij}%
\overline{g}^{jk}\equiv l_{i}^{k}\neq \delta _{i}^{k}$. But then, since $%
l_{i}^{k}$ cannot be determined from any physical considerations and at the
same time the important mathematical structure from a physical point of view
is the Gravitational Lagrangian, a natural question arises:\ \textbf{Is it
possible that in such a theory with a more general assumption in respect to
the contravariant metric tensor, the Gravitational Lagrangian is the same
(scalar density) as in the usual case, provided also that the usual
connection and the Ricci tensor are also given? From a physical point of
view, this is the central problem, treated in this paper, and the answer,
which is given, is affirmative. }Namely, it has been shown that if $e_{i}$
are the components of the covariant basic vectors, and $dX^{j}$ are the
components of a contravariant vector field (which, however, \textbf{are not }%
contravariant basic vectors and therefore $e_{i}dX^{k}\equiv \overline{l}%
_{i}^{k}\neq \delta _{i}^{k}$ ), then \textbf{they satisfy a cubic algebraic
equation}. Of course, if $dX^{i}$ are to be found from this equation, then
it can be shown that $\overline{g}^{jk}$ will also be known because of the
relation $\overline{g}^{jk}=dX^{j}dX^{k}$. Also, it has been assumed that
the affine connection $\Gamma _{ij}^{k}$ and the Ricci tensor $R_{ij}$,
determined in the standard way through the inverse metric tensor are known.

The obtained \textit{cubic algebraic equation } can be expressed in a very
simple form, but unfortunately it is not easy at all to solve it. That is
why a mathematical approach for dealing with such an equation has been
developed, on which any further physical application will be based. The
equation has been derived in two cases - when $d^{2}X^{i}\equiv 0$ and when $%
d^{2}X^{i}\neq 0$. As will be shown, the first assumption means that $dX$
has \textit{zero-vorticity components }(and non-zero divergency, however),
and in physical considerations this restriction can be imposed. The second
assumption would mean that $dX$ has both \textit{non-zero divergency and
non-zero vorticity} components. From the mathematical theory of cubic
equations, the investigation of the two cases will not be different, because
in the second case only the algebraic variety from the first case (with $%
dX^{i}$) will be supplemented with the components $d^{2}X^{i}$. It is worth
mentioning also that the derivation of the equation does not presume
zero-covariant derivatives of the covariant tensor field, thus leaving an
opportunity to unvestigate the different kinds of transports on the
space-time manifold. The algebraic equation may enable one to make a kind of
a classification (from an algebraic point of view) of the contravariant
tensors, satisfying the same gravitational Lagrangian.

So far, the problem investigated here may seem to be of pure
''mathematical'' interest, but it may have also numerous physical
applications. In supergravity theories, ADS/CFT, five-dimensional and brane
theories$^{15,16,17}$, one deals with an action, consisting of a
gravitational part, added to a (for example) string action of the kind $%
S_{str.}\equiv -\frac{T}{2}\int d^{2}\xi \sqrt{-h}h^{\alpha \beta }\partial
_{\alpha }X^{\mu }\partial _{\beta }X_{\mu }$, where $X^{\mu }$ are the
string coordinates, $h^{\alpha \beta }$-the world sheet metric tensor, $T$-
the string tension and the partial derivatives $\partial _{\alpha }$ are
taken in respect to the world sheet coordinates $\xi ^{\alpha }=(\tau
,\sigma )$. One can easily guess that the above described methodology can
easily be applied to the string part of the action. More concretely, $%
h^{\alpha \beta }$ may be expressed as $h^{\alpha \beta }=d\xi ^{\alpha
}d\xi ^{\beta }$, the gravitational metric tensor may be assumed to depend
on the string coordinates and the derivatives of the string coordinates will
be taken in respect to the world sheet coordinates $\xi ^{\alpha }$. Also,
the partial derivatives in the gravitational part of the action may also be
taken in respect to the coordinates $\xi ^{\alpha }$. As a result, taking
the gravitational and the string part of the action together and without
applying any variational principle, one would get the same kind of a cubic
algebraic equation as the one, which will be proposed in this paper. In a
sense, this dependence of $g_{ij}$ on the string coordinates is a sort of a
coupling between the gravitational part of the action and the string one,
and the resulting cubic equation may be called ''\textit{an algebraic
equation for the effective parametrization of the total Lagrangian in terms
of the string coordinates}''. The ''coupling'' between the two parts of the
action provides another interesting possibility, if the first variation of
the Lagrangian is performed, even without taking into account any equations
of motion. Provided that the gravitational Lagrangian depends also on the
\textit{first} and \textit{second differentials }of the metric tensor, the
first variation of the Lagrangian can be regarded also as an cubic algebraic
equation in respect to the differentials of the vector field$^{18}$. \

There is also another interesting problem, which is related to the current
trends in ADS/CFT correspondence$^{33}$ and the WZW theory of strings on a
curved background.$^{34,35}$ For example, in WZW theory it is not clear how
to relate the two-dimensional string world-sheet symmetries to the global
symmetries (global coordinates) of the three-dimensional ADS spacetime, in
terms of which the parametrization of the group element is presented. If one
has the ADS metric (and the ADS hyperboloid equation), then usually some
parametrization of the global ADS coordinates is chosen, in terms of which
the hyperboloid equation is satisfied. However, this is performed from the
viewpoint of maximum simplicity and convenience. The formalism, developed in
this paper on the base of cubic algebraic curves gives a possibility to find
some other parametrizations. For example, the (three-dimensional) ADS
spacetime has a boundary which can be found when one of the (global)
coordinates tends to infinity (for example $r\rightarrow \infty )$, and thus
a two-dimensional \ space is obtained, which can be identified with the one,
on which the world-sheet coordinates are defined. By ''identification'' one
may mean that not the two-dimensional coordinates, but only the Ricci
tensors and the Christoffel connections of the two space-times can be
identified. Then, and as will further be shown in the general case, the
obtained in this paper cubic algebraic equation will give an opportunity to
relate one of the differentials with the Weierstrass function and the other
differential - with its derivative. In such a way, one obtains a system of
two equations in partial derivatives, from where the parametrizations can be
found. In principle, 2+1 dimensional gravity and the WZW model of strings on
an ADS background are very convenient for application of the
algebro-geometric approach. Another interesting moment in these theories is
that very often one has to deal with the ADS metric, written in different
coordinates (including the two-dimensional coordinates of the worldsheet),
and it may be supposed that the transition from one system of coordinates to
another can probably be given by a linear-fractional transformation, which
will be invesigated in this paper. From this algebraic point of view, it is
interesting to investigate the coordinate transformations in Ref. 36 and in
Ref. 37.

Until now, only the physical aspects of the implementation of the algebraic
approach have been discussed. The mathematical theory of cubic algebraic
equations is also worth mentioning, and creating the relevent mathematical
methods will be the main purpose of this paper. In principle, the theory of
cubic algebraic equations and \ surfaces has been an widely investigated
subject for a long time. The mathematical theory of cubic hypersurfaces$%
^{19} $ puts the emphasis especially on the classification of points on the
cubic hypersurface, minimal cubic surfaces, two-dimensional birational
geometry and quasi-groups. But no concrete applications of cubic curves are
given. In well-known monographs,$^{20}$ the general theory of affine and
projective varieties and algebraic and projective plane curves is exposed,
and some examples are considered too, but the theory of cubic forms is
restricted only with the Pascal's theorem. A more comprehensive introduction
to the algebraic theory of second and third-rank curves, their normal forms,
turning points (where the second derivatives of the curves's equation equal
to zero), rational transformations and etc. is given in the book of Walker$%
^{21}$. Of \ particular relevence to the present research will be the theorem%
$^{21}$ that if $f(x,y)=0$ is a non-degenerate cubic curve, then by
introducing an affine set of coordinates $x_{1}=\frac{x}{z}$, $y_{1}=\frac{y%
}{z}$ and choosing the turning point at $(0,0,1)$, the curve can be brought
to the form $y^{2}=g(x)$, where $g(x)$ is a third-rank polynomial with
different roots. However, the situation is much more interesting in the
complex plane, where one may define the lattice $\Lambda =\{m\omega
_{1}+n\omega _{2}\mid m,n\in Z;$ $\omega _{1},\omega _{2}\in C,Im\frac{%
\omega _{1}}{\omega _{2}}>0\}.$ Let a mapping $f:$ $C/\Lambda \rightarrow
CP^{2}$ is defined of the factorized along the points of the lattice part of
the complex plane into the two-dimensional complex projective space $CP^{2}.$
If under this mapping the complex coordinates $z$ are mapped as $%
z\rightarrow (\rho (z),\rho ^{^{\prime }}(z),1)$ when $z\neq 0$ and $%
z\rightarrow (0,1,0)$ when $z=0$ $(\rho (z)$ is the \textit{Weierstrass
elliptic function}), then the mapping $\ f$ maps the torus $C/\Lambda $ into
the following \textit{affine curve }$y^{2}=4x^{3}-g_{2}x-g_{3}$, where $%
g_{2} $ and $g_{3}$ are complex numbers.$^{13}$The important meaning of this
statement is that excluding the points on the lattice which may be mapped
into one point of the torus (where the Weierstrass function has real
values), the mapping $z\rightarrow (x,y)=(\rho (z),\rho ^{^{\prime }}(z))$
parametrizes the cubic curve. The consequence from that is also essential
since the solution of the resulting differential equation can be obtained in
terms of elliptic functions.$^{5,13,22}$ In spite of the fact that the
parametrization can be presented in a purely algebraic manner, it is
inherently connected to basic notions from algebraic geometry such as
divisors and the Riemann-Roch theorem,$^{23}$ which reveals the dimension of
the vector space of meromorphic functions, having a pole of order at most $n
$ at the point $z=0$. This should be kept in mind because some results may
be obtained by algebraic methods only, but the explanation may probably be
found in algebraic geometry. In the present paper, a more general
parametrization of a cubic curve is considered , when $g_{2}$ and $g_{3}$
are not complex numbers (the so called \textit{Eisenstein series }$%
g_{2}=60\sum\limits_{\omega \subset \Gamma }\frac{1}{\omega ^{4}}%
=\sum\limits_{n,m}\frac{1}{(n+m\tau )^{4}}$; $g_{3}=140\sum\limits_{\omega
\subset \Gamma }\frac{1}{\omega ^{6}}=\sum\limits_{n,m}\frac{1}{(n+m\tau
)^{6}}$), but complex functions. \textbf{It has been proved that if the
Weierstrass function parametrizes again the cubic curve, then the infinite
sums (in pole number terms) }$\sum\limits_{\omega \subset \Gamma }\frac{1}{%
\omega ^{n}}$\textbf{\ for }$n=1$\textbf{\ and }$n=2$\textbf{\ turn out to
be finite (convergent) ones}\textit{, in spite of the fact that in the
general case they might be infinite ones (divergent)}. The explanation of
this fact from the point of view of algebraic geometry remains an open
problem, but it can be supposed that standard \textit{arithmetical theory}
of elliptic functions and algebraic equations is contained in some other,
more general theory, which may be called \textit{non-arithmetical theory},
and from this theory the standard parametrization should also follow. The
considered case of parametrization of a cubic curve with coefficient
functions of a complex variable, although performed in this paper in a
trivial algebraic manner, is the first step towards constructing such a
theory. At least, a certain motivation from a physical point of view is
evident for constructing such a theory.

The above-presented outlook on standard parametrization implied the use of
affine coordinates, which unfortunately exclude from consideration the
infinity point. But the infinity point cannot be ruled out not only from
mathematical grounds, but also from physical considerations. For example, in
the five-dimensional Randall-Sundrum model,$^{29,30}$ \ one has to assume a
compactification into a four-dimensional universe from an infinite extra
dimension, containing also the infinity point. From this point of view, the
more convenient transformation, chosen in the present paper, which brings
the cubic curve into a parametrizable form, is the \textit{linear-fractional
transformation.} This transformation allows one to parametrize with the
Weierstrass function the ratio of the two of the parameters, entering the
linear-fractional transformation and in the case the parameters in this
transformation represent complex functions. Of course, the other parameters
remain unfixed, leaving the opportunity to determine them in an appropriate
way. In a sense, from most general grounds the appearence of the Weierstrass
function in the linear-fractional transformation might be expected , since
according to a theorem in the well-known monograph of Courant and Hurwitz,$%
^{24}$ an algebraic curve of the kind \ $%
w^{2}=a_{0}v^{4}+a_{1}v^{3}+a_{2}v^{2}+a_{1}v+a_{4}$ can be parametrized as $%
v=\frac{a\rho (z)+b}{c\rho (z)+d}=\varphi (z)$ and $w=\varphi ^{^{\prime
}}(z)$ by means of the transformations $v=\frac{av_{1}+b}{cv_{1}+d}$ and $%
w=w_{1}\frac{ad-bc}{(cv_{1}+d)^{2}}$. In the case $a_{0}=0$ (which is the
present case of a third-rank polynomial), $\varphi (z)$ will be a linear
function of $\rho (z).$

In the present paper, however, the situation is quite different, since the
linear-fractional transformation is applied only in respect to one of the
variables $(v)$, and in order to get the standard parametrizable form $%
w^{2}=4v^{3}-g_{2}v-g_{3}$ (with $g_{2}$ and $g_{3}-$ complex functions), an
additional quadratic algebraic equation has to be satisfied. What is more
interesting is that after the parametrization is performed, the
linear-fractional transformation turns out to be of a more general kind $v=%
\frac{A(z)\rho (z)+\frac{b}{c}+B(z).\frac{d}{dt}(\rho ^{2}(z))}{C(z)\rho
^{^{\prime }}(z)+D(z)+\frac{d}{c}}$, where $A,B,C,D$ are functions of $z$,
and the expression for $v$ represents a\textit{\ rational transformation of
the kind }$v(z)=\frac{P(z)}{Q(z)}$. Now from another point of view it can
also be understood why it is justifiable to apply the rational
transformation only in respect to $v$ and not in respect to $w$. The reason
is in a well-known theorem$^{13}$ from algebraic geometry that \textit{%
''each non-degenerate cubic curve does not admit a rational parametrization''%
}. Since each non-degenerate cubic curve can be brought to the form $%
w^{2}=v(v-1)(v-\lambda )$, $(\lambda \neq 0,1),$the essence of the above
theorem is that this (algebraic) form \textit{cannot} be satisfied by a
rational parametrization of\textbf{\ \ both} $v=\frac{P(z)}{Q(z)}$ and $w=%
\frac{T(z)}{R(z)}$.

In respect to the problem about finiteness of the infinite sums $\sum \frac{1%
}{\omega ^{n}}$ for $n=1$ and $n=2$, the application of the
linear-fractional transformation has also turned out to be useful, when the
case of poles at infinity is considered. In principle, in this paper two
separate cases are distinguished - the first case of an infinite point of
the linear-fractional transformation and the second case of poles at
infinity, when in the infinite limit $\omega \rightarrow \infty $ the sum $%
\sum \frac{1}{\omega ^{n}}$ tends to the Riemann zeta function $\xi (n)$.
For this partial case and applying a different mathematical method, based on
the Tauber's theorem,$^{34}$ one comes again to the fact about the
finiteness of $G_{1}$, proved in the general case by performing a Loran
function decomposition.

The present paper is organized as follows:

Sect. II gives some basic formulaes about the so called gravitational theory
with contravariant and covariant metric tensors. In Sect. III the third-rank
algebraic equation has been derived, starting from the standard
gravitational Lagrangian. Also, the effective parametrization problem has
been formulated in an algebraic language. In Sect. IV the general
mathematical set-up for parametrization of the cubic equation has been
discussed, and some physical motivation for the application of the
linear-fractional transformation from the point of view of Randall-Sundrum
theory has been presented. Sect. V shows how the cubic algebraic equation
transforms under the action of the linear-fractional transformation. Sect.
VI shows how from the transformed cubic equation one can get the standard
parametrizable form of the cubic equation (with $g_{2}$ and $g_{3}$ -complex
numbers) and also the \textit{quadratic algebraic equation }$\ $is derived,
which has to be fulfilled if the parametrizable form holds. The approach is
valid also when $g_{2}(z)$ and $g_{3}(z)$ are complex functions. In Sect.
VII it was proved that the nonlinear and nonpolynomial transformation from
the ''unbar'' to the ''bar'' variables is also invertible, thus giving the
opportunity to write down two of the additionally imposed equations in terms
of the new ''bar'' variables. In Sect. VIII the Loran's decomposition has
been performed of \ the functions on the both sides of the algebraic
equation $(\frac{d\rho }{dz})^{2}=M(z)\rho ^{3}+N(z)\rho ^{2}+P(z)\rho +E(z)$%
, where $\rho (z)$\ is the Weierstrass function and $M,N,P,E$ are functions
of \ the complex variable $z$. A system of three iterative (depending on $n$%
) algebraic equations has been obtained, representing a necessary (but not
sufficient!) condition for parametrization of a cubic equation of a general
form with the Weierstrass function. It is not occasional that the condition
is called \ '' a necessary, but not sufficient one'', because in principle
more algebraic equations have to be solved in order to prove the existence
of such a parametrization. In Sect. IX the possible parametrization of the
more simplified cubic equation $\left[ \rho ^{^{\prime }}(z)\right]
^{2}=4\rho ^{3}-g_{2}(z)\rho -g_{3}(z)$ has been considered, and of course
the main motivation for considering such a case is the close analogy with
the well-known case, when $g_{2}$ and $g_{3}$ are complex numbers. By
calculating the coefficients in the negative power Loran expansion and
combining them, it has been proved that the sums $\sum \frac{1}{\omega }$
and $\sum \frac{1}{\omega ^{2}}$ represent finite (convergent) quantities.
The other equations for the other values of $m=-3,-1$ have been presented in
Appendix A; for values of \ $m=2k$ - in Appendix B and for $m=2k+1$ and $%
m=-k $ - in Appendix C. The equations in these Appendixes in fact complete
the proof that all the Loran coefficient functions can be uniquely expressed
through a combination of the finite sums $G_{n}$. The calculations are
purely technical but they serve as a strict mathematical motivation and a
proof of the new and basic qualitative fact that the Weierstrass function
can parametrize the simplified form of the cubic equation with coefficient
functions $g_{2}(z)$ and $g_{3}(z)$. This fact probably might represent one
of the starting points in the creation of the so called \textit{%
non-arithmetical theory of algebraic equations}. Sect.X investigates the
positive-power decomposition of the above-mentioned equation, and from the
convergency radius of the infinite sum the asymptotic behaviour of some of
the Loran coefficients was found to be $-\frac{n^{l+1}}{l+1}$. Sect. XI
considers the case of poles at infinity in the positive-power Loran
decomposition, and from the requirement to have a certain convergency
radius, expressions for some of the Loran coefficient functions are
obtained. In Sect. XII a split of the original cubic equation into two
equations is performed, and based on the fact from Sect. IX, it has been
proved that the parametrization of the first equation leads to a
parametrization of the second equation. For the two ''splitted-up''
equations, Sect. XIII presents an algebraic equation, defined on a Riemann
surface, which has to be satisfied if the so called $j-$invariants of the
two equations are to be equal. In Sect. XIV on the base of the Loran
function decomposition of $g_{2}(z)$ an infinite sum is obtained in which
the coefficient functions contain the sums $G_{n}$. In Sect. XV this
formulae has been combined with a proof \ that the Tauber's theorem can be
applied, and this combination resulted in an expression for $G_{1}$ in the
limit of poles at infinity. In this partial case, the expression again
confirms that the sum $G_{1}$is convergent. In Sect. XVI the case of the
infinite point of the linear-fractional transformation is considered, and
the approach essentially represents a combination of the ''split-up''
approach from Sect. XII and the approach from Sect. VI , based on the
derivation of the additional quadratic equation. In Sect. XVII the relation
between the two integration constants is found which appear in the process
of integration of the two splitted-up equations. A peculiarity of the
developed approach is the appropriate ''fixing'' of some of \ the functions
in the linear-fractional transformation so that the simplest and most
trivial form of the quadratic equation from Sect. VI is obtained. Sect.
XVIII starts with an algebraic equation of a fourth rank, derived from the
original equation in the case of an infinite point of the linear-fractional
transformation. The main result here is that the constant Weierstrass
function can parametrize this equation if it should be fulfilled in the
entire complex plane. For the same equation, Sect. XIX investigates the
second case, when the fourth-rank algebraic equation determines a Riemann
surface for the pair of variables $(\rho (z)=w_{1}(z)+iw_{2}(z);z)$, and six
values of $w_{1}$ are found, satisfying this equation. Sect. XX finds the
necessary and sufficient condition for parametrization with a constant
Weierstrass function, based again on the approach of Riemann surfaces. As a
result, an integrable nonlinear equation is obtained for the coefficient
functions of the algebraic equation. The coefficient functions in the
solution of this equation appear in powers of non-integer (fractional)
numbers.

\section*{\protect\bigskip II. Covariant and Contravariant Metric Tensor}

Usually in gravitational theory it is assumed that a local coordinate system
can be defined so that to each metric tensor $g_{ij}$ an inverse one $g^{jk}$
can be defined

\begin{equation}
g_{ij}g^{jk}\equiv \delta _{i}^{k}=\left\{ 0\text{ if }i\neq k\text{ and }1%
\text{ if }i=k\right\} .  \tag{1}
\end{equation}
However, the notion of a reference frame can be defined in different ways in
Ref. 25 - coordinate, tetrad and monad. In the last case the contravariant
vector field $dx^{i}$ of an observer, moving along a space-time trajectory,
\ represents a reference system. In such a case one may have instead of (2)

\begin{equation}
e_{i}dx^{j}\equiv f_{i}^{j}\neq \delta _{i}^{j}=S(e_{i},dx^{j}).  \tag{2}
\end{equation}
In the context of the so called dual algebraic spaces in Ref. 26, $%
S(e_{i},dx^{j})$ is called a \textit{contraction operator}. Assuming that an
inverse operator of contraction $f_{j}^{i}$ exists, it can easily be
obtained, as in Ref.14

\begin{equation}
e^{j}\equiv f_{i}^{j}dx^{i}.  \tag{3}
\end{equation}
Therefore, the metric tensor field $g$ can be decomposed in respect to the
conravariant basic eigenvectors in the following way

\begin{equation}
g\equiv g_{ij}(e^{i}\otimes e^{j})\equiv
g_{ij}f_{k}^{i}f_{l}^{j}dx^{k}dx^{l}(e_{k}\otimes e_{l})\equiv
(dx^{k}dx^{l})(e_{k}\otimes e_{l}),  \tag{4}
\end{equation}
and the contravariant components $\widetilde{g}^{ij}$ of the tensor field $g$
are represented as a contraction of the two vector fields $dx^{i}$ and $%
dx^{j}$:

\begin{equation}
\widetilde{g}^{ij}\equiv dx^{i}dx^{j}.  \tag{5}
\end{equation}
It is important to realize that this definition of a contravariant tensor
field is not related to any notion of infinitesimality. In order to
understand this, consider a set of global coordinates $X^{\mu }$, defined on
the given manifold and depending also on some other (local) coordinates.
Then the set of global coordinates, regarded as functions of the local ones,
can be considered as a system of equations, defining some algebraic surface.
Provided that the partial derivatives of the global coordinates in respect
to the local ones are non-zero, \textit{at each point of this surface} the
corresponding \textit{tangent space }can be determined, and the
differentials of the global coordinates are defined on this tangent space.
If one assumes that the global differentials are infinitesimally small, then
either the (partial) derivatives of the global coordinates or the ''local'
differentials should be small. However, the parial derivatives cannot be
small, because one considers arbitrary global and local coordinates on the
manifold. Also, if the local differentials are assumed to be small, then
they will not be allowed to take arbitrary values. But this will mean that a
large variety of integral curves on the manifold should be excluded from
consideration. This will be unacceptable since one would like to define
integral curves through each point of the manifold and moreover, it would
contradict to our initial assumption about existence of a tangent space at
each point of the surface (or manifold). Therefore, as a partial case, each
of the local differentials should be allowed to take arbitrary numerical
values and of course, they may be equal also to an arbitrary function of the
local coordinates.

It might be concluded, therefore, that since the partial derivatives and the
local differentials cannot be infinitesimally small, then the global
differentials cannot be also infinitesimally small.

Apart from the definition (5) of a contravariant tensor field, we have also
the definition of a length interval in Riemannian geometry

\begin{equation}
ds^{2}\equiv l^{2}(\overline{r})\equiv g_{ij}dx^{i}dx^{j}  \tag{6}
\end{equation}
If we would like to ''incorporate'' in this definition the standard
definition of an inverse metric tensor as $g_{ij}g^{jk}\equiv \delta
_{i}^{k} $, we can set up for the ordinary inverse metric tensor

\begin{equation}
g^{ij}\equiv \frac{1}{l^{2}}dx^{i}dx^{j}.  \tag{7}
\end{equation}
Therefore, in terms of the differentials, the ordinary inverse tensor $%
g^{ij} $ can be represented in the same way as in (5), but divided by the
length interval. However, usually the lenght interval is not known, so from
a physical point of view the definition (7) is undesirable and this is the
motivation to deal further with the definition (5) of a contavariant tensor
field. In order to distinguish the ''newly'' defined tensor in (5), a
''tilda'' sign has been placed.

From (3) and (7) it follows

\begin{equation}
\left[ \frac{1}{l^{2}}-g_{kl}f_{i}^{k}f_{j}^{l}\right] dx^{i}dx^{j}\equiv 0.
\tag{8}
\end{equation}
Clearly, the requirement for existence of an inverse contraction operator is
equivalent to putting \ $l=1$, i.e. assuming that there is a unit lenght
interval, which is again physically unacceptable, and it is more natural to
assume that the lenght interval is varying. Let us assume that $l^{2}$ and $%
f_{k}^{i}$ are known in advance, then it can be be investigated which is the
algebraic variety of values of $dx^{i}$, satisfying this quadratic form. The
main difficulty in this approach is that $f_{k}^{i}$ cannot be determined
from physical considerations. That is why the aim in the next section wil be
to derive an algebraic equation, in which known physical quantities will
enter - the metric tensor $g_{ij}$, the Christoffel connection $\Gamma
_{ij}^{k}$ and the Ricci tensor $R_{ij}$.

\section*{ III. CUBIC \ ALGEBRAIC \ EQUATION \\
NOT \ FOLLOWING \ FROM \\
A \ VARIATIONAL \ PRINCIPLE}

Further in this paper it shall be assumed that if $X^{i}$ are some
generalized coordinates, defined on a $n-$dimensional manifold with
coordinates on it $(x^{1},x^{2},.....,x^{n}),$ then the differential $dX^{i}$
is defined in the corresponding tangent space $T_{X}$ of the generalized
coordinates $X^{i}\equiv X^{i}(x^{1},x^{2},x^{3}...,x^{n})$. Even if written
with a small letter, it shall be understood that $x^{i}$ represent
generalized coordinates.

Our starting point for the derivation of the cubic equation will be the
assumption that \textbf{in spite of the choice for the contravariant metric
tensor, the gravitational Lagrangian }$L=-\sqrt{-g}R$\textbf{\ should be the
same, provided also that the Ricci tensor does not change under the
definition of the contravariant metric tensor. }The meaning of this
statement is the following.

Essentially, the gravitational Lagrangian will have two representations. The
\textbf{first representation} is based on the standardly defined
Christoffell connection $\Gamma _{ik}^{l}$
\begin{equation}
\Gamma _{ik}^{l}\equiv \frac{1}{2}g^{ls}(g_{ks,i}+g_{is,k}-g_{ik,s})  \tag{9}
\end{equation}
and the Ricci tensor
\begin{equation}
R_{ik}=\frac{\partial \Gamma _{ik}^{l}}{\partial x^{l}}-\frac{\partial
\Gamma _{il}^{l}}{\partial x^{k}}+\Gamma _{ik}^{l}\Gamma _{lm}^{m}-\Gamma
_{il}^{m}\Gamma _{km}^{l}\text{ \ \ \ .}  \tag{10}
\end{equation}
The \textbf{second representation} of the gravitational Lagrangian will be
based on the definition (5) of the contravariant metric tensor $\widetilde{g}%
^{jk}=dx^{j}dx^{k}$. Therefore, the Christoffell connection and the Ricci
tensor will be different from the previous ones and will be denoted
respectively by $\widetilde{\Gamma }_{ik}^{l}\ $and $\widetilde{R}_{ik}$
\begin{equation}
\widetilde{\Gamma }_{ik}^{l}\equiv \frac{1}{2}\widetilde{g}%
^{ls}(g_{ks,i}+g_{is,k}-g_{ik,s})=\frac{1}{2}dx^{l}dx^{s}g_{ks,i}+\frac{1}{2}%
dx^{l}dx^{s}g_{is,k}-\frac{1}{2}dx^{l}dg_{ik}\text{ \ \ \ ,}  \tag{11}
\end{equation}
\begin{equation}
\widetilde{R}_{ik}=\frac{\partial \widetilde{\Gamma }_{ik}^{l}}{\partial
x^{l}}-\frac{\partial \widetilde{\Gamma }_{il}^{l}}{\partial x^{k}}+%
\widetilde{\Gamma }_{ik}^{l}\widetilde{\Gamma }_{lm}^{m}-\widetilde{\Gamma }%
_{il}^{m}\widetilde{\Gamma }_{km}^{l}\text{ \ \ . }  \tag{12}
\end{equation}

\bigskip The gravitational Lagrangian in this \textbf{second representation}
is
\begin{equation*}
L_{2}\equiv -\sqrt{-g}R=-\sqrt{-g}\widetilde{g}^{ik}\widetilde{R}_{ik}=-%
\sqrt{-g}dx^{i}dx^{k}(\frac{\partial \widetilde{\Gamma }_{ik}^{l}}{\partial
x^{l}}-\frac{\partial \widetilde{\Gamma }_{il}^{l}}{\partial x^{k}})-
\end{equation*}
\begin{equation}
-\sqrt{-g}dx^{i}dx^{k}(\widetilde{\Gamma }_{ik}^{l}\widetilde{\Gamma }%
_{lm}^{m}-\widetilde{\Gamma }_{il}^{m}\widetilde{\Gamma }_{km}^{l})\text{ \
\ .}  \tag{13}
\end{equation}

Note that physical meaning of this Lagrangian will depend not only on the
properties of the (covariant) metric tensor, but also on the first and the
second differentials $dx^{l}$ and $d^{2}x^{l}$. It should be mentioned also
that the notion of a metric tensor, depending on generalized coordinates,
understood in the sense of a hypersurface (an infinite-dimensional manifold
of all space-like hypersurfaces, embedded in a given Riemannian spacetime),
has been introduced long time ako by Kuchar in Ref.31. In such an approach,
the description of the gravitational field essentially depends on the
\textit{tangential and normal deformations }of the embedded hypersurface. In
our case, we do not restrict to space-like hypersurfaces, but the notion of
the differentials begins to play a self-consistent role, similarly to the
dynamics and the deformations of the hypersurface in Kuchar's approach. Yet,
the standard gravitational physics with the usual inverse metric tensor is
contained in the proposed in this paper approach, because one can identify
the components of the usually known inverse metric tensor with the
components of the contravariant metric tensor, defined in terms of the
differentials. Thus one can obtain a \textit{system of \ first -order
nonlinear differential equations in partial derivatives. }The solution of
this system may enable one to chose such global (generalized) coordinates,
in terms of which the usual inverse tensor will be equivalent to the
contravariant one in terms of the differentials.

Let us now use expressions (5) for the contravariant metric tensor $%
\widetilde{g}^{ij}$ and \ (11) for the Christoffel connection $\widetilde{%
\Gamma }_{ij}^{k}$ in order to rewrite the gravitational Lagrangian in the
second representation. The first two terms in (13) can be calculated to be
\begin{equation*}
-\sqrt{-g}dx^{i}dx^{k}(\frac{\partial \widetilde{\Gamma }_{ik}^{l}}{\partial
x^{l}}-\frac{\partial \widetilde{\Gamma }_{il}^{l}}{\partial x^{k}})=\sqrt{-g%
}dx^{i}dx^{k}dx^{l}\{g_{is,l}\frac{\partial (dx^{s})}{\partial x^{k}}-\frac{1%
}{2}pg_{ik,l}+\frac{1}{2}g_{il,s}\frac{\partial (dx^{s})}{\partial x^{k}}\}=
\end{equation*}
\begin{equation}
=-\sqrt{-g}dx^{i}dx^{l}\{p\Gamma _{il}^{r}g_{kr}dx^{k}-\Gamma
_{ik}^{r}g_{lr}d^{2}x^{k}-\Gamma _{l(i}^{r}g_{k)r}d^{2}x^{k}\}\text{ \ \ ,}
\tag{14}
\end{equation}
where $p$ is the scalar quantity

\begin{equation}
p\equiv div(dx)\equiv \frac{\partial (dx^{l})}{\partial x^{l}}\text{,}
\tag{15}
\end{equation}
which ''measures'' the divergency of the vector field $dx$. It will be more
interesting to calculate the contribution of the second term in (12)
\begin{equation*}
-\sqrt{-g}dx^{i}dx^{k}(\widetilde{\Gamma }_{ik}^{l}\widetilde{\Gamma }%
_{lm}^{m}-\widetilde{\Gamma }_{il}^{m}\widetilde{\Gamma }_{km}^{l})\text{ }=-%
\frac{1}{2}\sqrt{-g}%
dx^{i}dx^{k}dx^{l}dx^{m}(-dg_{lm}dx^{s}g_{ks,i}-dg_{ik}dx^{r}g_{mr,l}+
\end{equation*}
\begin{equation}
+dg_{il}dx^{r}g_{mr,k}+dg_{km}dx^{s}g_{ls,i})-\sqrt{-g}%
dx^{i}dx^{k}dx^{l}dx^{m}dx^{s}dx^{r}[g_{ks,i}g_{mr,l}-g_{ls,i}g_{mr,k}]=0
\tag{16}
\end{equation}
and the $\mathit{first\ differential}$ $dg_{ij}$ is represented as $%
dg_{ij}\equiv \frac{\partial g_{ij}}{\partial x^{s}}dx^{s}\equiv \Gamma
_{s(i}^{r}g_{j)r}dx^{s}$ and $\Gamma _{si}^{r}$ is the standard Christoffell
connection. Therefore, the second two terms in (13) give no contribution to
the gravitational Lagrangian. This is not surprising, since the
''factorization'' of the contravariant metric tensor as $dx^{i}dx^{j}$
introduces an additional ''symmetry'', due to which all the terms in (16)
cancel. That is why the \textbf{second representation} of the gravitational
Lagrangian will be given only by the first two terms $-\sqrt{-g}dx^{i}dx^{k}(%
\frac{\partial \widetilde{\Gamma }_{ik}^{l}}{\partial x^{l}}-\frac{\partial
\widetilde{\Gamma }_{il}^{l}}{\partial x^{k}})$ in expression (13).

Concerning the \textbf{first representation} of the gravitational
Lagrangian, it was based on the standard Christoffell connection $\Gamma
_{ij}^{k}$ , the Ricci tensor $R_{ik}$ and the usual inverse metric tensor $%
g^{ij}$. The basic assumption at the beginning \ concerned the gravitational
Lagrangian and the Ricci tensor, which means that together with the inverse
metric tensor $g^{ij}$, \textbf{another contravariant tensor} $\widetilde{g}%
^{ij}=dx^{i}dx^{j}$ exists, which enters the expression for the \textbf{%
first representation} of the gravitational Lagrangian
\begin{equation}
L_{1}=-\sqrt{-g}\widetilde{g}^{ik}R_{ik}=-\sqrt{-g}dx^{i}dx^{k}R_{ik}\text{
\ .}  \tag{17}
\end{equation}
Comparing this representation with the \textbf{second} one, given by
expression (13)
\begin{equation}
L_{2}=-\sqrt{-g}\widetilde{g}^{il}\widetilde{R}_{il}=-\sqrt{-g}%
dx^{i}dx^{l}\{p\Gamma _{il}^{r}g_{kr}dx^{k}-\Gamma
_{ik}^{r}g_{lr}d^{2}x^{k}-\Gamma _{l(i}^{r}g_{k)r}d^{2}x^{k}\}\text{ }
\tag{18}
\end{equation}
\ and remembering the initial assumption, acccording to which the Lagrangian
should be \textbf{one and the same in both the representations (i.e. }$%
L_{1}=L_{2})$, one arrives at the following algebraic equation in respect to
the first differential $dx^{k}$ and the second differential $d^{2}x^{k}$
\begin{equation}
dx^{i}dx^{l}\left( p\Gamma _{il}^{r}g_{kr}dx^{k}-\Gamma
_{ik}^{r}g_{lr}d^{2}x^{k}-\Gamma _{l(i}^{r}g_{k)r}d^{2}x^{k}\right)
-dx^{i}dx^{l}R_{il}=0\text{ \ \ \ \ .}  \tag{19}
\end{equation}
In the limit $d^{2}x_{k}=0$ this equation assumes the form of a \textbf{%
manifestly cubic in respect to }$dx^{i}$\textbf{\ algebraic equation }
\begin{equation}
dx^{i}dx^{j}dx^{k}p\Gamma _{j(i}^{r}g_{k)r}-R_{ij}dx^{i}dx^{j}=0\text{.}
\tag{20}
\end{equation}
Equation (20) is the basic equation, which shall be investigated further in
this paper. Most importantly, it is manifestly cubic in the differentials $%
dx^{i}$. Due to this reason, one qualitative argument can be given in favour
of such a Lagrangian. In 1988, Witten derived the Lagrangian for 2+1
dimensional gravity in Ref. 32, which is also manifestly cubic in the chosen
gauge variables $A_{\mu }$. The Lagrangian was obtained under the assumption
that there is an isomorphism between an abstractly introduced ($d-$%
dimensional) vector bundle with a structure group $SO(d-1,d)$ and the
tangent bundle of the given manifold, on which the metric is the induced one
from the metric on the vector bundle. Besides, the verbein was assumed to be
invertible, but as Witten remarks ''permitting the verbein to not be
invertible seems like a minor change.'' In the present case, we don't have
at all any symmetry on the tangent bundle, neither is anything supposed
about the dimensionality of spacetime or even about the existence of the
usual inverse tensor, but yet the Lagrangian exibits the same cubic
structure. Therefore, it may be concluded that the cubic structure of
Chern-Simons theory$^{32}$ is inherent in the structure of the gravitational
Lagrangian itself, and not in the additional assumptions in Ref.32, which
affect the choice of the gauge variables. In view of this, it might be
interesting to investigate whether there is a transition from the Lagrangian
in our case to the Lagrangian for 2+1 dimensional gravity, presented in Ref.
32.

Of course, one might slightly modify the basic assumption, concerning the
first representation of the gravitational Lagrangian. For example, instead
of assuming that the Ricci tensor will be the same in both representations,
one might instead assume that the \textbf{Ricci tensor }should not change.
In such a case again in the limit $d^{2}x^{k}=0$ the cubic algebraic
equation will be in a form without the quadratic in $dx^{i}$ term
\begin{equation}
dx^{i}dx^{j}dx^{k}p\Gamma _{j(i}^{r}g_{k)r}-R=0\text{ \ \ \ \ .}  \tag{21}
\end{equation}
One can write down also the vacuum Einstein's equations when the
contravariant tensor is defined as $\widetilde{g}^{ij}=dx^{i}dx^{j}$
\begin{equation*}
0=\widetilde{R}_{ij}-\frac{1}{2}g_{ij}\widetilde{R}=\widetilde{R}_{ij}-\frac{%
1}{2}g_{ij}dx^{m}dx^{n}\widetilde{R}_{mn}=
\end{equation*}
\begin{equation*}
=-\frac{1}{2}pg_{ij}\Gamma _{mn}^{r}g_{kr}dx^{k}dx^{m}dx^{n}+\frac{1}{2}%
g_{ij}(\Gamma _{km}^{r}g_{nr}+\Gamma
_{n(m}^{r}g_{k)r})d^{2}x^{k}dx^{m}dx^{n}+
\end{equation*}
\begin{equation}
+p\Gamma _{ij}^{r}g_{kr}dx^{k}-(\Gamma _{ik}^{r}g_{jr}+\Gamma
_{j(i}^{r}g_{k)r})d^{2}x^{k}\text{ \ \ \ .}  \tag{22}
\end{equation}
Note the following subtle moment : \ since we have an expression equal to
zero, this time \textbf{it is not necesssary }to assume that the above
algebraic equation is valid under the assumption that the Ricci tensor does
not change. Therefore, equation (22) provides the interesting possibility
for classification of all solutions of the vacuum Einsteins equations with a
given metric tensor $g_{ij}$ and unknown contravariant tensor $\widetilde{g}%
^{ij}=dx^{i}dx^{j}$. In spite of the presence of the second differentials $%
d^{2}x^{k}$ , equation (22) can be treated on an equal footing as an
algebraic equation simply by ''extending'' the algebraic variety for the $%
\{dx^{k}\}$ variables with the new variable $dy^{k}=d^{2}x^{k}$ . However,
if \textbf{additionally} it is assumed that the Ricci tensor does not change
under the definition of the contravariant tensor $(i.e.$ $\widetilde{R}%
_{ij}=R_{ij})$, then one has
\begin{equation}
\left( \Gamma _{ik}^{r}g_{jr}+\Gamma _{j(i}^{r}g_{k)r}\right)
d^{2}x^{k}=p\Gamma _{ij}^{r}g_{kr}dx^{k}-R_{ij}  \tag{23}
\end{equation}
and consequently all the terms with $d^{2}x^{k}$ in the Einstein's vacuum
equations (22) drop out and the algebraic equation becomes a cubic one in
respect to the variables $dx^{k}$ only. The above analyses has the purpose
to demonstrate that depending on the initial assumptions about the Ricci
tensor or scalar curvature, the structure of the algebraic equation also
changes.

In an algebraic language,$^{20,27,28}$ the investigated problem can be
formulated in the following way:

\begin{proposition}
Let the differentials $dx^{i}(i=1,....,n;n$ is the space-time dimension)
represent elements of an algebraic variety $\overline{X}%
=(dx^{1},dx^{2}.....,dx^{n}).$ For different metric tensors (and therefore -
different connections $\Gamma _{ij}^{k}$ and Riemannian tensors $R_{ik}$), a
set of polynomials \ (cubic algebraic equations) $F(\overline{X})\equiv 0$
may be obtained, which are defined on the algebraic variety $\overline{X}$
and belong to the ring $R[dx^{1},dx^{2},...,dx^{n}]$ of all third-rank
polynomials. Then finding all the possible parametrizations of some
introduced generalized coordinates $X^{i}(x^{1},x^{2},x^{3},..,x^{n})$ is
equivalent \ to: 1. Finding all \ the elements $dX^{i}$ of the algebraic
variety $\overline{X},$ satisfying the equation $F(\overline{X})\equiv 0.$
These elements will be represented in the following way

\begin{equation}
dx^{i}=\Phi ^{i}(x^{1},...x^{n},g_{ij}(x^{1},x^{2}...x^{n}),\Gamma
_{ij}^{k}(x^{1},x^{2},...x^{n}),R_{ij}(x^{1},x^{2},..x^{n})).  \tag{24}
\end{equation}
2. Finding all the solutions of the above system of partial differential
equations.
\end{proposition}

\bigskip\ In the present case, the algebraic equation is obtained \textit{%
before performing} the variation of the Lagrangian, unlike the considered in
Ref.18 another case, when again a cubic algebraic equation had been obtained
\textit{after performing} a variation.

Let us comment briefly on the important from a physical point of view
assumption $d^{2}x^{i}\equiv 0,$ under which the cubic equation (20) was
derived. Suppose that for the set of generalized coordinates $X^{i}\equiv
X^{i}(x^{1},x^{2}...x^{n})$ one has

\begin{equation}
dX\equiv a_{i}dx^{i}  \tag{25}
\end{equation}
and let us assume that the Poincare's theorem is fulfilled in respect to $%
dx^{i}$, i.e. $d^{2}x^{i}=0$. Then

\begin{equation}
d^{2}X=da_{i}dx^{i}+a_{i}d^{2}x^{i}=\frac{\partial a_{i}}{\partial x^{j}}%
dx^{j}\wedge dx^{i}=(\frac{\partial a_{i}}{\partial x^{j}}-\frac{\partial
a_{j}}{\partial x^{i}})dx^{i}dx^{j}\text{.}  \tag{26}
\end{equation}
Clearly, $d^{2}X=0$ only in the following two cases: 1. $a_{i}=const.,$ i.e.
$dX^{i}$ is a full differential.

2. ($rota)_{ij}\equiv \frac{\partial a_{i}}{\partial xj}-\frac{\partial a_{j}%
}{\partial xi}\equiv 0.$ The last means that if $dx^{i}$ are considered to
be basic eigenvectors, then $dX^{i}$ have \textit{zero-vorticity components}%
. Throughout the whole paper $dX^{i}$ shall be considered as vector field's
components in the tangent space $T_{X}$.

\bigskip Note also that the algebraic equation (19) with the first and the
second differentials $dx^{i}$ and $d^{2}x^{i}$ takes into account \textit{%
two important physical characteristics of the vector field }$dx^{i}$\textit{%
\ - the divergency }$p$\textit{\ and the vorticity (through the term }$%
d^{2}x^{i}).$ It might be required that these characteristics vanish, i.e. $%
p=d^{2}x^{i}=0.$ In such a case one is left only with the equation

\begin{equation}
R_{ik}dx^{i}dx^{k}\equiv 0\text{.}  \tag{27}
\end{equation}
If additionally the requirement for the existence of \ the (usual) inverse
metric tensor is imposed then the intersection variety of the quadratic form
(27) with the quadratic forms (one-when $\delta _{i}^{j}=0,$and the other -
when $\delta _{i}^{i}=1$).

\begin{equation}
g_{ik}dx^{k}dx^{j}\equiv \delta _{i}^{j}\text{ .}  \tag{28}
\end{equation}
has to be found. From the two last equations one easily obtains

\begin{equation}
(R_{ik}-\frac{1}{2}g_{ik}R)dx^{k}dx^{j}\equiv -\frac{1}{2}R\delta _{i}^{j}%
\text{ ,}  \tag{29}
\end{equation}
in which the left-hand side is identically zero for every $dx^{i}$ in view
of the Einstein' equations $R_{ik}-\frac{1}{2}g_{ik}R\equiv 0$, but the
right-hand side is zero only for $i\neq j$, but not also when $i=j$.
Therefore, the Einstein's equations are obtained only in one case and not in
the other case. In fact, it shouldn't be surprising that the Einstein's
equations cannot be obtained for both the cases $i\neq j$ and $i=j$. One
should remember that the usual variational procedure in general relativity
takes into account also the variation of the volume factor $\sqrt{-g}$,
while in our purely algebraic treatment and without any variation this
volume factor was not subjected to any changes at all. Moreover, it is one
standard procedure to perform the variational procedure with the usual
gravitational Lagrangian \ and the inverse metric tensor (when the
Einstein's equations are obtained) and its quite a different procedure to
start from the other representation of the gravitational Lagrangian (where
the variables to be variated are $g_{ij},\Gamma _{ij}^{k}$ (or $g_{ij,k}$)
and $dx^{i}$ and $d^{2}x^{i}$) and afterwards to impose the requirement for
identification of the contravariant metric tensor with the inverse one in
the form of another, additional equation. So one should not even hope to
obtain anything similar to the Einstein's equations. However, as already
shown, if one has the Einstein's equations, one may still ask the question
are they satisfied under the new definition of the contravariant tensor.

\section*{ IV. A \ GENERAL \ MATHEMATICAL \ SETUP \\
FOR \ TREATING \ THE CUBIC \\
ALGEBRAIC \ EQUATION (20)}

The subsequent investigation of equation (20) will be restricted to the case
of a 5-dimensional space-time, although the approach of course may be
applicable to any dimensions. The main reason for chosing a 5-d spacetime is
related to the widely discussed Randall-Sundrum (R - S) model,$^{29,30}$ in
which the process of compactification of the five-dimensional universe to
our present four-dimensional universe is related to the existence of a
\textit{large extra dimension}. In the original R - S scenario the metric
was chosen to be

\begin{equation}
ds^{2}=e^{-2kr_{c}r_{5}}\eta ^{\mu \nu }dx^{\mu }dx^{\nu
}+r_{c}^{2}dx_{5}^{2}\text{,}  \tag{30}
\end{equation}
where $r_{c}$ is a compactification radius, $\eta ^{\mu \nu }$ is the
ordinary Minkowski metric, $x_{5}\subset \left[ 0,\pi \right] $ is a
periodic coordinate, $\mu \nu $ are four dimensional indices and $k$ is a
scale of order of the Planck scale. Instead of the coordinate $x_{5},$ one
may chose for example a fifth coordinate $X_{5}=kr_{c}x_{5}$, which in view
of the largeness of the scale factor $k$ may be assumed to range to
infinity. \textit{But the infinity point, from a purely mathematical point
of view, may be treated on an equal footing with all other points in the
framework of projective geometry.}$^{7,21,38,39}$\textit{\ } In the present
case the infinity point shall be realized in respect to $dx^{5}$ after
performing the \textit{linear-fractional transformation}

\begin{equation}
dx^{5}\equiv \frac{a\widetilde{dx}^{5}+b\text{ }}{c\widetilde{dx}^{5}\text{ }%
+d\text{ }}\text{ ,}  \tag{31}
\end{equation}
where $a$, $b$ , $c$ and $d$ will be assumed to be functions, depending on
the complex variable $z$ (or on two complex variables). Also, the remaining
four-dimensional space-time with coordinates $(x^{1},x^{2},x^{3},x^{4})$ may
be complexified in the following way

\begin{equation}
z_{1}=x_{1}+ix_{2};\text{ \ \ \ \ \ \ \ \ \ \ \ \ \ \ \ \ }z_{2}=x_{3}+ix_{4}%
\text{ .}  \tag{32}
\end{equation}
It is easily seen that the infinity point in respect to $dx^{5}$ is situated
at $d\widetilde{x}^{5}\equiv -\frac{d}{c}$ and it is a zero point for the
complex plane $d\widetilde{x}^{5}$. The convenience of the linear-fractional
transformation from a physical point of view matches also the mathematical
requirements of the problem. In order to parametrize the third-rank
algebraic equation (15), written in a two-dimensional form, one has to bring
it to the form

\begin{equation}
(d\widetilde{x}^{5})^{2}\equiv 4(d\widetilde{x}^{4})^{3}-g_{2}(d\widetilde{x}%
^{4})-g_{3}\text{ .}  \tag{33}
\end{equation}
In the case when $g_{2}$ and $g_{3}$ are complex numbers $%
g_{2}=60G_{4}=60\sum\limits_{\omega }\frac{1}{\omega ^{4}}$ and $%
g_{3}=140G_{6}=\sum\limits_{\omega }\frac{1}{\omega ^{6}}$, standard
algebraic geometry contains a well-known prescription how to parametrize
this algebraic curve$^{13}$ by introducing the variables

\begin{equation}
d\widetilde{x}^{4}\equiv \rho (z)\text{ \ \ \ \ \ \ \ \ \ \ \ \ \ \ \ \ \ \
\ \ \ \ \ \ \ \ }d\widetilde{x}^{5}\equiv \rho ^{^{\prime }}(z)\text{\ \ \ \
\ \ \ \ \ \ ,\ \ \ }  \tag{34}
\end{equation}
\ $\ $where  $z$ \ is \ a complex variable and $\rho (z)$ is a complex
meromorphic function - the \textit{Weierstrass function}

\begin{equation}
\rho (z)\equiv \frac{1}{z^{2}}+\sum_{\varpi }\left[ \frac{1}{(z-\varpi )^{2}}%
-\frac{1}{\varpi ^{2}}\right] \text{ ,}  \tag{35}
\end{equation}
and the summation is over all non-null elements \hspace{0.5cm}
$$
\varpi \subset \Lambda =
\left\{(m \varpi _1 + n\varpi_2 )\mid m,n\subset Z \
{(\mbox{integer numbers}),}\,\, \varpi_1\,,\varpi_2\, \subset C, Im  > 0
\right\}.$$
Since further in the text the parametrization (34) will be repeatedly used,
it is instructive to give just an idea how in classical textbooks it is
proven that the parametrization (34) satisfies eq.(33). Let us take for
example the proof, given in Ref.13, where the basic idea is to compare the
Loran expansions for the non-positive degrees of $z$ for the function $\left[
\rho ^{^{\prime }}(z)\right] ^{2}$ and for the polynomial  $a\rho
^{3}(z)+b\rho ^{2}(z)+c\rho (z)+d$, where $a$, $b$, $c$ and $d$ are complex
numbers. If the corresponding coefficients in the Loran expansion of these
two expressions are equal, this would mean that the expressions themselves
are equal. Also, it should be accounted that the function $\left[ \rho
^{^{\prime }}(z)\right] ^{2}$ is an \textit{even one, }and consequently only
the even (non-positive) powers of $z$ in the Loran decomposition of the two
expressions should be taken into account. After performing the Loran \
decomposition, it becomes evident that equality of the two expressions is
possible only if $a=4,$ $b=0,$ $c=-60G_{4},$ $d=-140G_{6}$. \ Since these
coefficients give exactly the algebraic equation (33), it follows that the
Weierstrass function and its derivative (34) satisfy equation (33), thus
representing ''uniformization variables'' for the equation (33), i.e.
variables, which are functions of the complex variable $z$ and at the same
time satisfy the given algebraic relation.

It is important to stress that the ''tilda'' differentials $d\widetilde{x}%
^{4}$ and $d\widetilde{x}^{5},$ which are related through the algebraic
relation (33) and the parametrization (34) with the Weierstrass function, do
not result in any dependence between the original differentials $dx^{4}$ and
$dx^{5}$, which should remain independent since are related to the
independent coordinates in the gravitational Lagrangian. The reason for this
independence between the tilda and the non-tilda differentials is that the
linear-fractional transformation (31), which relates $d\widetilde{x}^{5}$
and $dx^{5}$, introduces an additional arbitrariness in the non-tilda
differentials due to the arbitrary complex functions $a$, $b$, $c$ and $d$.

In the present case, however, there are some specific facts about the
parametrization of the curve: 1. The parametrization shall be carried out
not in respect to two of the variables, but in respect to $\widetilde{dx}%
^{5} $ and another variable, which is $\frac{a}{c}$ - the two of the
parameters, entering the linear-fractional transformation. The rest of the
variables, entering the cubic algebraic equation shall be ''hidden'' in the
free term, which is a function. So actually the final result will be for $%
\widetilde{dx}^{5}$ (or $\frac{a}{c})$, expressed through the Weierstrass
function, but in order the parametrization to be consistent, the remaining
four differentials $(dx^{1},dx^{2},dx^{3},dx^{4})$ should be related to $a$,
$b$, $c$ and $d$ in a complicated way. 2. After performing the
transformation (31) with the purpose of chosing $a,b,c$ and $d$ in to
eliminate the highest (third) power of $\widetilde{dx}^{5},$the obtained
equation will be like equation (33), but with $g_{2}$ and $g_{3}-$ functions
and not complex numbers. On the other hand, the standard parametrization \
(34) with the Weiestrass function and its derivatives is valid \textit{only}
for $g_{2}$ and $g_{3}$ complex numbers. However, it will be proved in the
next sections, that in such a case the formalism and the parametrizable
equation can also be used. It will be shown that all the coefficient
functions (those standing before the pole terms) in the Loran expansion can
be found if the sums $G_{n}$ are known.

\section*{\protect\bigskip V. TRANSFORMED \ CUBIC \ EQUATION \\
WITH \ THE \ \ HELP \ OF  \ THE \\
LINEAR - FRACTIONAL \ \ TRANSFORMATION}
\bigskip In order to derive this equation, all the terms with $dx^{5}$ in
equation (20) shall be singled out and it can be written in the following way

\begin{equation}
A(dx^{5})^{3}+B(dx^{5})^{2}+C(dx^{5})+G^{(4)}(dx^{4},...dx^{1},g_{ij},\Gamma
_{ij}^{k},R_{ik})\equiv 0\text{,}  \tag{36}
\end{equation}
where $A$, $B$ and $C$ are the following functions, depending on $g_{ij}$, $%
\Gamma _{ij}^{k}$, $R_{ij}$ and the differentials $dx^{\alpha }$, $dx^{\beta
}$ ; the indices $\alpha ,\beta =1,2,3,4;r=1,2,...5.$

\begin{equation}
A\equiv 2p\Gamma _{55}^{r}g_{5r}  \tag{37}
\end{equation}
\begin{equation}
B\equiv 6p\Gamma _{\alpha 5}^{r}g_{5r}dx^{\alpha }  \tag{38}
\end{equation}
and

\begin{equation}
C\equiv -2R_{\alpha 5}dx^{\alpha }+2p(2\Gamma _{\alpha \beta
}^{r}g_{5r}+\Gamma _{5\alpha }^{r}g_{\beta r})dx^{\alpha }dx^{\beta }\text{.}
\tag{39}
\end{equation}
The function $G^{(4)}(...)$ is of the following form

\begin{equation}
G^{(4)}(dx^{4},...dx^{1},g_{ij},\Gamma _{ij}^{k},R_{ik})\equiv -R_{\alpha
\beta }dx^{\alpha }dx^{\beta }+pdx^{\gamma }dx^{\alpha }dx^{\beta }\Gamma
_{\gamma (\alpha }^{r}g_{\beta )r}\text{.}  \tag{40}
\end{equation}
In (40) the indice $\gamma =1,2,3,4$ and $(\alpha ,\beta )$ means
symmetrization in respect to the two indices. Further, after performing the
linear-fractional transformation (31), one easily obtains the new cubic
algebraic equation, written in terms of the new variables $\widetilde{dx^{5}}
$:

\begin{equation*}
(G^{(4)}c^{3}+aQ)(\widetilde{dx}^{5})^{3}+(bQ+aT+3c^{2}dG^{(4))}(\widetilde{%
dx}^{5})^{2}+
\end{equation*}
\begin{equation}
+(aS+bT+3cd^{2}G^{(4)})(\widetilde{dx}^{5})+(bS+G^{(4)}d^{3})\equiv 0\text{ ,%
}  \tag{41}
\end{equation}
where $Q,T,S$ denote the following expressions

\begin{equation}
Q\equiv Aa^{2}+Cc^{2}+Bac+2cdC\text{,}  \tag{42}
\end{equation}
\begin{equation}
T\equiv 2Aab+Bbc+Bad+2cdC\text{,}  \tag{43}
\end{equation}
\begin{equation}
S\equiv Ab^{2}+Bbd+Cd^{2}\text{.}  \tag{44}
\end{equation}
In fact, the linear-fractional transformation is performed with the purpose
of setting up to zero the expression before $(\widetilde{dx}^{5})^{3}$, from
where $G^{(4)}$ is expressed as

\begin{equation}
G^{(4)}=-\frac{aQ}{c^{3}}\text{.}  \tag{45}
\end{equation}
This equation is the \textbf{first} additional equation, which is imposed in
order to receive the parametrizable form of the cubic equation. Let us write
down in more details equation (45) , in order to understand its meaning.
Making use of the expressions for $G^{(4)}$ and $Q,$ it can be written in
the form again of a \textbf{cubic algebraic equation} in respect to the
remaining four differentials
\begin{equation}
p\Gamma _{\gamma (\alpha }^{r}g_{\beta )r}dx^{\gamma }dx^{\alpha }dx^{\beta
}+K_{\alpha \beta }^{(1)}dx^{\alpha }dx^{\beta }+K_{\alpha }^{(2)}dx^{\alpha
}+2p\left( \frac{a}{c}\right) ^{3}\Gamma _{55}^{r}g_{5r}=0\text{ \ \ \ , }
\tag{46}
\end{equation}
where $K_{\alpha \beta }^{(1)}$ and $K_{\alpha }^{(2)}$ are the
corresponding quantities
\begin{equation}
K_{\alpha \beta }^{(1)}\equiv -R_{\alpha \beta }+2p\frac{a}{c}(1+2\frac{d}{c}%
)(2\Gamma _{\alpha \beta }^{r}g_{5r}+\Gamma _{5\alpha }^{r}g_{\beta r})\text{
\ \ \ }  \tag{47}
\end{equation}
and
\begin{equation}
K_{\alpha }^{(2)}\equiv 2\frac{a}{c}\left[ 3p\frac{a}{c}\Gamma _{\alpha
5}^{r}g_{5r}-(1+2\frac{d}{c})R_{\alpha 5}\right] \text{ \ \ .}  \tag{48}
\end{equation}
The indices $\alpha ,\beta ,\gamma =1,2,3,4$ $($ but $r=1,2,....,5)$ and $%
(\alpha ,\beta )$ means symmetrization in respect to the two indices. In
other words, the imposed (''by hand'') equation (45) \textbf{simply fixes
the cubic algebraic equation in respect to the remaining four differentials,
if one would like to parametrize the differential of the fifth coordinate
with the Weierstrass function. No ratios }$\frac{a}{c}$ and $\frac{d}{c}$
are to be determined from this equation - later on from the equation in
respect to the fifth coordinate they will be determined.

Using expression (45), the functions standing before $(\widetilde{dx}%
^{5})^{2}$, $\widetilde{dx}^{5}$ in (41) and also the free term in the same
equation can be written in a form of an algebraic expression in respect to $%
\frac{a}{c},\frac{b}{c}$ and $\frac{b}{d}$

\begin{equation*}
bQ+aT+3c^{2}dG^{(4)}=d^{3}\{-3A\frac{a}{c}(\frac{a}{d})^{2}+C\frac{b}{d}(%
\frac{c}{d})^{2}+2C\frac{b}{d}\frac{c}{d}-6C\frac{a}{d}+
\end{equation*}

\begin{equation}
+3A\frac{b}{d}(\frac{a}{d})^{2}+B\frac{a}{d}\frac{b}{d}\frac{c}{d}-2B(\frac{a%
}{d})^{2}-C\frac{a}{d}\frac{c}{d}\}\text{,}  \tag{49}
\end{equation}
\begin{equation*}
aS+bT+3cd^{2}G^{(4)}=d^{3}\{-3A(\frac{a}{c})^{2}\frac{a}{d}+B\frac{c}{d}(%
\frac{b}{d})^{2}+2C\frac{c}{d}\frac{b}{d}-
\end{equation*}
\begin{equation}
-3B\frac{a}{c}\frac{a}{d}-6C\frac{a}{c}+2B\frac{a}{d}\frac{b}{d}+3A\frac{a}{d%
}(\frac{b}{d})^{2}-C\frac{a}{d}\}\text{,}  \tag{50}
\end{equation}
and

\begin{equation*}
bS+G^{(4)}d^{3}=d^{3}\{-A(\frac{a}{c})^{3}+A(\frac{b}{d})^{3}+B(\frac{b}{d}%
)^{2}-
\end{equation*}
\begin{equation}
-B(\frac{a}{c})^{2}+C\frac{b}{d}-C\frac{a}{c}-2C\frac{a}{c}\frac{d}{c}\}%
\text{.}  \tag{51}
\end{equation}
Let us now introduce the notations

\begin{equation}
\frac{a}{c}\equiv m\text{ \ \ \ \ \ \ \ \ \ \ \ \ \ \ \ \ \ \ \ \ \ \ \ }%
\widetilde{dx}^{5}\equiv n\text{ .}  \tag{52}
\end{equation}
Equations (49-51) shall be rewritten in such a way so that the terms with
powers of $m$ will be singled out. The rest of the terms will be denoted by $%
\overline{F}$, $\overline{M}$ and $\overline{N}$ and they will contain only
powers of $\frac{c}{d}$ and $\frac{b}{d}$ only. The transformed equations
(49-51), if substituted back into equation (41), allow one to write the
equation into the following form

\begin{equation*}
-3A(\frac{c}{d})^{2}m^{3}n^{2}-3A(\frac{c}{d})m^{3}n+[3A(\frac{c}{d})^{2}%
\frac{b}{d}-2B(\frac{c}{d})^{2}]m^{2}n^{2}-3B\frac{c}{d}m^{2}n+
\end{equation*}

\begin{equation*}
+[-6C\frac{c}{d}+B\frac{b}{d}(\frac{c}{d})^{2}-C(\frac{c}{d}%
)^{2}]mn^{2}+[-6C+2B\frac{c}{d}\frac{b}{d}+3A\frac{c}{d}(\frac{b}{d})^{2}-C%
\frac{c}{d}]mn+
\end{equation*}
\begin{equation}
+\overline{F}n^{2}+\overline{N}n+[\overline{M}-Am^{3}-Bm^{2}-Cm-2\frac{d}{c}%
Cm]\equiv 0\text{ .}  \tag{53}
\end{equation}
The terms $\overline{F}$, $\overline{M}$ and $\overline{N}$ have the
following form

\bigskip
\begin{equation}
\overline{F}\equiv C\frac{b}{d}(\frac{c}{d})^{2}+2C\frac{b}{d}\frac{c}{d}%
\text{ ,}  \tag{54}
\end{equation}
\begin{equation}
\overline{M}=A(\frac{b}{d})^{3}+B(\frac{b}{d})^{2}+C\frac{b}{d}\text{ ,}
\tag{55}
\end{equation}
\begin{equation}
\overline{N}=B\frac{c}{d}(\frac{b}{d})^{2}+2C\frac{c}{d}\frac{b}{d}\text{ .}
\tag{56}
\end{equation}
\ In other words, we have transformed the original third-rank algebraic
equation of five variables $dx^{1},dx^{2},dx^{3},dx^{4},dx^{5}$ into an
algebraic equation of two variables only ($m$ and $n$), but with a higher
rank (in the case it's five).

\section*{VI. A\ \ PROPOSAL \ FOR \ STANDARD \\
PARAMETRIZATION \ OF \ THE \ CUBIC \\
ALGEBRAIC \ EQUATION \ WITH \\
THE \ WEIERSTRASS \ FUNCTION}

By standard parametrization it shall be meant that the cubic algebraic
equation should be brought to its standard parametrizable form
\begin{equation}
\widetilde{n}^{2}=4m^{3}-g_{2}m-g_{3}\text{ ,}  \tag{57}
\end{equation}
where $g_{2}$ and $g_{3}$ are the already known complex numbers. Then one
has the right to set up
\begin{equation}
\widetilde{n}=\rho ^{^{\prime }}(z)=\frac{d\rho }{dz}\text{ \ \ \ \ \ \ \ \
\ \ \ \ \ \ \ \ \ \ \ \ \ \ \ \ \ \ \ \ \ \ \ \ \ \ \ \ \ \ \ }m=\rho (z)%
\text{\ \ \ \ .}  \tag{58}
\end{equation}
In order to obtain the parametrizable form (57), it is instructive to write
down the obtained algebraic equation in the form of a third-rank polynomial
of $m$ with coefficient functions $P_{1}(n)$ , $P_{2}(n)$, $P_{3}(n)$ and $%
P_{4}(n)$, representing quadratic forms of $n$ and at the same time cubic
algebraic expressions in respect to $\frac{c}{d}$ and $\frac{b}{d}:$

\begin{equation}
P_{1}(n)m^{3}+P_{2}(n)m^{2}+P_{3}(n)m+P_{4}(n)\equiv 0\text{ ,}  \tag{59}
\end{equation}
where

\begin{equation}
P_{1}(n)\equiv r_{1}n^{2}+r_{2}n+r_{3}=-3A(\frac{c}{d})^{2}n^{2}-3A\frac{c}{d%
}n-A\text{ ,}  \tag{60}
\end{equation}
\begin{equation}
P_{2}(n)\equiv q_{1}n^{2}+q_{2}n+q_{3}=[3A(\frac{c}{d})^{2}\frac{b}{d}-2B(%
\frac{c}{d})^{2}]n^{2}-3B\frac{c}{d}n-B\text{ ,}  \tag{61}
\end{equation}
\begin{equation*}
P_{3}(n)\equiv p_{1}n^{2}+p_{2}n+p_{3}=\left[ -6C\frac{c}{d}+B\frac{b}{d}%
\frac{c}{d}-C(\frac{c}{d})^{2}\right] n^{2}+
\end{equation*}

\begin{equation}
+\left[ -6C+2B\frac{c}{d}\frac{b}{d}+3A\frac{c}{d}(\frac{b}{d})^{2}-C\frac{c%
}{d}\right] n-C-2\frac{d}{c}C\text{ ,}  \tag{62}
\end{equation}
\begin{equation}
P_{4}(n)\equiv \overline{F}n^{2}+\overline{N}n+\overline{M}\text{ .}
\tag{63}
\end{equation}
Let us write down the last expression in the following form

\begin{equation}
P_{4}(n)\equiv \overline{F}\left[ (n+\frac{\overline{N}}{2\overline{F}})^{2}+%
\frac{\overline{M}}{\overline{F}}-\left( \frac{\overline{N}}{2\overline{F}}%
\right) ^{2}\right] \equiv \widetilde{n}^{2}+\overline{M}-\frac{\overline{N}%
^{2}}{4\overline{F}}\text{ ,}  \tag{64}
\end{equation}
where $\widetilde{n}$ denotes

\begin{equation}
\widetilde{n}\equiv \sqrt{\overline{F}}\left( n+\frac{\overline{N}}{2%
\overline{F}}\right) \text{ .}  \tag{65}
\end{equation}
In terms of $\widetilde{n}$, the transformed equation (59) can be written as

\begin{equation}
\widetilde{n}^{2}=\overline{P}_{1}(\widetilde{n})\text{ }m^{3}+\overline{P}%
_{2}(\widetilde{n})\text{ }m^{2}+\overline{P}_{3}(\widetilde{n})\text{ }m+%
\overline{P}_{4}(\widetilde{n})\text{ ,}  \tag{66}
\end{equation}
where the coefficient function $\overline{P}_{1}(\widetilde{n})$ is

\begin{equation*}
\overline{P_{1}}(\widetilde{n})\equiv \overline{r}_{1}\text{ }\widetilde{n}%
^{2}+\overline{r}_{2}\text{ }\widetilde{n}+\overline{r}_{3}=
\end{equation*}
\begin{equation}
=-\frac{r_{1}}{\overline{F}}\widetilde{n}^{2}+\left[ \frac{\overline{N}}{%
\overline{F}^{\frac{3}{2}}}r_{1}-\frac{r_{2}}{\overline{F}^{\frac{1}{2}}}%
\right] \widetilde{n}+\left[ -r_{1}\frac{\overline{N}^{2}}{4\overline{F}^{2}}%
+r_{2}\frac{\overline{N}}{2\overline{F}}-r_{3}\right]  \tag{67}
\end{equation}
and

\begin{equation}
\overline{P}_{4}(\widetilde{n})\equiv \frac{\overline{N}^{2}}{4\overline{F}}-%
\overline{M}\text{ .}  \tag{68}
\end{equation}
The other coefficient functions $\overline{P}_{2}(\widetilde{n})$ and $%
\overline{P}_{3}(\widetilde{n})$ can be written analogously, but with $%
(q_{1},q_{2},q_{3})$ and $(p_{1},p_{2},p_{3})$ in (67) instead of $%
(r_{1},r_{2},r_{3}).$ Note that unlike the expressions for $r,q$ and $p,$
representing \ cubic algebraic expressions in respect to $\frac{b}{d}$ and $%
\frac{c}{d}$, the corresponding ''bar'' quantities represent more
complicated expressions, which are no longer polynomials. It is also not
correct to consider the transformation from $(p,q,r)$ to $(\overline{p},%
\overline{q},\overline{r})$ as a linear affine transformation. The
expressions $\overline{N}$ and $\overline{F}$, entering the coefficient
functons of the transformation depend also on $\frac{b}{d}$ and $\frac{c}{d}%
, $so presumably they could also be expressed through $(p,q,r)$. The above
transformation shall be investigated further.

\textbf{Our purpose will be to identify the investigated equation (54) }$%
\widetilde{n}^{2}=\overline{P}_{1}(\widetilde{n})$ $m^{3}+\overline{P}_{2}(%
\widetilde{n})$ $m^{2}+\overline{P}_{3}(\widetilde{n})$ $m+\overline{P}_{4}(%
\widetilde{n})$\textbf{\ } \textbf{\ with equation (57) }$\widetilde{n}%
^{2}=4m^{3}-g_{2}m-g_{3}$\textbf{, for which we already know that the
substitution (58) can be performed.} In order to obtain the standard
parametrizable form of the cubic equation, one has to require that the two
equations are to be made equal, which means that the polynomials $\overline{P%
}_{1}(\widetilde{n})$ $,\overline{P}_{2}(\widetilde{n}),$ $\overline{P}_{3}(%
\widetilde{n})$ \ and $\overline{P}_{4}(\widetilde{n})$ (depending on the
variable $\widetilde{n})$ are to be made equal to the numerical coefficients
$4,0,-g_{2},$ and $-g_{3}$ respectively. Therefore, the following system of
equations should be fulfilled

\begin{equation}
4=\overline{r}_{1}\text{ }\widetilde{n}^{2}+\overline{r}_{2}\text{ }%
\widetilde{n}+\overline{r}_{3}\text{ ,}  \tag{69}
\end{equation}
\begin{equation}
0=\overline{q}_{1}\text{ }\widetilde{n}^{2}+\overline{q}_{2}\text{ }%
\widetilde{n}+\overline{q}_{3}\text{ ,}  \tag{70}
\end{equation}
\begin{equation}
-g_{2}=\overline{p}_{1}\text{ }\widetilde{n}^{2}+\overline{p}_{2}\text{ }%
\widetilde{n}+\overline{p}_{3}\text{ ,}  \tag{71}
\end{equation}
\begin{equation}
-g_{3}=\frac{\overline{N}^{2}}{4\overline{F}}-\overline{M}\text{ .}  \tag{72}
\end{equation}
The last equation (72)\ represents the \textbf{second} additional equation,
imposed in order to obtain the parametrizable form of the cubic equation.
Note that this equation has an extremely complicated structure: since $%
\overline{N},\overline{F}$ and $\overline{M}$ are \textbf{third-rank
polynomials in respect to }$\frac{b}{d}$ and $\frac{c}{d}$, the equation
will be of \textbf{sixth order}! This causes inconvenience in investigating
such equations, therefore it is appropriate to search another variables, in
terms of which the algebraic treatment will be comparatively more convenient.

Let us try to find such variables. From the first and the second equations
(70,71) the terms with $\widetilde{n}^{2}$ can be excluded, and also from
the second and the third equations. The obtained equations are

\begin{equation}
4\overline{q}_{1}=(\overline{r}_{2}\overline{q}_{1}-\overline{r}_{1}%
\overline{q}_{2})\widetilde{n}+(\overline{r}_{3}\overline{q}_{1}-\overline{r}%
_{1}\overline{q}_{3})\text{ ,}  \tag{73}
\end{equation}
\begin{equation}
-g_{2}\overline{q}_{1}=(\overline{p}_{2}\overline{q}_{1}-\overline{p}_{1}%
\overline{q}_{2})\widetilde{n}+(\overline{p}_{3}\overline{q}_{1}-\overline{p}%
_{1}\overline{q}_{3})\text{ .}  \tag{74}
\end{equation}
From the last two equations the terms with $\widetilde{n}$ can also be
excluded and a fourth-rank algebraic equation is obtained in respect to $%
p_{i},$ $q_{i}$ and $r_{i}$ ($i=1,2,3)$%
\begin{equation*}
(\overline{p}_{2}\overline{q}_{1}-\overline{p}_{1}\overline{q}_{2})(4%
\overline{q}_{1}-\overline{r}_{3}\overline{q}_{1}+\overline{r}_{1}\overline{q%
}_{3})+
\end{equation*}
\begin{equation}
+(\overline{r}_{2}\overline{q}_{1}-\overline{r}_{1}\overline{q}_{2})(g_{2}%
\overline{q}_{1}+\overline{p}_{3}\overline{q}_{1}-\overline{p}_{1}\overline{q%
}_{3})=0\text{ .}  \tag{75}
\end{equation}
The above equation represents the \textbf{third} additional equation,
imposed in order to obtain the parametrizable form of the cubic equation.
This equation is difficult to deal with, but there is a way to rewrite it in
a more convenient and simple form. Let us introduce the \textbf{''angular''
type} variables $l$ and $f$ with the corresponding components

\begin{equation*}
l=(l^{1},l^{2},l^{3})=(l_{12},l_{23},l_{31})=
\end{equation*}
\begin{equation}
=(\overline{p}_{1}\overline{q}_{2}-\overline{p}_{2}\overline{q}_{1},\text{ }%
\overline{p}_{2}\overline{q}_{3}-\overline{p}_{3}\overline{q}_{2},\text{ }%
\overline{p}_{3}\overline{q}_{1}-\overline{p}_{1}\overline{q}_{3})\text{ ,}
\tag{76}
\end{equation}
\begin{equation*}
f=(f^{1},f^{2},f^{3})=(f_{12},f_{23},f_{31})=
\end{equation*}
\begin{equation}
=(\overline{r}_{1}\overline{q}_{2}-\overline{r}_{2}\overline{q}_{1},\text{ }%
\overline{r}_{2}\overline{q}_{3}-\overline{r}_{3}\overline{q}_{2},\text{ }%
\overline{r}_{3}\overline{q}_{1}-\overline{r}_{1}\overline{q}_{3})\text{ .}
\tag{77}
\end{equation}
In terms of these variables, the fourth-rank algebraic equation (75) will be
reduced to the following quadratic equation
\begin{equation}
4\overline{q}_{1}l^{1}+g_{2}f^{1}\overline{q}_{1}+l^{1}f^{3}+f^{1}l^{3}=0%
\text{ .}  \tag{78}
\end{equation}
Having found the algebraic variety for $(\overline{q}%
_{1},l^{1},l^{3},f^{1},f^{3})$, one can go back to find the algebraic
variety for $(\overline{p},\overline{q},\overline{r})$. From there by means
of the inverse transformation of (67)
\begin{equation}
r_{1}=-\overline{F}\overline{r}_{1}\text{ \ \ \ ; \ \ \ \ \ \ \ \ \ \ }%
r_{2}=-\overline{F}^{\frac{1}{2}}\overline{r}_{2}-\overline{N}\overline{r}%
_{1}\text{\ \ }  \tag{79}
\end{equation}
\begin{equation}
r_{3}=-\frac{\overline{N}^{2}}{4\overline{F}}\overline{r}_{1}-\frac{%
\overline{N}}{2\overline{F}^{\frac{1}{2}}}\overline{r}_{2}-\overline{r}_{3}
\tag{80}
\end{equation}
(the same for $(p,q)$), one can obtain the ''non-bar'' variables $(p,q,r)$.
As already mentioned, the coefficient functions of the above transformation
depend in a complicated way on $\overline{N}$ and $\overline{F}$ and
therefore on $\frac{b}{d}$ and $\frac{c}{d}$. Therefore, if the dependence
of $\frac{b}{d}$ and $\frac{c}{d}$ on the ''bar'' variables is known, one
would have a well-determined transformation (although it is a nonpolynomial
and nonlinear one) from the ''non-bar'' variables to the ''bar'' variables $(%
\overline{p},\overline{q},\overline{r})$. That is why the purpose in the
next section will be to find this nonpolynomial transformation.

Finally, it may be noted that if the parametrization of (57) with the
Weierstrass function is performed and $\widetilde{n}=\rho ^{^{\prime }}(z)=%
\frac{d\rho }{dz}$ and $m=\rho (z)$, then the parametrized cubic equation
can be written in an integral form

\begin{equation}
\int \frac{d\rho (z)}{\sqrt{4\rho ^{3}-g_{2}\rho -g_{3}\rho }}=\int dz\text{
\ \ \ .}  \tag{81}
\end{equation}
The variable $z$ is a complex one and it may be different from the variables
$z_{1}$ and $z_{2}$, ''performing'' the complexification of the
four-dimensional manifold $(\alpha =1,..4)$, and the integration is
performed along some contour in the complex plane.

\section*{VII. FINDING \ THE \ NONLINEAR \\
AND \ NONPOLYNOMIAL \\
INVERTIBLE \ TRANSFORMATION}

\bigskip We shall start from expressions (60-62), from where one can find
\begin{equation}
r_{3}=-A\text{ \ \ \ \ \ \ \ \ \ \ \ \ \ \ \ \ \ \ }r_{2}=3\frac{c}{d}r_{3}%
\text{ \ \ \ \ \ \ \ \ \ \ \ \ \ \ \ \ \ \ \ }r_{1}=\frac{r_{2}^{2}}{3r_{3}}%
\text{ \ \ \ \ ,}  \tag{82}
\end{equation}

\begin{equation}
q_{3}=-B\text{ \ \ \ \ \ \ \ \ \ \ \ \ \ \ }q_{2}=3\frac{c}{d}q_{3}\text{ \
\ \ \ \ \ \ \ \ \ \ \ \ \ \ }q_{1}=-r_{2}\frac{q_{2}}{3q_{3}}\frac{b}{d}+2%
\frac{q_{2}^{2}}{9q_{3}}\text{ \ \ \ \ ,}  \tag{83}
\end{equation}
where it has been used that $\frac{q_{2}}{q_{3}}=\frac{r_{2}}{r_{3}}$. If
expressions (83) for $q=(q_{1},q_{2},q_{3})$ are substituted into the
defined by (67) expressions for $\overline{q}_{1},$ it can be obtained
\begin{equation}
\overline{q}_{1}=-\frac{q_{1}}{\overline{F}}=-\frac{1}{\overline{F}}\left(
\frac{c}{d}\right) ^{2}\left[ 2-3r_{3}\frac{b}{d}\right]  \tag{84}
\end{equation}
\begin{equation}
\overline{q}_{2}=\frac{\overline{N}}{\overline{F}^{\frac{3}{2}}}q_{1}-\frac{%
q_{2}}{\overline{F}^{\frac{1}{2}}}=\frac{\overline{N}}{\overline{F}^{\frac{3%
}{2}}}\left( \frac{c}{d}\right) ^{2}\left[ 2-3r_{3}\frac{b}{d}\right] -3q_{3}%
\frac{c}{d}\frac{1}{\overline{F}^{\frac{1}{2}}}  \tag{85}
\end{equation}
\begin{equation*}
\overline{q}_{3}=-q_{1}\frac{\overline{N}^{2}}{4\overline{F}^{2}}+q_{2}\frac{%
\overline{N}}{2\overline{F}}-q_{3}=
\end{equation*}
\begin{equation}
=q_{3}\left[ \frac{3\overline{N}}{2\overline{F}}\frac{c}{d}-1\right] +r_{3}%
\left[ \frac{3\overline{N}^{2}}{4\overline{F}^{2}}\left( \frac{c}{d}\right)
^{2}\frac{b}{d}\right] -\frac{\overline{N}^{2}}{\overline{F}^{2}}\left(
\frac{c}{d}\right) ^{2}\text{ \ \ \ \ \ .}  \tag{86}
\end{equation}
The corresponding equations for $r=(\overline{r}_{1},\overline{r}_{2},%
\overline{r}_{3})$ are
\begin{equation}
\overline{r}_{1}=-\frac{3r_{3}}{\overline{F}}\left( \frac{c}{d}\right) ^{2}%
\text{ \ \ \ \ \ \ ,\ \ \ \ \ }  \tag{87}
\end{equation}
\begin{equation}
\overline{r}_{2}=\frac{\overline{N}}{\overline{F}^{\frac{3}{2}}}3r_{3}\left(
\frac{c}{d}\right) ^{2}-\frac{3r_{3}\frac{c}{d}}{\overline{F}^{\frac{1}{2}}}%
\text{ \ \ \ ,}  \tag{88}
\end{equation}
\begin{equation}
\overline{r}_{3}=\overline{r}_{1}\left[ -\frac{\overline{N}}{2\frac{c}{d}}+%
\frac{\overline{N}^{2}}{4\overline{F}}+\frac{\overline{F}}{3\left( \frac{c}{d%
}\right) ^{2}}\right] \text{ \ \ .}  \tag{89}
\end{equation}
From the first and the second two equations it can be obtained respectively
\begin{equation}
\overline{N}=\frac{1}{\overline{r}_{1}}[-\overline{r}_{2}\overline{F}^{\frac{%
1}{2}}+\overline{F}\frac{1}{\frac{c}{d}}  \tag{90}
\end{equation}
and
\begin{equation}
\frac{(3+4\overline{r}_{1}^{2}-6\overline{r}_{1})}{12\overline{r}_{1}}Y^{2}+%
\frac{\left( \overline{r}_{1}-1\right) \overline{r}_{2}}{2\overline{r}_{1}}%
Y+(\frac{\overline{r}_{2}^{2}}{4\overline{r}_{1}^{2}}-\overline{r}_{3})=0%
\text{ \ \ \ \ \ \ \ \ ,}  \tag{91}
\end{equation}
where \ $Y\equiv \frac{\overline{F}^{\frac{1}{2}}}{\frac{c}{d}}$ . It is
important to note that $Y$ can be found as a solution \ of the above
quadratic equation with coefficient functions, which consist only of $%
\overline{r}$. Therefore
\begin{equation}
\frac{c}{d}=\frac{1}{Y}\overline{F}^{\frac{1}{2}}=Z\overline{F}^{\frac{1}{2}}%
\text{ \ \ \ \ \ \ \ \ \ \ \ \ }\overline{N}=O\overline{F}^{\frac{1}{2}}=%
\left[ \frac{1}{\overline{r}_{1}}(-\overline{r}_{2}+\frac{1}{Y})\right]
\overline{F}^{\frac{1}{2}}\text{ \ \ \ .}  \tag{92}
\end{equation}
Now let us write down the corresponding equations for $p$ from (62)
\begin{equation}
p_{1}=-6C\frac{c}{d}+B\frac{b}{d}\frac{c}{d}-C\left( \frac{c}{d}\right) ^{2}%
\text{ \ \ ,}  \tag{93}
\end{equation}
\begin{equation}
p_{2}=-6C+\left( \frac{c}{d}\right) \left[ 2B\frac{b}{d}+3A\left( \frac{b}{d}%
\right) ^{2}-C\right] \text{ \ \ \ \ \ \ \ ,}  \tag{94}
\end{equation}
\begin{equation}
p_{3}=-2C\frac{d}{c}-C\text{ \ \ \ .}  \tag{95}
\end{equation}
If from the last expression $C$ is expressed and is substituted into (93),
an expression for $\frac{b}{d}$ can be obtained in the form
\begin{equation}
\frac{b}{d}=-\frac{p_{1}}{q_{3}\overline{F}^{\frac{1}{2}}Z}+\frac{p_{3}Z%
\overline{F}^{\frac{1}{2}}[6+Z\overline{F}^{\frac{1}{2}}]}{q_{3}[2+Z%
\overline{F}^{\frac{1}{2}}]}\text{ \ ,}  \tag{96}
\end{equation}
where the derived expressions (92) have also been used. In order to obtain
an expression for $\frac{b}{d}$ in terms of the ''bar'' variables $\overline{%
p}=(p_{1},p_{2},p_{3})$, the ''non-bar'' variables $p_{1},p_{2}$ and $p_{3}$
should be expressed from the system of equations for $\overline{p}$ :
\begin{equation}
\overline{p}_{1}=-\frac{p_{1}}{\overline{F}}\text{ \ \ \ \ ;\ \ \ \ \ \ \ }%
\overline{p}_{2}=\text{\ }\frac{\overline{N}}{\overline{F}^{\frac{3}{2}}}%
p_{1}-\frac{p_{2}}{\overline{F}^{\frac{1}{2}}}\text{\ \ }  \tag{97}
\end{equation}
\begin{equation}
\overline{p}_{3}=-p_{1}\frac{\overline{N}^{2}}{4\overline{F}^{2}}+p_{2}\frac{%
\overline{N}}{2\overline{F}}-p_{3}  \tag{98}
\end{equation}
and substituted into (96). The result is
\begin{equation}
\frac{b}{d}=\frac{1}{q_{3}}\left[ \frac{\overline{F}^{\frac{1}{2}}}{Z}%
\overline{p}_{1}-\frac{(6+Z\overline{F}^{\frac{1}{2}})}{(2+Z\overline{F}^{%
\frac{1}{2}})}\left( \frac{O^{2}\overline{F}^{\frac{1}{2}}}{4}Z\overline{p}%
_{1}+\frac{O\overline{F}^{\frac{1}{2}}Z}{2}\overline{p}_{2}+Z\overline{F}^{%
\frac{1}{2}}\overline{p}_{3}\right) \right] \text{ \ \ .}  \tag{99}
\end{equation}
The only ''unbar' variable $q_{3}$ can be expressed from the first two
equations (84-85) for $\overline{q}_{1}$ and $\overline{q}_{2}$
\begin{equation}
q_{3}=-\frac{O\overline{q}_{1}+\overline{q}_{2}}{3Z}\text{ \ \ \ \ .}
\tag{100}
\end{equation}
Also, from the third equation (86) for $\overline{q}_{3}$ it can be obtained
\begin{equation}
3r_{3}\frac{b}{d}=\frac{8\left[ 6Z\overline{q}_{3}+3O^{2}Z^{3}+(3OZ-2)(O%
\overline{q}_{1}+\overline{q}_{2})\right] }{3O^{2}Z^{3}}  \tag{101}
\end{equation}
and from the first equatiom (3) the same expression can be found to be
\begin{equation}
3r_{3}\frac{b}{d}=\frac{2Z^{2}+\overline{q}_{1}}{Z^{2}}\text{ \ \ \ .}
\tag{102}
\end{equation}
From the equality of the above two formulaes one relation between the 'bar'
variables can be found. More concretely, since $O$ and $Z$ depend only on $%
\overline{r}$, the relation will concern how $\overline{q}_{3}$ can be
expressed through $O,Z$ and $\overline{q}_{1},\overline{q}_{2}.$ This will
not be used further in the text, since our main purpose will be to find the
ratio $\frac{r_{3}\overline{F}}{q_{3}^{2}}$, which is to be used in the
subsequent formulaes
\begin{equation}
\frac{r_{3}\overline{F}}{q_{3}^{2}}=-\frac{\overline{F}^{\frac{1}{2}}(2Z^{2}+%
\overline{q}_{1})(2+Z\overline{F}^{\frac{1}{2}})}{(O\overline{q}_{1}+%
\overline{q}_{2})K_{1}}\text{ \ \ \ ,}  \tag{103}
\end{equation}
where $K_{1}$ is the expression
\begin{equation}
K_{1}\equiv \overline{p}_{1}(2+Z\overline{F}^{\frac{1}{2}})-Z^{2}(6+Z%
\overline{F}^{\frac{1}{2}})(\frac{O^{2}\overline{p}_{1}}{4}+\frac{O}{2}%
\overline{p}_{2}+\overline{p}_{3})\text{ \ \ .}  \tag{104}
\end{equation}
At this moment the only equation not yet used is the one, which can be
derived from (93-95) for $p_{2}$
\begin{equation*}
p_{2}=\frac{2p_{3}\frac{c}{d}(6+\frac{c}{d})}{2+\frac{c}{d}}(-\frac{c}{d}+%
\frac{r_{3}}{q_{3}^{2}}3p_{1}+\frac{1}{2})+(2p_{1}-\frac{r_{3}}{q_{3}^{2}}%
\frac{3p_{1}^{2}}{\frac{c}{d}})-
\end{equation*}
\begin{equation}
-\frac{r_{3}}{q_{3}^{2}}\frac{3p_{3}^{2}\left( \frac{c}{d}\right) ^{3}(6+%
\frac{c}{d})^{2}}{(2+\frac{c}{d})^{2}}\text{ \ \ .}  \tag{105}
\end{equation}
If the $p$-variables are expressed from (97-98) through their ''bar'
counterparts and all preceeding expressions are used, the following \textbf{%
cubic algebraic equation} in respect to $\overline{F}^{\frac{1}{2}}\equiv T$
is obtained
\begin{equation}
N_{1}T^{3}+N_{2}T^{2}+N_{3}T+N_{4}=0\text{ \ \ \ ,}  \tag{106}
\end{equation}
where $N_{1},N_{2},N_{3}$ and $N_{4}$ are complicated expressions of the
''bar'' quantities only. These expressions will be presented in Appendix D.
Therefore, the roots of this cubic equation in respect to $T$ can be found
and consequently, the quantities $\overline{N},\overline{F},\overline{M}$,
entering the second additional equation (72) also can be expressed in terms
of the ''bar'' variables. This in fact proves that 1. The two additional
equations (72) and (78), imposed in order to obtain the parametrizable form
of the cubic equation, can be expressed in terms of the ''bar' variables
only. 2. The nonlinear and non-polynomial transformation from the $(r,q,p)$
to the $(\overline{r},\overline{q},\overline{p})$ variables is an invertible
one. This is an important fact, since one first may study the properties of
the algebraic equations, given by (72) and (78), and then chose the most
convenient form for the ratios $\frac{b}{d}$ and $\frac{c}{d}.$

\section*{VIII. PARAMETRIZATION \ OF \ A \ GENERAL \\
CUBIC \ CURVE \ WITH \ COEFFICIENT \\
FUNCTIONS OF\ A\ COMPLEX VARIABLE}

\noindent\ \ \ \ In this section an attempt will be made to deal with a
cubic curve of a more general kind

\begin{equation}
\widetilde{n}^{2}=M(z)m^{3}+N(z)m^{2}+P(z)m+E(z)\text{ ,}  \tag{107}
\end{equation}
where $M,N,P$ and $E$ are functions of the complex variable $z$ and
therefore \textbf{not} \textbf{complex numbers}, as usually accepted in
standard complex analyses$^{5}$ and algebraic geometry$^{13}$. In other
words, the main problem is whether it is possible to parametrize with the
Weierstrass function the above equation, i.e. when does the Weierstrass
function satisfy the equation
\begin{equation}
(\frac{d\rho }{dz})^{2}=M(z)\rho ^{3}+N(z)\rho ^{2}+P(z)\rho +E(z)\text{ ?}
\tag{108}
\end{equation}
As already briefly discussed in Sect. IV for the standard and usually
investigated case of $M$, $N$, $P$, $E$ - constants, the Weierstrass
function parametrizes the cubic equation (107) \textit{only if }$M=4$, $N=0$%
, $P=-60G_{4}$ and $Q=-140G_{6}$, but evidently in the present case of
\textbf{functions}, the situation will be quite different.

Let us first decompose $\rho (z)$ into an infinite sum, assuming that $\mid
\varpi \mid $ is a large number and therefore
\begin{equation*}
\rho (z)=\frac{1}{z^{2}}+\sum \left[ \frac{1}{\varpi ^{2}(\frac{z}{\varpi }%
-1)^{2}}-\frac{1}{\varpi ^{2}}\right] =
\end{equation*}
\begin{equation}
=\frac{1}{z^{2}}+\sum \frac{1}{\varpi ^{2}}(2\frac{z}{\varpi }+3(\frac{z}{%
\varpi })^{2}+...+(n+1)\frac{z^{n}}{\varpi ^{n}}+....)\text{ .}  \tag{109}
\end{equation}
The first derivative of the Weierstrass function is

\begin{equation}
\rho ^{^{\prime }}(z)=\frac{d\rho }{dz}=-\frac{2}{z^{3}}+\sum \frac{n(n+1)}{%
\varpi ^{2+n}}z^{n-1}  \tag{110}
\end{equation}
and its square degree is

\begin{equation}
\left[ \rho ^{^{\prime }}(z)\right] ^{2}=\frac{4}{z^{6}}-4\sum_{n=1}^{\infty
}\frac{n(n+1)}{\varpi ^{2+n}}z^{n-4}+\sum_{n=1}^{\infty }\frac{n^{2}(n+1)^{2}%
}{\varpi ^{2(n+2)}}z^{2(n-1)}\text{ .}  \tag{111}
\end{equation}
Note that in the strict mathematical sense, the second sum in the last
expression is in fact a double sum over $m$ and $n$ \ \ $\sum\limits_{m=1}^{%
\infty }\sum\limits_{n=1}^{\infty }\frac{mn(m+1)(n+1)}{\varpi ^{m+n}}%
z^{m+n-2}$, obtained as a result of the multiplication of the two infinite
sums (110) for $\rho ^{^{\prime }}(z)$ with different summation indices. Of
course, since the two sums are equal and infinite ones, the representation
in the form of a single sum is also correct. The appearence of the double
sum should be kept in mind, since the idea further will be to compare the
coefficient functions in the Loran power expansion of the functions on the
left- and on the right - hand sides of (108), and naturally a double sum
will appear in the R.H.S. of \ (108).

In reference to this, an important remark follows. Suppose one works in the
framework of \textit{standard arithmetical theory of elliptic functions},
when $M,N,P$ and $E$ are assumed to be just \textit{number coefficients}.
Since the Weierstrass functions $\rho (z)$ is an \textit{even function of
the complex variable }$z$ (see Ref.5 and Ref. 7), the whole expression on
the right-hand side (R.H.S.) of (108) will be an \textit{even} one too. On
the other hand, the function $\rho ^{^{\prime }}(z)$ in the left-hand side
(L.H.S.) of (108) is an \textit{odd} one, but its square again gives an
\textit{even function. }Therefore, comparing the coefficients in front of
the powers in $z$ means that only the \textit{even} powers should be
included in the infinite sum decomposition
\begin{equation}
\left[ \rho ^{^{\prime }}(z)\right] ^{2}=\frac{4}{z^{6}}-76G_{6}-24G_{4}%
\frac{1}{z^{2}}+...\text{ ,}  \tag{112}
\end{equation}
where $G_{n}$ will denote the following infinite sum of the complex pole
numbers
\begin{equation}
G_{n}=\sum \frac{1}{\varpi ^{n}}\text{ .}  \tag{113}
\end{equation}
\textit{Note also another very important fact the proof of which is given in
Ref.5: the infinite (in numbers of }$\varpi )$\textit{\ sum (113) is always
convergent (i.e.finite) when }$n>2$\textit{, but for }$n\leq 2$\textit{\ the
finiteness is not guaranteed! }$\ $In the presently investigated case of $%
M,N,P,E$ - complex functions, no information is available whether the R.H.S.
of (108) is an even or an odd function in $z$. Consequently, one should not
use formulae (112), but just start with the more general expression (111)
for $\left[ \rho ^{^{\prime }}(z)\right] ^{2}$.

In order to find the Loran decomposition \ of the functions on the
right-hand side of (108), one should first find the second and the third
powers of $\rho (z)$, which may be written as

\begin{equation}
\rho ^{2}(z)=\frac{1}{z^{4}}+2\sum_{n=1}^{\infty }(n+1)\frac{z^{n-2}}{\varpi
^{n}}+\sum_{n=1}^{\infty }(n+1)^{2}\frac{z^{2n}}{\varpi ^{2n}}\text{ \ ,}
\tag{114}
\end{equation}
\begin{equation*}
\rho ^{3}(z)=\frac{1}{z^{4}}+2\sum_{n=1}^{\infty }(n+1)\frac{z^{n-4}}{\varpi
^{n}}+\sum_{n=1}^{\infty }(n+1)^{2}\frac{z^{2n-2}}{\varpi ^{2n}}+
\end{equation*}
\begin{equation}
+\sum_{n=1}^{\infty }(n+1)\frac{z^{n-2}}{\varpi ^{n}}+2\sum_{n=1}^{\infty
}(n+1)\frac{z^{2n-2}}{\varpi ^{2n}}+\sum_{n=1}^{\infty }(n+1)^{2}\frac{z^{3n}%
}{\varpi ^{3n}}\text{ .}  \tag{115}
\end{equation}
Since these two expressions are to be multiplied by another infinite sums,
here in (114-115) we have retained the single-sum representaion.

The function $E(z)$ has the following Loran expansion around the zero point
\begin{equation}
E(z)=\sum_{m=-\infty }^{\infty }c_{m}^{(0)}z^{m}=\sum_{m=0}^{\infty
}a_{m}^{(0)}z^{m}+\sum_{m=1}^{\infty }\frac{b_{m}^{(0)}}{z^{m}}\text{ ,}
\tag{116}
\end{equation}
where $a_{m}^{(0)}$ and $b_{m}^{(0)}$ can be represented as integrals along
some contour in the complex plane ($w$ is a complex integration variable)

\begin{equation}
a_{m}^{(0)}=\frac{1}{2\pi i}\int \frac{E(w)}{w^{m-1}dw}\text{ \ \ \ \ \ \ \
\ \ \ \ \ \ \ \ }b_{m}^{(0)}=\frac{1}{2\pi i}\int E(w)w^{m-1}dw\text{ .}
\tag{117}
\end{equation}
The coefficient functions in the Loran expansion of the functions $N(z),P(z)$
and $Q(z)$ will be denoted respectively by $c_{m}^{(1)}$, $c_{m}^{(2)}$ and $%
c_{m}^{(3)}$. Each term of the expression for the right-hand side of (108)
is a product of two infinite sums, and the final result is

\begin{equation*}
M(z)\rho ^{3}+N(z)\rho ^{2}+P(z)\rho +E(z)=\sum_{m=-\infty }^{\infty
}\sum_{n=1}^{\infty }\{c_{m+4}^{(3)}+2(n+1)G_{n}c_{m+4-n}^{(3)}+
\end{equation*}
\begin{equation*}
+(n+1)^{2}G_{2n}c_{m+2-2n}^{(3)}+(n+1)G_{n}c_{m+2-n}^{(3)}+2(n+1)^{2}G_{2n}c_{m+2-2n}^{(3)}+
\end{equation*}
\begin{equation*}
+(n+1)^{3}G_{3n}c_{m-3n}^{(3)}+c_{m+4}^{(2)}+2(n+1)G_{n}c_{m-n+2}^{(2)}+
\end{equation*}
\begin{equation}
+(n+1)^{2}G_{2n}c_{m-2n}^{(2)}+c_{m+2}^{(1)}+c_{m-n}^{(1)}G_{n}+c_{m}^{(0)}%
\}z^{m}\text{ .}  \tag{118}
\end{equation}
In principle, the general case for an arbitrary $m$ may also be considered.
Then the above expression should be put equal to formulae (111) \ for $\left[
\rho ^{^{\prime }}(z)\right] ^{2},$ where in the first sum one should set up
$2(n-1)=m$ and in the second sum $n-4=m$. In the first sum in (111) the
summation will be over values of $m=0,2,4,6,...2k,.....$ and in the second
sum over $m=-3,-2,-1,0,1,2,......$. In such a case and for a given $n$, one
would have to consider a recurrent (in $n$) set of \textit{seven algebraic
equations} in respect to the \textit{four Loran expansion coefficients} $%
c_{n}^{(0)},c_{n}^{(1)},c_{n}^{(2)},c_{n}^{(3)}$ and for the seven values of
$m=-6,-3,-2,-1,0,$ $2k,$ $2k+1$ $(k=1,2...)$. Therefore, the system of
equations is predetermined, which enables one to find not only the the
unknown variables (the coefficient functions), but also certain relations
about the ''coefficient'' expressions, represented in the case by the sums $%
G_{n}$. This is an important moment, which shall be worked out further in
this paper, and indeed certain interesting relations will be found.
Moreover, since the summation over $m$ in the R.H.S. of (118) ranges from $%
-\infty $ to $+\infty $, terms with values of $m$, different from the above
written shall be present also in the R.H.S. of (81), but not in the L.H.S.
of (111). Therefore, \textit{two additional algebraic equations} may be
obtained by putting $m=-2k$ $(k\neq 0,1,3)$ $\ $and then $m=-(2k+1)$ $(k\neq
0,1)$ in the R.H.S. of (118) and then setting up the whole expression equal
to zero. In fact, effectively instead of two additional equations one may
have just one additional equation by putting $m=-k$ $(k>3$ and $k\neq 6)$,
so the total number of equations will be \textit{eight}. This complicated
calculation for the general case has not been performed in the present
paper, because due to considerable technical difficulties it would be
impossible to reconstruct analytically the whole set of \textit{Loran
coefficients} $c_{n}^{(0)},c_{n}^{(1)},c_{n}^{(2)},c_{n}^{(3)}$ as solutions
of the above system of \textit{eight algebraic equations}. However, the
calculation will be performed in Appendix A for the simplified case, which
will be described also below.

In this Section we shall restrict ourselves to the case of negative-power
expansion terms in the decomposition of $\left[ \rho ^{^{\prime }}(z)\right]
^{2}$, obtained for values of $m=-6,-2,0$, and the main motivation for this
is the analogy with the standard parametrization of the cubic curve. In the
next sections the case of positive-power expansion will be considered too.
Unfortunately, even under this additional assumption it is impossible to
resolve analytically the corresponding system of algebraic equations, if
some other simplifying assumption is not added. This assumption will be
given in the next Section.

The first recurrent relation for $m=-6$ is

\begin{equation*}
4=c_{-2}^{(3)}+2(n+1)G_{n}c_{-n-2}^{(3)}+(n+1)^{2}G_{2n}c_{-4-2n}^{(3)}+(n+1)G_{n}c_{-4-n}^{(3)}+
\end{equation*}
\begin{equation*}
+2(n+1)^{2}G_{2n}c_{-4-2n}^{(3)}+(n+1)^{3}G_{3n}c_{-6-3n}^{(3)}+c_{-2}^{(2)}+2(n+1)G_{n}c_{-n-4}^{(2)}+
\end{equation*}
\begin{equation}
+(n+1)^{2}G_{2n}c_{-6-2n}^{(2)}+c_{-4}^{(1)}+c_{-6-n}^{(1)}G_{n}+c_{-6}^{(0)}%
\text{ .}  \tag{119}
\end{equation}
For $m=-2$ the relation is

\begin{equation*}
-76G_{6}=c_{4}^{(3)}+2(n+1)G_{n}c_{4-n}^{(3)}+(n+1)^{2}G_{2n}c_{2-2n}^{(3)}+(n+1)G_{n}c_{2-n}^{(3)}+
\end{equation*}

\begin{equation*}
+2(n+1)^{2}G_{2n}c_{2-2n}^{(3)}+(n+1)^{3}G_{3n}c_{-3n}^{(3)}+c_{-4}^{(2)}+2(n+1)G_{n}c_{-n+2}^{(2)}+
\end{equation*}
\bigskip
\begin{equation}
+(n+1)^{2}G_{2n}c_{-2n}^{(2)}+c_{2}^{(1)}+c_{-n}^{(1)}G_{n}+c_{0}^{(0)}\text{
.}  \tag{120}
\end{equation}
The last relation for $m=0$ is

\begin{equation*}
-24G_{4}=c_{2}^{(3)}+2(n+1)G_{n}c_{2-n}^{(3)}+(n+1)^{2}G_{2n}c_{-2n}^{(3)}+(n+1)G_{n}c_{-n}^{(3)}+
\end{equation*}
\begin{equation*}
+2(n+1)^{2}G_{2n}c_{-2n}^{(3)}+(n+1)^{3}G_{3n}c_{-2-3n}^{(3)}+c_{2}^{(2)}+2(n+1)G_{n}c_{-n}^{(2)}+
\end{equation*}
\begin{equation}
+(n+1)^{2}G_{2n}c_{-2-2n}^{(2)}+c_{0}^{(1)}+c_{-2-n}^{(1)}G_{n}+c_{-2}^{(0)}%
\text{ .}  \tag{121}
\end{equation}
To avoid the possible confusion why $n$ appears in the R.H.S of (119-121)
but not in the L. H.S., let us remind that the left-hand sides for $\left[
\rho ^{^{\prime }}(z)\right] ^{2}$in these three equations have been
obtained by fixing both summation indices $n$ $(n=m)$ and also $m$ $%
(m=-6,-2,0)$, while in the right - hand sides only the indice $m$ is fixed
and \textit{the indice }$n$\textit{\ is left unfixed! }What will be
performed in the next Section will be for each value of \ $m=-6,-2,0$ to fix
in an appropriate way the possible values of $n$. Therefore more than three
algebraic equations will be obtained, in which there will be no summation
left.

From the above system of \textit{three recurrent algebraic equations}
(119-121), the infinite sequence of coefficient functions $%
c_{n}^{(0)},c_{n}^{(1)},c_{n}^{(2)}$ and $c_{n}^{(3)}$ should be found and
moreover, it should be proved that this sequence is convergent in the limit $%
n\rightarrow \pm \infty $. Still, because of the restriction to three values
of $m$ only, even if is possible to find $%
c_{n}^{(0)},c_{n}^{(1)},c_{n}^{(2)} $ $c_{n}^{(3)}$, it would not be correct
to assert that the Weierstrass function parametrizes an arbitrary cubic
curve with coefficient functions \ of a complex variable . This problem
probably may be resolved by means of computer simulations only.

In the next section the system of equations (119-121) shall be used for
parametrizing a more simplified cubic curve (without the quadratic in $\rho
(z)$ term).

\section*{IX. PARAMETRIZATION \ WITH \ THE \\
WEIERSTRASS \ FUNCTION \\
OF \ THE \ CUBIC \ CURVE $\ $ \\
$\left[ \protect\rho ^{^{\prime }}(z)\right] ^{2}=4\protect\rho
^{3}-g_{2}(z)\protect\rho -g_{3}(z)$}

\bigskip The form of the cubic curve is the same as the parametrizable cubic
curve in standard algebraic geometry,$^{5,10,11,13,24}$ but here it will be
with $g_{2}$ and $g_{3}$ - functions of a complex variable. The key problem,
which can be raised is: \textit{does there exist an algorithm for finding
out the sequence of coefficient functions in the Loran decomposition of }$%
g_{2}(z)$\textit{\ and }$g_{3}(z),$\textit{\ satisfying the above algebraic
equation, provided that its more simple form will result in the following
restrictions on the coefficient functions of the already considered general
cubic equation (108)}

\begin{equation}
M(z)=4=\sum_{m=-\infty }^{m=+\infty }c_{m}^{(3)}z^{m}\ \ \ \text{;}\ \ \ \ \
\ \ N(z)=\sum_{m=-\infty }^{m=+\infty }c_{m}^{(2)}z^{m}=0\text{ ,}  \tag{122}
\end{equation}
\begin{equation}
N(z)=-g_{2}(z)=-\sum_{m=-\infty }^{m=+\infty }c_{m}^{(1)}z^{m}\text{ \ ;\ \
\ \ \ \ \ \ \ }E(z)=\text{\ \ }-g_{3}(z)=-\sum_{m=-\infty }^{m=+\infty
}c_{m}^{(0)}z^{m}\text{\ \ .\ \ \ }  \tag{123}
\end{equation}
From the first sequence of equations one obtains for the coefficients $%
c_{m}^{(3)}$ and $c_{m}^{(2)\text{ }}$

\begin{equation}
c_{0}^{(3)}=4\text{ \ \ \ \ \ \ \ \ \ \ \ \ \ \ \ }c_{m}^{(2)}=0\text{ \ \ \
for all }m\text{ ,}  \tag{124}
\end{equation}
\begin{equation}
c_{m}^{(3)}=0\text{ \ \ \ \ \ \ for all \ \ \ }m\neq 0\text{ .}  \tag{125}
\end{equation}
Taking the above relations into consideration, equation (120) for $m=-2$ can
be written as

\begin{equation}
-24G_{4}=2(n+1)G_{n}c_{2-n}^{(3)}-c_{0}^{(1)}-G_{n}c_{-2-n}^{(1)}-c_{-2}^{(0)}%
\text{ .}  \tag{126}
\end{equation}
For values of $n=2$ and $n=1$ from the above equation the following
equations are obtained
\begin{equation}
-24G_{4}=24G_{2}-c_{0}^{(1)}-c_{0}^{(1)}-c_{-4}^{(1)}G_{2}-c_{-2}^{(0)}\text{
,}  \tag{127}
\end{equation}
\begin{equation}
24G_{4}=c_{0}^{(1)}+G_{1}c_{-3}^{(1)}+c_{-2}^{(0)}\text{ .}  \tag{128}
\end{equation}
From the above two equations $c_{-4}^{(1)}$ and $c_{-3}^{(1)}$ can be found

\begin{equation}
c_{-4}^{(1)}=\frac{1}{G_{2}}\left[ 24G_{4}+24G_{2}-c_{0}^{(1)}-c_{-2}^{(0)}%
\right] \text{ ,}  \tag{129}
\end{equation}
\begin{equation}
c_{-3}^{(1)}=\frac{1}{G_{1}}\left[ c_{0}^{(1)}+c_{-2}^{(0)}-24G_{4}\right]
\text{ . }  \tag{130}
\end{equation}
From (89), the general recurrent relation for $n=p>2$ can be obtained

\begin{equation}
c_{-2-p}^{(1)}=\frac{1}{G_{p}}\left[ 24G_{4}-c_{0}^{(1)}-c_{-2}^{(0)}\right]
\text{ .}  \tag{131}
\end{equation}
It is clear that for the determination of $c_{-4}^{(1)},c_{-3}^{(1)}$ and $%
c_{-2-p}^{(1)}$ one has to know $c_{-2}^{(0)}$ and $c_{0}^{(1)}$. There is,
however, one exception - in (129) $G_{2}=\sum \frac{1}{\varpi ^{2}}$ may be
a divergent sum, so then one has $c_{-4}^{(1)}=24$ (since $G_{2}$ is in the
denominator, when $G_{2}\rightarrow \infty $, the corresponding part of the
expression will tend to zero).

Further, for $m=0$ and keeping in mind (124-125), equation (121) will give

\begin{equation*}
-76G_{6}=2(n+1)G_{n}c_{4-n}^{(3)}+(n+1)^{2}G_{2n}c_{2-2n}^{(3)}+(n+1)G_{n}c_{2-n}^{(3)}+
\end{equation*}
\begin{equation}
+2(n+1)^{2}G_{2n}c_{2-2n}^{(3)}-c_{2}^{(1)}-c_{-n}^{(1)}G_{n}-c_{0}^{(0)}%
\text{ .}  \tag{132}
\end{equation}
For values of $n=4,2,1,$ when there are non-vanishing values among the
coefficients $c_{m}^{(3)},$ the corresponding equations are
\begin{equation}
-76G_{6}=40G_{4}-c_{2}^{(1)}-c_{-4}^{(1)}G_{4}-c_{0}^{(0)}\text{ ,}
\tag{133}
\end{equation}
\begin{equation}
-76G_{6}=12G_{2}-c_{2}^{(1)}-c_{-2}^{(1)}G_{2}-c_{0}^{(0)}\text{ ,}
\tag{134}
\end{equation}
\begin{equation}
-76G_{6}=48G_{2}-c_{2}^{(1)}-c_{-1}^{(1)}G_{1}-c_{0}^{(0)}\text{ .}
\tag{135}
\end{equation}
The above linear algebraic equations can be solved trivially linear to find
the coefficients $c_{-4}^{(1)}$, $c_{-2}^{(1)}$ and $c_{2}^{(1)}$, which
depend on $c_{-1}^{(1)}$ and $c_{0}^{(0)}$%
\begin{equation}
c_{-4}^{(1)}=\frac{1}{G_{4}}[c_{-1}^{(1)}G_{1}-48G_{2}+40G_{4}]\text{ ,}
\tag{136}
\end{equation}
\begin{equation}
c_{-2}^{(1)}=-36+\frac{G_{1}}{G_{2}}c_{-1}^{(1)}\text{ ,}  \tag{137}
\end{equation}
\begin{equation}
c_{2}^{(1)}=76G_{6}+48G_{2}-c_{0}^{(0)}-c_{-1}^{(1)}G_{1}\text{ .}  \tag{138}
\end{equation}
Note that these coefficients can be divergent if $G_{2}$ and $G_{1}$ are
divergent. Taking into account equation (138) and also (125) for the case $%
m=0$, but for a general value of $n=p\neq 1,2,4$, an expression for $%
c_{-k}^{(1)}$ can easily be found

\begin{equation}
c_{-k}^{(1)}=\frac{1}{G_{k}}(-48G_{2}+c_{-1}^{(1)}G_{1})\text{ .}  \tag{139}
\end{equation}
This formulae should be compared to the previously derived formulae (131),
setting up $\ -2-p=-k$. From the two expressions $c_{-2}^{(0)}$ can be
expressed
\begin{equation}
c_{-2}^{(0)}=24G_{4}-c_{0}^{(1)}-\frac{G_{k-2}}{G_{k}}%
(-48G_{2}+c_{-1}^{(1)}G_{1})\text{ .}  \tag{140}
\end{equation}
However, $c_{-2\text{ }}^{(0)}$can be expressed also from the two formulaes
(129) and (136) for $c_{-4}^{(1)}:$%
\begin{equation}
c_{-2}^{(0)}=24G_{4}+24G_{2}-c_{0}^{(1)}+\frac{G_{2}}{G_{4}}%
(48G_{2}-c_{-1}^{(1)}G_{1}-40G_{4})\text{ .}  \tag{141}
\end{equation}
Comparing (140-141), an expression for $c_{-1}^{(1)}$ can be found, \textit{%
which does not depend on any Loran coefficient functions}
\begin{equation}
c_{-1}^{(1)}=\frac{16G_{2}}{G_{1}}\frac{%
[3G_{2}G_{k}-G_{4}G_{k}-3G_{4}G_{k-2]}}{[G_{2}G_{k}-G_{4}G_{k-2}]}\text{ .}
\tag{142}
\end{equation}
Substituting this expression into the formulae (139) for $c_{-k}^{(1)}$, one
obtains the convergent expression $(k>2,k\neq 4)$

\begin{equation}
c_{-k}^{(1)}=-\frac{16G_{4}}{G_{k}-\frac{G_{4}}{G_{2}}G_{k-2}}\text{ . }
\tag{143}
\end{equation}
The obtained expression (142) for $c_{-1}^{(1)}$ can be substituted into
(140) to find a formulae for $c_{-2}^{(0)},$ which will depend on $G_{k}$
and only on the Loran coeffficient function $c_{0}^{(1)}$

\begin{equation}
c_{-2}^{(0)}=-c_{0}^{(1)}+24G_{4}+\frac{16G_{k-2}}{\frac{G_{k}}{G_{4}}-\frac{%
G_{k-2}}{G_{2}}}\text{ .}  \tag{144}
\end{equation}
The above expression is well defined also when $G_{2}\rightarrow \infty .$
It shall be proved subsequently that such a case will turn out to be
impossible.

Further, from (144) and expression (130)\ for $c_{-3}^{(1)}$ it follows

\begin{equation}
c_{-3}^{(1)}=\frac{G_{4}}{G_{2}}\frac{16G_{k-2}}{\left[ G_{k}-\frac{G_{4}}{%
G_{2}}G_{k-2}\right] }\text{ .}  \tag{145}
\end{equation}
But since $k$ in expression (143) can take a value $k=3$, it follows also

\begin{equation}
c_{-3}^{(1)}=-\frac{16G_{4}}{G_{3}-\frac{G_{4}}{G_{2}}G_{1}}\text{ .}
\tag{146}
\end{equation}
The comparison of the two expressions gives the following formulae for the
infinite sum $G_{k}:$%
\begin{equation*}
G_{k}=\gamma G_{k-2}=\gamma ^{s}G_{k-2s}=.......=\gamma ^{\frac{2p-1}{2}%
}G_{1}\text{ for }k=2p
\end{equation*}
\begin{equation}
=\gamma ^{p}G_{1}\text{ for }k=2p+1\text{ ,}  \tag{147}
\end{equation}
where
\begin{equation}
\gamma =2\frac{G_{4}}{G_{2}}-\frac{G_{3}}{G_{1}}\text{ .}  \tag{148}
\end{equation}
Another recurrent relation for $G_{k}$ can be found also from (145)
\begin{equation}
G_{k}=\frac{(G_{k-2})^{2}}{G_{k-4}}=\frac{(G_{k-4})^{3}}{(G_{k-6})^{2}}=...=%
\frac{(G_{k-2s})^{s+1}}{(G_{k-2s-2)^{2}}}\text{ . }  \tag{149}
\end{equation}
This formulae for values of $k=2p$ and $k=2p+1$, combined with the previous
formulae (147), allows one to find an expression for $G_{1}:$%
\begin{equation}
G_{1}=\frac{G_{3}^{\frac{2p-1}{2p-3}}}{G_{2p}^{\frac{2}{2p-3}}}=\frac{G_{3}^{%
\frac{p}{p-1}}}{G_{2p+1}^{\frac{1}{p-1}}}\text{ .}  \tag{150}
\end{equation}
\textit{The last formulae is interesting, because it shows that the
divergent in the general case quantity }$G_{1}$\textit{\ in the present case
is expressed through convergent quantities only - }$G_{3}$\textit{\ and }$%
G_{2p+1}(p$\textit{\ is of course a finite number!)}. Substituting (150)
into (149), one can get expressions for $G_{2p}$ and $G_{2p+1}:$%
\begin{equation}
G_{2p}=G_{3}\left( \frac{G_{4}}{G_{2}}\right) ^{\frac{2p-3}{2}}\text{ \ ; \
\ \ \ \ \ \ \ \ \ \ }G_{2p+1}=G_{3}\left( \frac{G_{4}}{G_{2}}\right) ^{p-1}%
\text{ .}  \tag{151}
\end{equation}
It is seen that $G_{2}$ is also expressed through convergent quantities.

\bigskip\ From equation (130) for $m=-6$ one obtains

\begin{equation}
4=c_{-4}^{(1)}+c_{-6-n}^{(1)}G_{n}+c_{-6}^{(0)}  \tag{152}
\end{equation}
($n=1,.....\infty )$. Since $c_{-4}^{(1)}$ and $c_{-6-n}^{(1)}$ can be
found, $c_{-6}^{(0)}$ can also be determined. It is clear that among the
coefficients $c_{m}^{(0)}$ two of them - $c_{-6}^{(0)}$ $\ $\ and $%
c_{-2}^{(0)}$ can be determined \ from (143). The other coefficients will be
determined \ in the Appendixes.

Let us summarize the obtained results in this Section and in Appendixes A, B
and C by formulating the following

\begin{proposition}
Let $g_{2}(z)$ and $g_{3}(z)$ are functions of a complex variable, which
have a Loran function decomposition $g_{2}(z)=\sum\limits_{m=-\infty
}^{\infty }c_{m}^{(1)}z^{m}$ and $g_{3}(z)=\sum\limits_{m=-\infty }^{\infty
}c_{m}^{(0)}z^{m}$ and satisfy the algebraic equation $\left[ \rho
^{^{\prime }}(z)\right] ^{2}=4\rho ^{3}-g_{2}(z)\rho -g_{3}(z),$ where $\rho
(z)$ is the Weierstrass (elliptic) function. Then the following statements
represent (only) necessary conditions for the fulfillment of the above
equation:

1. The poles of the Weierstrass function (even if they are infinite in
number), must be situated in such a way so that the sums $G_{1}=\sum \frac{1%
}{\varpi }$ and $G_{2}=\sum \frac{1}{\varpi ^{2}}$ are convergent (i.e. \
finite). The sum $G_{1}$ can be expressed through formulae (150).

2. All the coefficients $c_{m}^{(1)}$ and $c_{m}^{(0)}$ in the Loran
positive- and negative- power expansion can be expressed uniquely from the
finite sums $G_{n}$.

3. The sum $G_{1}$ is proportional to the sum $G_{3}$ with a coefficient of
proportionality, equal to the ratio of the sums $G_{2}$ and $G_{4}$, i.e $%
G_{1}=\frac{G_{3}}{G_{4}}G_{2}$ (from A22). This formulae follows also from
the more general one $G_{2p+1}=G_{3}(\frac{G_{4}}{G_{2}})^{p-1}$ (151) for $%
p=0$.

4. As a consequence from the above relation and formulaes (147-148), the sum
$G_{2}$ can be uniquely expressed as $G_{2}=\sqrt{G_{1}G_{3}\text{ }}$.

5. All the even-number sums $G_{6},G_{8},G_{10}....$equal to zero.

6. The following relation is fulfilled $G_{k}^{2}=20^{2}\frac{G_{k-2}G_{k+2}%
}{G_{2(k+1)}}$, which can be obtained from (B15) and (B16). In order this
relation to comply with statement 5, additionally one should have that $%
G_{5},G_{7},G_{13},G_{15},G_{21},G_{23}..$ should be zero. However, $G_{9},$
$G_{11},G_{17},G_{19}$ are different from zero.
\end{proposition}

Finally, with the help of (152), a check can also be made for the
consistency of the obtained results. Substracting the two equations (152)
for values of $\ n$ and $n+l$, one obtains

\begin{equation}
\frac{G_{n}}{G_{n+6}-\frac{G_{4}}{G_{2}}G_{n+4}}-\frac{G_{n+l}}{G_{n+6+l}-%
\frac{G_{4}}{G_{2}}G_{n+4+l}}=0  \tag{153}
\end{equation}
(it's more appropriate to divide everywhere by $G_{n}$). From (151) for
values of $n=2p$ and $l=2q$, for example, it can be found

\begin{equation}
\frac{G_{n+l}}{G_{n}}=\left( \frac{G_{4}}{G_{2}}\right) ^{2q}\text{ .}
\tag{154}
\end{equation}
For other combinations (even and odd) of $n$ and $l$ the calculation is
similar. Using the above formulae, it can easily be verified that equation
(153) is \textit{identically satisfied}. This confirms that the investigated
in this paragraph system of equations gives consistent and noncontradictory
results.

\section*{X. POSITIVE - POWER \ TERMS \ IN \ THE \\
INFINITE \ SUM \ DECOMPOSITION \ OF \\
EQUATION  $\left[ \protect\rho ^{^{\prime }}(z)\right] ^{2}=4%
\protect\rho ^{3}-g_{2}(z)\protect\rho -g_{3}(z)-$\\
-THE \ CASE \ OF \ POLES \ NOT \ AT \ INFINITY}

\bigskip For the purpose, the expansion (111) for $\left[ \rho ^{^{\prime
}}(z)\right] ^{2}$ has to be used, and a change in the summation indices is
performed so that positive-power terms starting from $n=1$ (without the \
free term) are taken into account
\begin{equation}
\left[ \rho ^{^{\prime }}(z)\right] ^{2}=\sum_{n=1}^{\infty
}(n+1)^{2}(n+2)^{2}G_{2(n+3)}\text{ }z^{2n}-4\sum_{n=1}^{\infty
}(n+4)(n+5)G_{n+6}z^{n}\text{ .}  \tag{155}
\end{equation}
Using the Loran decomposition $g_{2}(z)=\sum\limits_{m=-\infty }^{\infty
}c_{m}^{(1)}z^{m}$, it can be obtained for $-g_{2}(z)\rho (z)$ for the
positive terms only
\begin{equation}
-g_{2}(z)\rho (z)=-\sum_{n=1}^{\infty }c_{n+2}^{(1)}z^{n}-\sum_{n=1}^{\infty
}\sum_{k=-\infty }^{0}(n-k+1)G_{n-k+2}c_{k}^{(1)}z^{n}\text{ .}  \tag{156}
\end{equation}
By means of the relevent expressions for $\rho ^{3}(z)$ from (115) and for $%
g_{3}(z)=\sum\limits_{m=-\infty }^{\infty }c_{m}^{(0)}z^{m}$, the following
expression can be obtained for the positive - power terms of
\begin{equation*}
0=\left[ \rho ^{^{\prime }}(z)\right] ^{2}-4\rho ^{3}+g_{2}(z)\rho +g_{3}(z)=
\end{equation*}
\begin{equation}
=\sum_{n=1}^{\infty }\left[ A_{0}(n,G)z^{n}+A_{1}(n,G)z^{2n}+A_{2}(n,G)z^{3n}%
\right] \text{ ,}  \tag{157}
\end{equation}
where $A_{0}(n,G)$, $A_{1}(n,G)$ and $A_{2}(n,G)$ are the following
expressions
\begin{equation*}
A_{0}(n,G)=8(n+5)G_{n+4}+4(n+3)G_{n+2}+4(n+4)(n+5)G_{n+6}-
\end{equation*}

\bigskip
\begin{equation}
-c_{n+2}^{(1)}-c_{n}^{(0)}-\sum_{k=0}^{\infty }(n+k+1)G_{n+k+2}c_{-k}^{(1)}%
\text{ ,}  \tag{158}
\end{equation}
\begin{equation}
A_{1}(n,G)=4(n+2)^{2}G_{2(n+1)}+8(n+2)G_{2(n+1)}-(n+1)^{2}(n+2)^{2}G_{2(n+3)}%
\text{ ,}  \tag{159}
\end{equation}
\begin{equation}
A_{2}(n,G)=4(n+1)^{3}G_{3n}\text{ .}  \tag{160}
\end{equation}
Two important observations can be made from these expressions.

\textit{First}, the unknown coefficient functions $c_{n+2}^{(1)}$, $%
c_{n}^{(0)}$ and $c_{-k}^{(1)}$ are singled out only in $A_{0}(n,G).$
Therefore, it is more appropriate to write down \ (157) in the form
\begin{equation}
\sum_{n=1}^{\infty }A_{0}(n,G)z^{n}=-\sum_{n=1}^{\infty
}A_{1}(n,G)z^{2n}-\sum_{n=1}^{\infty }A_{2}(n,G)z^{3n}\text{ .}  \tag{161}
\end{equation}
The above mentioned coefficient functions can be determined in such a way so
that the characteristics of the infinite sum on the left-hand side (L.H.S)
would correspond to the characteristics of the infinite sum on R.H.S. Such a
characteristic would be for example the ''\textit{convergency radius}'',
defined in standard complex analyses (for the infinite sum on the L.H.S.) as

\begin{equation}
R_{0}=lim_{n\rightarrow \infty }\mid A_{0}(n,G\mid ^{-\frac{1}{n}}\text{ .}
\tag{162}
\end{equation}

\textit{Second}, since the infinite sums in (158-160) are finite, it can be
seen that

\begin{equation}
lim_{n\rightarrow \infty }\mid A_{1}(n,G\mid ^{-\frac{1}{n}%
}=lim_{n\rightarrow \infty }\mid A_{2}(n,G\mid ^{-\frac{1}{n}}=\infty \text{
.}  \tag{163}
\end{equation}
Let us find the convergency radius of the first infinite sum on the R.H.S.
of (158)
\begin{equation*}
R_{1}=lim_{n\rightarrow \infty }\mid A_{1}(n,G\mid ^{-\frac{1}{n}%
}=lim_{n\rightarrow \infty }exp\{-\frac{ln\mid A_{1}(n,G)\mid }{n}\}=
\end{equation*}
\begin{equation}
=exp\{-lim_{n\rightarrow \infty }\frac{\frac{d\mid A_{1}\mid }{dn}}{\mid
A_{1}\mid }\}  \tag{164}
\end{equation}
(the Lopital's rule has been used, since $n\rightarrow \infty $ and $\mid
A_{1}(n,G)\mid \rightarrow \infty $. It can be found that
\begin{equation*}
\frac{d\mid A_{1}(n,G)\mid }{dn}=8(n+2)G_{2(n+1)}+8G_{2(n+1)}+
\end{equation*}
\begin{equation}
+8G_{2(n+1)}-2(n+1)(n+2)(2n+3)G_{2(n+3)}\text{ .}  \tag{165}
\end{equation}
This expression is a third-rank polynomial in $n$, while $\mid A_{1}\mid $
is a second-rank polynomial. Therefore the ratio $\frac{\frac{d\mid
A_{1}\mid }{dn}}{\mid A_{1}\mid }$ will evidently tend to infinity when $%
n\rightarrow \infty $ and consequently

\begin{equation}
R_{1}=exp[-\infty ]=0\text{ .}  \tag{166}
\end{equation}
The same can also be proved in an analogous way for $R_{2}.$

Now, since on the R.H.S. of (161) one has an infinite sum with a zero
convergency radius, it's natural to suppose that the same holds also for the
L.H.S. A simple calculation shows that this may happen only if
\begin{equation*}
lim_{n\rightarrow \infty }\frac{d\mid A_{0}(n,G)\mid }{dn}=
\end{equation*}
\begin{equation}
=lim_{n\rightarrow \infty }\frac{\left[ \frac{dc_{n+2}^{(2)}}{dn}+\frac{%
dc_{n}^{(3)}}{dn}\frac{1}{4G_{n+6}}+\sum G_{n+k+2}\text{ }c_{-k}^{(2)}\right]
}{n^{2}}=\infty \text{ .}  \tag{167}
\end{equation}
This means that either $c_{n+2}^{(1)},$ or $c_{n}^{(0)}$, or $c_{-k}^{(1)}$ $%
\ $\ have to be proportional to $\frac{n^{l}}{(l+1)}$, where $l>2.$ In such
a way, we acquired information how the coefficient functions $c_{n}^{(1)%
\text{ }}$behave both in the positive-power $(n>0)$ and the negative power
decomposition $(n<0)$.

\section*{XI. POLES \  AT \ INFINITY  $\left( \protect\varpi \rightarrow
\infty \right) $ IN \ THE \\
 POSITIVE - POWER \ DECOMPOSITION \\
OF \ $\left[\protect\rho ^{^{\prime }}(z)\right] ^{2}=4\protect\rho ^{3}-g_{2}(z)\protect%
\rho -g_{3}(z)$}

\bigskip If \ the period of the Weierstrass function can be represented as $%
\varpi =q\varpi _{1}+p$, then in the limit $\varpi \rightarrow \infty $ one
has
\begin{equation}
G_{n}(\varpi )=\sum \frac{1}{\varpi ^{n}}=\sum \frac{1}{(q\varpi _{1}+p)^{n}}%
\rightarrow _{\varpi \rightarrow \infty }\sum_{p\neq 0}\frac{1}{p^{n}}%
=2\zeta (n)\text{ ,}  \tag{168}
\end{equation}
where $\zeta (n)$ denotes the Riemann zeta-function. Therefore, in the
asymptotic limit $n\rightarrow \infty $ (when this limit is used, as in the
preceeding section), one should also take into account the asymptotic limit
of the zeta-function as compared to the other power-like terms of $n.$

For the present particular case, let us calculate for example $R_{2}.$ It
can be found that
\begin{equation}
lim_{n\rightarrow \infty }\frac{\frac{d\mid A_{2}\mid }{dn}}{\mid A_{2}\mid }%
=lim_{n\rightarrow \infty }\frac{3}{n+1}\frac{\mid \zeta (3n)-n(n+1)\zeta
(3n+1)\mid }{\zeta (3n)}=\infty \text{ ,}  \tag{169}
\end{equation}
and consequently,
\begin{equation}
R_{2}=lim_{n\rightarrow \infty }exp\{-\frac{ln\mid A_{2}\mid }{n}%
\}=lim_{n\rightarrow \infty }exp\{-\infty \}=0\text{ .}  \tag{170}
\end{equation}
After performing similar calculations, the same result can also be obtained
for $R_{1}$ .

Let us find now $R_{0}.$ Some more lenghty calculations will give
\begin{equation}
lim_{n\rightarrow \infty }\frac{\frac{d\mid A_{0}\mid }{dn}}{\mid A_{0}\mid }%
=lim_{n\rightarrow \infty }\frac{\mid C^{(1)}(n,k,p)\mid }{\mid
C^{(2)}(n,k,p)\mid }\text{ ,}  \tag{171}
\end{equation}
where $C^{(1)}(n,k,p)$ and $C^{(2)}(n,k,p)$ will be the following
expressions
\begin{equation*}
C^{(1)}(n,k,p)=-\frac{1}{4(n+4)(n+5)(n+6)}\{\frac{d}{dn}[%
c_{n+2}^{(1)}+c_{n}^{(0)}+\sum_{k=0}^{\infty }\zeta (n+k+2)c_{-k}^{(1)}]+
\end{equation*}
\begin{equation}
+\zeta (n+7)+\sum_{k=0}^{\infty }(n+k+1)\left[ \zeta (n+k+2)\frac{%
dc_{-k}^{(1)}}{dn}-(n+k+2)c_{-k}^{(1)}\zeta (n+k+3)\right] \text{ ,}
\tag{172}
\end{equation}
\begin{equation}
C^{(2)}(n,k,p)=-\frac{1}{4(n+4)(n+5)(n+6)}\left[ c_{n+2}^{(1)}+c_{n}^{(0)}+%
\sum_{k=0}^{\infty }(n+k+1)\zeta (n+k+2)c_{-k}^{(1)}\right] \text{ .}
\tag{173}
\end{equation}
It is easily seen that $R_{0}$ will again tend to zero if the nominator $%
C^{(1)}(n,k,p)$ in (172) tends to $\pm \infty .$ This may happen if at least
one of the equations below is fulfilled
\begin{equation}
\frac{dc_{n+2}^{(1)}}{dn}\text{ }\symbol{126}\text{ }const.n^{l};\text{ \ \
\ \ \ \ \ \ \ }\frac{dc_{n}^{(0)}}{dn}\text{ }\symbol{126}\text{ \ }%
const.n^{l}\text{ \ \ \ \ \ }(l>3)\text{ ,}  \tag{174}
\end{equation}
\begin{equation*}
\frac{d}{dn}[\sum_{k=0}^{\infty }\zeta (n+k+2)c_{-k}^{(1)}]\text{ }\symbol{%
126}\text{ }const.n^{r}\text{ \ \ \ }(l>3)\text{ \ \ \ }
\end{equation*}
\begin{equation}
\Longrightarrow c_{-k}^{(1)}\text{ }\symbol{126}\text{ }const.\int
n^{l}p^{n+k+2}dn\text{ \ \ \ \ \ \ }for\text{ \ }some\text{ }k\text{ \ }and%
\text{ }some\text{ }p\text{ ,}  \tag{175}
\end{equation}
\begin{equation*}
\sum_{k=0}^{\infty }\zeta (n+k+3)c_{-k}^{(1)}\text{ }=\sum_{k=0}^{\infty
}\sum_{p\neq 0}c_{-k}^{(1)}\frac{1}{p^{n+k+3}}\text{\ \ }\symbol{126}\text{
\ }const.n^{r}\text{ \ \ }(r>1)\text{ \ \ \ \ \ \ }
\end{equation*}
\begin{equation}
\Longrightarrow c_{-k}^{(1)}\text{ }\symbol{126}\text{ }%
const.n^{r}p^{n+k+3}dn\text{ \ \ \ \ \ \ }for\text{ \ }some\text{ }k\text{ \
}and\text{ }some\text{ }p\text{ ,}  \tag{176}
\end{equation}
\begin{equation*}
\sum_{k=0}^{\infty }\sum_{p\neq 0}\frac{1}{p^{n+k+3}}\frac{dc_{-k}^{(1)}}{dn}%
\sim \text{ }const.n^{r}\text{ \ \ }(r>2)
\end{equation*}
\begin{equation}
\Rightarrow c_{-k}^{(1)}\sim \text{ }const.\int n^{r}p^{n+k+2}dn\text{ .}
\tag{177}
\end{equation}
It is interesting to note that after performing a change of variables
\begin{equation}
tn=x\text{ \ \ \ \ \ \ \ \ \ \ \ \ \ }exp(-t)=p\text{ ,}  \tag{178}
\end{equation}
the integral (175) for $c_{-k}^{(1)}$ can be transformed as follows
\begin{equation}
\int\limits_{1}^{\infty }n^{l}p^{n+k+2}dn=\frac{exp[-t(k+2)]}{t^{l+1}}\text{
}\int\limits_{t}^{\infty }exp(-x)x^{l}dx\text{ .}  \tag{179}
\end{equation}
Keeping in mind the generally known formulae for the Gamma function
\begin{equation}
\Gamma (l)\equiv \int\limits_{0}^{\infty
}exp(-x)x^{l-1}dx=\int\limits_{1}^{\infty
}exp(-x)x^{l-1}dx+\sum\limits_{n=0}^{\infty }\frac{(-1)^{n}}{n!(l+n)}\text{ ,%
}  \tag{180}
\end{equation}
it can be derived for $c_{-k}^{(1)}$%
\begin{equation}
c_{-k}^{(1)}\text{ }\symbol{126}\text{ }const.\frac{exp[-t(k+2)]}{t^{l+1}}%
\{-\int\limits_{1}^{t}exp(-x)x^{l}dx+\Gamma (l+1)-\sum\limits_{n=0}^{\infty }%
\frac{(-1)^{n}}{n!(l+n+1)}\}\text{ .}  \tag{181}
\end{equation}
Let us denote the integral in the above expression by
\begin{equation}
I(l)\equiv \int\limits_{1}^{t}exp(-x)x^{l}dx\text{ .}  \tag{182}
\end{equation}
Then the following relation can be found after performing multiple
integration by parts
\begin{equation*}
I(l)=-\mid x^{l}exp(-x)\mid _{x=1}^{x=t}+I(l-1)=....-\mid x^{l}exp(-x)\mid
_{x=1}^{x=t}-
\end{equation*}
\begin{equation*}
-l\mid x^{l-1}exp(-x)\mid _{x=1}^{x=t}.......-l(l-1)..(l-k+2)\mid
x^{l-k+1}exp(-x)\mid _{x=1}^{x=t}+
\end{equation*}
\begin{equation}
+l(l-1)(l-2)...(l-k+1)I(l-k)\text{ .}  \tag{183}
\end{equation}
Continuing in the same way, one can derive
\begin{equation*}
I(l)=-exp(-t)[t^{l}+lt^{l-1}+..+l!t)+
\end{equation*}
\begin{equation*}
+exp(-1)[1+l+l(l-1)+..+l!]+l!\int\limits_{1}^{t}exp(-x)dx=
\end{equation*}
\begin{equation}
=-exp(-t)\sum\limits_{k=0}^{l-1}\frac{d}{dt^{k}}t^{l}+exp(-1)\sum%
\limits_{k=0}^{l-1}l(l-1)..(l-k)-exp(-1)l![exp(1-t)+1]\text{ .}  \tag{184}
\end{equation}
Substituting this formulae into (181) and returning to the original
variables $(n,p)$, finally an expression is derived for $c_{-k}^{(1)}:$%
\begin{equation*}
c_{-k}^{(1)}\text{ }\symbol{126}\text{ }const.(-1)^{l+1}\frac{p^{k+2}}{%
(lnp)^{l+1}}\{l!(p+exp(-1))+p\sum\limits_{k=0}^{l-1}\frac{d}{dt^{k}}t^{l}-
\end{equation*}
\begin{equation}
-exp(-1)\sum\limits_{k=0}^{l-1}l(l-1)..(l-k)+\Gamma
(l+1)-\sum\limits_{n=0}^{\infty }\frac{(-1)^{n}}{n!(l+n+1)}\}\text{ .}
\tag{185}
\end{equation}
The above expression is of course not an exact one, since other
representations of $c_{-k}^{(1)}$ may exist, when the convergency radius of
the infinite sum tends to $\infty .$ Nevertheless, it can be used as a
possible model representation.

It may happen also that the denominator $C^{(2)}(n,k,p)$ tends to zero, and
then again the covergency radius will tend to infinity. For the purpose, the
following two equations have to be fulfilled together
\begin{equation}
c_{n+2}^{(1)}+c_{n}^{(0)}\text{ }\symbol{126}\text{ }const.n^{r}\text{ \ \ \
\ \ \ }(r=1,2,3)  \tag{186}
\end{equation}
and
\begin{equation}
\zeta _{n+k+2}\text{ }c_{-k}^{(1)}\text{ }\symbol{126}\text{ }n\text{ \ }%
\Longrightarrow \text{ }c_{-k}^{(1)}\text{ }\symbol{126}\text{ }const.n\text{
}p^{n+k+2}\text{ .}  \tag{187}
\end{equation}
The last possible choice is when the denominator is finite, but the
nominator is infinite (equations 174-177). The denominator is finite when
\begin{equation}
c_{n+2}^{(1)}\text{ }\symbol{126}\text{ }const.n^{3}\text{ \ \ \ \ \ \ \ \ \
}c_{-k}^{(1)}\text{ }\symbol{126}\text{ }const.n^{3}  \tag{188}
\end{equation}
.
\begin{equation}
c_{-k}^{(1)}\zeta (n+k+2)\text{ }\symbol{126}\text{ }const.n^{2}\text{ .}
\tag{189}
\end{equation}
But since these equations contradict equations (175-176), such a case of
finiteness of the denominator and tending to zero nominator has to be
excluded from consideration.

\section*{ XII. APPLICATION \ OF \ THE \ EQUATION \\
$\left[ \protect\rho^{^{\prime }}(z)\right] ^{2}=4\protect\rho ^{3}-g_{2}(z)\protect\rho
-g_{3}(z)-$ \\
- ANOTHER \ WAY \ FOR \ PARAMETRIZATION \\
OF \ THE \ CUBIC \ EQUATION \ (53)}

\bigskip In Sect. VI a parametrization of the cubic equation (53) was
proposed, based on presenting the equation in the form of a cubic equation
in respect to one of the variables, and more importantly, applying the cubic
equation for the Weierstrass function $\rho (z)$ and \ setting up $%
g_{2},g_{3}-$complex numbers. In this Section it shall be demonstrated how
one can \ obtain such an equation with $g_{2},g_{3}$ - functions of a
complex variable.

It is instructive first to note that the last terms in the square
parenthesis in (53) depend only on the variable $n.$ Therefore, let us
denote the term in the parenthesis by $\Pi ^{2}$ and thus let us represent
(53) in the form of two cubic equations
\begin{equation}
\Pi ^{2}=\overline{M}-A\widetilde{m}^{3}-B\widetilde{m}^{2}-C(1+2\frac{d}{c})%
\widetilde{m}  \tag{190}
\end{equation}
and
\begin{equation}
\Pi ^{2}=\widetilde{P}_{1}(n)\widetilde{m}^{3}+\widetilde{P}_{2}(n)%
\widetilde{m}^{2}+\widetilde{P}_{3}(n)\widetilde{m}+\widetilde{P}_{4}(n)%
\text{ .}  \tag{191}
\end{equation}
The quadratic forms $\widetilde{P}_{i}(n)(i=1..4)$ are the same as in
(60-63), but with a reversed sign of $n.$ Also, in the first three forms the
last terms $(-A),$ $(-B)$ and $(-C-2\frac{d}{c}C)$ are absent.

In order to bring the two equations to a parametrizable form, let us perform
the \textit{linear transformation}
\begin{equation}
\widetilde{m}=r\overline{m}+s\text{ .}  \tag{192}
\end{equation}
The first equation (190) transforms to the following equation
\begin{equation*}
\Pi ^{2}=[\overline{M}-As^{3}-Bs^{2}-Cs-2\frac{b}{d}Cs]+[-3rs^{2}A-Cr-2Brs-2%
\frac{b}{d}Cr]\overline{m}+
\end{equation*}
\begin{equation}
+[-3r^{2}sA-Br^{2}]\overline{m}^{2}-Ar^{3}\overline{m}^{3}\text{ .}
\tag{193}
\end{equation}
In order to obtain the parametrizable form
\begin{equation}
\left[ \rho ^{^{\prime }}(z)\right] ^{2}=4\rho ^{3}-g_{2}(z)\rho -g_{3}(z)
\tag{194}
\end{equation}
of the cubic equation, one should require
\begin{equation}
s=-\frac{B}{3A}\text{ \ \ \ \ \ \ \ \ \ \ \ \ \ \ \ \ \ \ \ \ \ \ \ \ \ \ \
\ \ \ \ \ }r=(-\frac{4}{A})^{\frac{1}{3}}=i^{\frac{2}{3}}(\frac{4}{A})^{%
\frac{1}{3}}\text{ ,}  \tag{195}
\end{equation}
\begin{equation}
g_{3}(z)=As^{3}+Bs^{2}+C(1+2\frac{b}{d})s-\overline{M}=-\overline{M}-\frac{B%
}{3A}C(1+2\frac{b}{d})\text{ ,}  \tag{196}
\end{equation}
\begin{equation}
g_{2}(z)=3rs^{2}A+Cr+2Brs+2\frac{b}{d}Cr=\frac{i^{\frac{2}{3}}2^{\frac{2}{3}}%
}{A^{\frac{1}{3}}}[C(1+2\frac{b}{d})-\frac{B^{2}}{3A}\text{ .}  \tag{197}
\end{equation}
Since in the preceeding sections it has been proved that the Weierstrass
function $\rho (z)$ parametrizes equation (194), one has the right to set up
\begin{equation}
\overline{m}=\rho (z)\text{ \ \ \ \ \ \ \ \ \ \ \ \ \ \ \ \ \ }\Pi =\rho
^{^{\prime }}(z)  \tag{198}
\end{equation}
and consequently, the transformation (192) acquires the form
\begin{equation}
\widetilde{m}=r\rho (z)+s\text{ .}  \tag{199}
\end{equation}
After performing this transformation, the second algebraic equation (191)
assumes the following form
\begin{equation}
\left[ \rho ^{^{\prime }}(z)\right] ^{2}=\overline{Q}_{1}(n)\rho ^{3}(z)+%
\overline{Q}_{2}(n)\rho ^{2}(z)+\overline{Q}_{3}(n)\rho (z)+\overline{Q}%
_{4}(n)\text{ ,}  \tag{200}
\end{equation}
where
\begin{equation}
\overline{Q}_{1}(n)\equiv \widetilde{P}_{1}(n)r^{3}\text{ ,}  \tag{201}
\end{equation}
\begin{equation}
\overline{Q}_{2}(n)\equiv 3r^{2}s\widetilde{P}_{1}(n)+r^{2}\widetilde{P}%
_{2}(n)\text{ ,}  \tag{202}
\end{equation}
\begin{equation}
\overline{Q}_{3}(n)\equiv 3rs^{2}\text{ }\widetilde{P}_{1}(n)+r\widetilde{P}%
_{3}(n)+2rs\widetilde{P}_{2}(n)\equiv -\overline{g}_{2}\text{ ,}  \tag{203}
\end{equation}
\begin{equation}
\overline{Q}_{4}(n)\equiv \widetilde{P}_{4}(n)+\widetilde{P}_{1}(n)s^{3}+%
\widetilde{P}_{2}(n)s^{2}+\widetilde{P}_{3}(n)s\equiv -\overline{g}_{3}\text{
.}  \tag{204}
\end{equation}
The first equation (194) and the second equation (200) will be identical in
respect to the cubic and the quadratic in $\rho $ terms if
\begin{equation}
s=-\frac{\widetilde{P}_{2}\left( n\right) }{3\widetilde{P}_{1}(n)}=-\frac{B}{%
3A}\text{ \ \ \ \ \ \ \ \ \ \ \ }r^{3}=\frac{4}{\widetilde{P}_{1}(n)}=-\frac{%
4}{A}\text{ .}  \tag{205}
\end{equation}
In fact, this means that the first two terms (cubic and quadratic) in the
transformed equations (201) and (202) are identical with the corresponding
ones in the original equations (190-191):
\begin{equation}
\widetilde{P}_{1}(n)=-A\text{ \ \ \ \ \ \ \ \ \ \ \ \ \ \ \ \ \ \ \ \ \ \ }%
\widetilde{P}_{2}(n)=-B\text{ ,}  \tag{206}
\end{equation}
\begin{equation}
\overline{Q}_{1}(n)=4\text{ \ \ \ \ \ \ \ \ \ \ \ \ \ \ \ \ \ \ \ \ }%
\overline{Q}_{2}(n)=0\text{ .}  \tag{207}
\end{equation}
In other words, the linear transformation is such that it transforms the
straight line
\begin{equation}
\lbrack \widetilde{P}_{3}(n)+C(1+2\frac{d}{c})]\widetilde{m}+\widetilde{P}%
_{4}(n)-\overline{M}=0\text{ ,}  \tag{208}
\end{equation}
defined on the points of the original cubic equation (53) (therefore -
intersecting it), into the straight line
\begin{equation}
(\overline{Q}_{3}(n)+g_{2}(z))\rho (z)+(\overline{Q}_{4}(n)+g_{3}(z))=0\text{
,}  \tag{209}
\end{equation}
defined on the algebraic variety ''points'' of the transformed cubic curve
(obtained from (201) and (206))
\begin{equation*}
(\overline{Q}_{1}(n)-4)\rho ^{3}(z)+\overline{Q}_{2}(n)\rho ^{2}(z)+
\end{equation*}
\begin{equation}
+(\overline{Q}_{3}(n)+g_{2}(z))\rho (z)+(\overline{Q}_{4}(n)+g_{3}(z))=0%
\text{ .}  \tag{210}
\end{equation}
Note also that it cannot be assumed \ that
\begin{equation}
\overline{Q}_{2}(n)=-g_{2}(z)\text{ \ \ \ \ \ \ \ \ \ \ \ \ \ }\overline{Q}%
_{4}(n)=-g_{3}(z)\text{ ,}  \tag{211}
\end{equation}
because then it will turn out that the linear transformation is a degenerate
one and maps the straight line into the zero point. Of course, the choice of
variables (206) does not mean that an additional and restrictive assumption
has been imposed. The variable identification (206) simply helps to ''fix''
the transformation so that an one-dimensional submanifold (the straight
line) is mapped again into an one-dimensional submanifold. In such a way,
the original cubic equations (191-192) are replaced with the transformed
ones (194) and (200), which differ yet in their last two terms.

Now, using expressions (203-204) for $\overline{g}_{2}(z)$ and $\overline{g}%
_{3}(z)$ and also equations (205) for $r$ and $s$, the following expressions
are obtained, which will be used in the next section:
\begin{equation}
\overline{g}_{2}(z)=4C(1+2\frac{b}{d})\frac{1}{\widetilde{P}_{1}}+\frac{4%
\widetilde{P}_{2}^{2}}{3\widetilde{P}_{1}^{2}}\text{ ,}  \tag{212}
\end{equation}
\begin{equation}
\overline{g}_{3}(z)=-C(1+2\frac{b}{d})\frac{\widetilde{P}_{2}}{3\widetilde{P}%
_{1}}-\overline{M}-\frac{2\widetilde{P}_{2}^{3}}{27\widetilde{P}_{1}^{2}}%
\text{ .}  \tag{213}
\end{equation}

\section*{ XIII. J - INVARIANT \ IN \ THE \ CASE \ OF \\
ARBITRARY \  RATIO \ $\ \frac{a}{c}$ \ AND \\
IN \ THE \ GENERAL \ CASE}

\bigskip The purpose in this section will be to see how the so called $j$ or
\textbf{modular invariant of an elliptic curve, } defined as:
\begin{equation}
j(E)=1728\frac{g_{2}^{3}}{g_{2}^{3}-27g_{3}^{2}}\text{ ,}  \tag{214}
\end{equation}
will change under some assumptions, for example when the ratio $\widetilde{m}%
=\frac{a}{c}$ of the parameter functions is an arbitrary one. In a broader
sense, the idea is to see if there is any relation between the possible
motions and group transformations on the complex plane and the $j-$%
invariant. This will not be considered in this section.

If the cubic equation (53) is satisfied under arbitrary $\widetilde{m}$,
then all the coefficient functions in front of $\widetilde{m}$ and its
powers should equal to zero.Therefore, the following identification holds:

\begin{equation}
\overline{M}=\widetilde{P}_{4}(n)\text{ \ \ \ \ \ \ \ \ \ \ \ \ \ \ \ \ \ \ }%
-C(1+2\frac{b}{d})=\widetilde{P}_{3}(n)\text{ ,}  \tag{215}
\end{equation}
\begin{equation}
\widetilde{P}_{1}(n)=-A\text{ \ \ \ \ \ \ \ \ \ \ \ \ \ \ \ \ \ \ \ \ \ \ \
\ \ }\widetilde{P}_{2}(n)=-B\text{ .}  \tag{216}
\end{equation}
Taking this into account, equations (196-197) for $g_{2}$ and $g_{3}$ can be
written as

\bigskip
\begin{equation}
g_{2}(z)=\frac{2^{\frac{2}{3}}}{\widetilde{P}_{1}^{\frac{1}{3}}}^{\prime }[%
\frac{\widetilde{P}_{2}^{2}}{3\widetilde{P}_{1}}-\widetilde{P}_{3}]\text{ ,}
\tag{217}
\end{equation}
\begin{equation}
g_{3}(z)=\widetilde{P}_{3}\frac{\widetilde{P}_{2}}{\widetilde{P}_{1}}-%
\widetilde{P}_{4}\text{ .}  \tag{218}
\end{equation}
Eliminating $\widetilde{P}_{3}(n)$ from the two equations and expressing $%
\widetilde{P}_{4}(n)$ gives
\begin{equation}
\widetilde{P}_{4}=\frac{g_{2}}{2^{\frac{2}{3}}}\frac{\widetilde{P}_{2}}{%
\widetilde{P}_{1}^{\frac{2}{3}}}+\frac{\widetilde{P}_{2}^{3}}{3\widetilde{P}%
_{1}^{2}}-g_{3}\text{ .}  \tag{219}
\end{equation}
Substituting $\widetilde{P}_{4}(n)$ from (219) and $\widetilde{P}_{3}(n)$
from (218) into equation (204) for $\overline{Q}_{4}(n),$ and taking into
account also (195), one obtains
\begin{equation}
-\overline{g}_{3}\equiv \overline{Q}_{4}(n)=-g_{3}+\frac{8\widetilde{P}%
_{2}^{3}}{27\widetilde{P}_{1}^{2}}+\frac{2^{\frac{4}{3}}}{3}\frac{\widetilde{%
P}_{2}}{\widetilde{P}_{1}^{\frac{2}{3}}}g_{2}\text{ .}  \tag{220}
\end{equation}
Performing the same substitutions in respect to equation (203) for $%
\overline{Q}_{3}(n)$, a rather unexpected result is obtained
\begin{equation}
-\overline{g}_{2}\equiv \overline{Q}_{3}(n)=\frac{2^{\frac{5}{3}}}{3}\frac{%
\widetilde{P}_{2}^{2}}{\widetilde{P}_{1}^{\frac{4}{3}}}-g_{2}-\frac{2^{\frac{%
5}{3}}}{3}\frac{\widetilde{P}_{2}^{2}}{\widetilde{P}_{1}^{\frac{4}{3}}}%
=-g_{2}\text{ .}  \tag{221}
\end{equation}
Therefore, $g_{2}$ for the cubic equation (190) does not change if assuming
that $\widetilde{m}$ is arbitrary. It follows also from (220) that $%
\overline{g}_{3}=g_{3}$ if and only if
\begin{equation}
g_{2}^{3}=-\frac{2^{5}}{9^{3}}\frac{\widetilde{P}_{2}^{6}}{\widetilde{P}%
_{1}^{4}}\text{ .}  \tag{222}
\end{equation}
Since
\begin{equation}
\widetilde{P}_{1}=-A=-2p\Gamma _{55}^{r}g_{5r}\text{ \ \ \ \ \ \ \ \ \ \ and
\ \ \ \ \ \ \ }\widetilde{P}_{2}=-B=-6p\Gamma _{\alpha
5}^{r}g_{5r}dx^{\alpha }\text{ ,}  \tag{223}
\end{equation}
clearly (222) will be a rather complicated algebraic equation of sixth rank
in respect to the sub-algebraic variety of the variables $dx^{\alpha }$ ($%
\alpha =1,2,3,4).$

However, it is more interesting to see when the $j$-invariants for the cubic
equations in the case of arbitrary $\frac{a}{c}=\widetilde{m}$ are equal,
i.e.

\begin{equation}
\overline{j}(E)=j(E)\text{ .}  \tag{224}
\end{equation}
Making use of the defining equation (214) for the $j-$invariant and of
equations (220-221), one easily finds
\begin{equation}
\left[ g_{3}-\frac{8\widetilde{P}_{2}^{3}}{27\widetilde{P}_{1}^{2}}-\frac{2^{%
\frac{4}{3}}}{3}\frac{\widetilde{P}_{2}}{\widetilde{P}_{1}^{\frac{2}{3}}}%
g_{2}\right] ^{2}=g_{3}^{2}\text{ .}  \tag{225}
\end{equation}
This is an algebraic equation on a Riemann surface. If one denotes
\begin{equation}
g_{3}(z)=w\text{ ,}  \tag{226}
\end{equation}
then in respect to $w$ the surface has two sheaves - for $+w$ and for $-w$.
If the positive sign is taken, then the algebraic equation is satisfied for $%
\overline{g}_{3}=g_{3}-$ a case already considered (equations (223-224)).

For the case of the minus sign $-w$, equation (225) will give another
possible relation between the functions $g_{2}(z)$ and $g_{3}(z)$ :

\begin{equation}
g_{3}(z)=\frac{4\widetilde{P}_{2}^{3}}{27\widetilde{P}_{1}^{2}}+\frac{2^{%
\frac{1}{3}}}{3}\frac{\widetilde{P}_{2}}{\widetilde{P}_{1}^{\frac{2}{3}}}%
g_{2}\text{ \ \ \ \ \ ,}  \tag{227}
\end{equation}
following from the equality of the $j$-invariants of the cubic equations for
the case of arbitrary $m$. If \ the Loran decomposition of $g_{2}(z)$ and $%
g_{3}(z)$
\begin{equation}
g_{2}(z)=\sum\limits_{m=-\infty }^{\infty }c_{m}^{(1)}z^{m}\text{ \ \ \ \ \
\ \ \ \ \ \ \ \ \ \ \ \ \ \ \ \ \ \ }g_{3}(z)=\sum\limits_{m=-\infty
}^{\infty }c_{m}^{(0)}z^{m}  \tag{228}
\end{equation}
is substituted into (227), then some additional relations may be obtained
between the Loran function coefficients.

One should keep in mind, however that the original cubic equation has been
''splitted up'' into two parts, so it should be seen how the $j$-invariants
of the two parts are related to the $j-$invariant of the original equation.

\section*{ XIV. INTEGRAL \ REPRESENTATION \ IN \\
THE \ LORAN  \ FUNCTION \ DECOMPOSITION \\
OF  $\ g_{2}(z)$ \ AND \ A \ SUMMATION  \ FORMULAE \\
WITH \ THE \ \ FINITE \ SUMS  $G_{n}$}

\bigskip In order to derive the summation formulae, let us remember that the
coefficient functions $c_{0}^{(1)}$ and $c_{m\text{ }}^{(1)}$ in the Loran
infinite sum decomposition possess also an integral representation
\begin{equation}
c_{m}^{(1)}\equiv \frac{1}{2\pi i}\int\limits_{C}\frac{g_{2}(w)}{w^{n+1}}dw%
\text{ \ \ \ \ \ \ \ \ \ \ \ \ \ \ \ \ \ \ \ \ \ \ }m=0,1,2.......\text{ .}
\tag{229}
\end{equation}
For the purpose of obtaining a formulae for $g_{2}(z)$, one should
substitute the already found coefficient functions $c_{-1}^{(1)}$ (eq.142), $%
c_{-2}^{(1)}$ (eq.137), $c_{-4}^{(1)}$ (eq.136) \ and $c_{-m}^{(1)}$
(eq.143) into the Loran's decomposition formulae (228). The obtained
expression is
\begin{equation*}
g_{2}(z)=\sum\limits_{m=1}^{\infty }c_{m}^{(1)}z^{m}+c_{0}^{(1)}+c_{-1}^{(1)}%
\frac{1}{z}+(-36+\frac{G_{1}}{G_{2}}c_{-1}^{(1)})\frac{1}{z^{2}}-\frac{%
16G_{4}}{G_{3}-\frac{G_{4}}{G_{2}}G_{1}}\frac{1}{z^{3}}-
\end{equation*}

\bigskip
\begin{equation}
-\frac{1}{G_{4}}(c_{-1}^{(1)}G_{1}-48G_{2}+40G_{4})\frac{1}{z^{4}}%
-\sum\limits_{m=5}^{\infty }\frac{16G_{4}}{G_{m}-\frac{G_{4}}{G_{2}}G_{m-2}}%
\frac{1}{z^{m}}\text{ .}  \tag{230}
\end{equation}
Let us see whether this will remain an expression for $g_{2}(z)$, or (which
will turn out to be the case) the term $g_{2}(z)$ on both sides of the
equality will cancell out. For the purpose, let us rewrite the first two
terms on the R.H.S. as
\begin{equation}
\sum\limits_{m=1}^{\infty
}c_{m}^{(1)}z^{m}+c_{0}^{(1)}=\sum\limits_{m=1}^{\infty }\left( \frac{1}{%
2\pi i}\int \frac{g_{2}(w)}{w^{m+1}}dw\right) z^{m}+\frac{1}{2\pi i}\int
\frac{g_{2}(w)}{w}dw\text{ .}  \tag{231}
\end{equation}
The integral and the sum in the first term can be interplaced, and also the
formulae for the infinite geometric progression will be taken into account
\begin{equation}
\sum\limits_{m=1}^{\infty }\frac{1}{\xi ^{m}}=\frac{\frac{1}{\xi }}{1-\frac{1%
}{\xi }}=\frac{1}{\xi -1}\text{ .}  \tag{232}
\end{equation}
So, expression (230) acquires the form
\begin{equation*}
\frac{1}{2\pi i}\int \frac{g_{2}(w)}{w}\left[ \sum\limits_{m=1}^{\infty }%
\frac{1}{(\frac{w}{z})^{m}}+1\right] dw=\frac{1}{2\pi i}\int \frac{g_{2}(w)}{%
w}\left[ \frac{z}{w-z}+1\right] dw=
\end{equation*}

\bigskip
\begin{equation}
=\frac{1}{2\pi i}\int \frac{g_{2}(w)}{w-z}dw=g_{2}(z)\text{ }  \tag{233}
\end{equation}
according to Coushie's formulae. Therefore, $g_{2}(z)$ on the two sides of
(230) cancels out. Changing the summation index in the last term in (230)
from $m$ to $m^{^{\prime }}=m-3$, one obtains the following expression for
the last infinite sum in (230)
\begin{equation*}
\sum\limits_{m=3}^{\infty }\frac{d_{m}}{z^{m+2}}=\frac{c_{-1}^{(1)}}{16G_{4}}%
z^{-1}+(\frac{c_{-1}^{(1)}}{16G_{4}}\frac{G_{1}}{G_{2}}-\frac{9}{4G_{4}}%
)z^{-2}+\frac{1}{G_{1}\frac{G_{4}}{G_{2}}-G_{3}}z^{-3}+
\end{equation*}
\begin{equation}
+\left[ \frac{3G_{2}}{G_{4}^{2}}-\frac{5}{2G_{4}}-\frac{c_{-1}^{(1)}}{16G_{4}%
}\frac{G_{1}}{G_{4}}\right] z^{-4}\text{ ,}  \tag{234}
\end{equation}
where
\begin{equation}
d_{m}\equiv \frac{1}{G_{m+2}-\frac{G_{4}}{G_{2}}G_{m}}\text{ .}  \tag{235}
\end{equation}

\bigskip The essence of the above presented proof is the following: The
Loran's decomposition (ranging from $-\infty $ to $+\infty $) is known to be
convergent only in a segment. On the other hand, for the positive - power
terms in $m$ in Loran's sum one has the Coushie's formulae, where the
integration is performed around a closed contour (a circle) in the complex
plane. As for the other part with the negative-power in $m$ terms, if $z$ is
not restricted in a segment area, in the general case there is no guarantee
that the sum will be convergent. However, in the present case the
negative-power part is represented by the infinite sum in (234), and also by
terms with inverse powers of $z$, which are convergent in the limit $%
z\rightarrow \infty $. The infinite sum $\sum d_{m}z^{-m}$ is also
convergent, since $G_{m}$ and $G_{m+2}$, entering $d_{m}$, are finite when $%
m\rightarrow \infty $ (according to a well-known theorem from complex
analyses), and therefore the whole sum converges to zero in this limit (and
also in the limit $z\rightarrow \infty $). This is entirely consistent with
the R.H.S. of (234), which is convergent to zero in the limit $z\rightarrow
\infty $ and naturally does not contain a free term.

\section*{ XV. AN \ EXPRESSION \ FOR $G_{1}$ FROM \\
TAUBER'S \ THEOREM  IN  THE \\
CASE \ OF  INFINITE \ POLES}

\bigskip Let us consider again the case of infinite poles, when $%
G_{k}\rightarrow \zeta (k)$. Note that for $k=1$ the zeta-function $\zeta
(k) $ is not defined, so further the notation $G_{1}$ shall be preserved.

\begin{proposition}
The coefficient function $d_{m}$ in the infinite sum (234) in the limit $%
m\rightarrow \infty $ and under the assumption of infinite poles has the
following behavior:
\begin{equation}
d_{m}=O(\frac{1}{m})  \tag{236}
\end{equation}
\end{proposition}

\begin{proof}
\bigskip :
\end{proof}

The proo\bigskip f is based on the following representation of the Riemann
zeta-function, which can be found in Ref.40:
\begin{equation*}
\zeta (m)=\frac{1}{m-1}+\frac{1}{2}+\sum%
\limits_{k=1}^{n}B_{k}m(m+1)....(m+k-2)\frac{1}{k!}-
\end{equation*}
\begin{equation}
-\frac{1}{n!}m(m+1)......(m+n-1)\int\limits_{1}^{\infty }\overline{B}%
_{n}(x)x^{-m-n}dx\text{ .}  \tag{237}
\end{equation}
Then $d_{m}$ can be represented as follows
\begin{equation*}
d_{m}=\left[ \zeta (m+2)-\frac{\zeta (4)}{\zeta (2)}\zeta (m)\right]
^{-1}=N^{-1}[1+\left( \frac{1}{N(m+1)}-\frac{\zeta (4)}{N\zeta (2)(m-1)}%
\right) +
\end{equation*}
\begin{equation}
+\sum\limits_{k=2}^{n}\frac{B_{k}}{N}\frac{F_{k}}{k!}(m+2)(m+3)...(m+k-2)+%
\frac{1}{N}\frac{P_{m}}{n!}(m+1)...(m+n-1)]^{-1}\text{ ,}  \tag{238}
\end{equation}
where $F_{k},P_{m}$ and $N$ are the following expressions
\begin{equation}
F_{k}=(m+k-1)(m+k)-m(m+1)\frac{\zeta (4)}{\zeta (2)}\text{ ,}  \tag{239}
\end{equation}
\begin{equation}
P_{m}=\int\limits_{1}^{\infty }\overline{B}%
_{n}(x)x^{-m-n}[-x^{-2}(m+2)(m+n)(m+n+1)+\frac{\zeta (4)}{\zeta (2)}m]dx%
\text{ ,}  \tag{240}
\end{equation}
\begin{equation}
N=\frac{1}{2}-\frac{\zeta (4)}{2\zeta (2)}\text{ .}  \tag{241}
\end{equation}
In order to estimate $d_{m},$ one needs the following inequalities for the
last two terms in (238):
\begin{equation}
\sum\limits_{k=2}^{n}\frac{B_{k}}{N}\frac{F_{k}}{k!}(m+2)(m+3)..(m+k-2)>\sum%
\limits_{k=2}^{n}\frac{B_{k}}{N}\frac{F_{k}}{k!}m^{k-3}  \tag{242}
\end{equation}
\begin{equation}
\frac{1}{N}\frac{P_{m}}{n!}(m+1)...(m+n-1)>\frac{1}{N}\frac{P_{m}}{n!}m^{n-1}%
\text{ .}  \tag{243}
\end{equation}
From these inequalities, an inequality for $d_{m}$ also follows
\begin{equation}
d_{m}<N^{-1}\left[ 1+(...)+\sum \frac{B_{k}}{N}\frac{F_{k}}{k!}m^{k-3}+\frac{%
1}{N}\frac{P_{m}}{n!}m^{n-1}\right] ^{-1}  \tag{244}
\end{equation}
and $(...)$ denotes the second term in the brackets in (238). It should be
kept in mind also that in the limit $m\rightarrow \infty $ due to $%
x^{-m-n-2} $ the ratio $\frac{F_{k}}{P_{m}}$ is a very small term.

Further, inequality (244) may be rewritten as
\begin{equation}
d_{m}<(\frac{P_{m}}{n!}m^{n-1})^{-1}\left[ 1+O(\frac{1}{m^{n}})+O(\frac{1}{%
m^{n-1}})+\sum\limits_{k=2}^{n}B_{k}\frac{n!}{k!}\frac{F_{k}}{P_{m}}m^{k-n-2}%
\right] ^{-1}\text{ .}  \tag{245}
\end{equation}
Since the last three terms in the squire brackets are small, the expression
in the brackets can be decomposed (when $X\ll 1)$ according to the formulae
\begin{equation}
\left[ 1+X\right] ^{-1}=1-X+X^{2}.....\text{ .}  \tag{246}
\end{equation}
Recall also from (240) that
\begin{equation}
P_{m}^{-1}\equiv O(\frac{1}{m^{3}})\text{ \ \ \ \ \ \ \ \ \ \ \ \ \ \ \ \ \ }%
F_{k}\sim m^{2}\text{ .}  \tag{247}
\end{equation}
Therefore, inequality (244) for $d_{m}$ can be rewritten as
\begin{equation}
d_{m}<n!\left( O(\frac{1}{m^{n-4}})-O(\frac{1}{m^{3}})-O(\frac{1}{m^{4}}%
)-\sum\limits_{k=2}^{n}B_{k}\frac{n!}{k!}\frac{F_{k}}{P_{k}}%
m^{k-2n-1}\right) \text{ .}  \tag{248}
\end{equation}
Neglecting all the small terms $O(..)$ and keeping in mind that in the last
term in (248) the powers of $m$ range from $-2n$ $(k=2)$ to $-n-2$ $(k=n)$, $%
\mid d_{m}\mid $can be estimated
\begin{equation*}
\mid d_{m}\mid <(n!)^{2}\sum\limits_{k=2}^{n}\mid \frac{B_{k}}{k!}\mid
m^{k-2n-2}<(n!)^{2}(n-1)\mid \frac{B_{k}}{k!}\mid m^{-n-2}=
\end{equation*}
\begin{equation}
=O(\frac{1}{m^{n+2}})<(\frac{1}{m})\text{ .}  \tag{249}
\end{equation}
This precludes the proof that $d_{m}=O(\frac{1}{m})$.

The purpose of the above proposition is to demostrate the opportunity to
apply the Tauber's theorem in respect to the infinite sum (234), which for
the presently investigated case and in terms of $\overline{z}=\frac{1}{z}$
can be rewritten as
\begin{equation*}
\sum\limits_{m=3}^{\infty }\widetilde{d}_{m}\overline{z}^{m}=F(\overline{z}%
)=\{\left( \frac{\overline{c}_{-1}^{(1)}}{16\zeta (4)\zeta (2)}-\frac{9}{%
4\zeta (4)}\right) +(\frac{3\zeta (2)}{\zeta ^{2}(4)}-
\end{equation*}
\begin{equation}
-\frac{5}{2\zeta (4)}-\frac{\overline{c}_{-1}^{(1)}}{16\zeta (4)\zeta (2)})%
\overline{z}^{2}+\left( \frac{\overline{c}_{-1}^{(1)}}{16\zeta (4)G_{1}}+%
\overline{z}^{2}\frac{1}{G_{1}\frac{\zeta (4)}{\zeta (2)}-\zeta (3)}\right)
\overline{z}^{-1}\}\text{ ,}  \tag{250}
\end{equation}
where $\widetilde{d}_{m}$ and $\widetilde{c}_{-1}^{(1)}$ denote
\begin{equation}
\widetilde{d}_{m}\equiv \frac{1}{\zeta (m+2)-\frac{\zeta (4)}{\zeta (2)}%
\zeta (m)}\text{ ,}  \tag{251}
\end{equation}
\begin{equation}
\overline{c}_{-1}^{(1)}\equiv \zeta (2)\frac{48\zeta (2)\zeta (k)-16\zeta
(4)\zeta (k)-48\zeta (4)\zeta (k-2)}{\zeta (2)\zeta (k)-\zeta (4)\zeta (k-2)}%
\text{ .}  \tag{252}
\end{equation}
Let us remind the formulation of Tauber's theorem in Ref. 41 in terms of our
notations:

\begin{theorem}
If the infinite sum (250) $\sum\limits_{m=3}^{\infty }\widetilde{d}_{m}%
\overline{z}^{m}=F(\overline{z})$ converges to $S$ when $\overline{z}%
\rightarrow 1$ and also $d_{m}=O(\frac{1}{m})$, then the infinite sum $%
\sum\limits_{m=3}^{\infty }\widetilde{d}_{m}$ also converges to $S.$
\end{theorem}

\bigskip Applying this theorem to the infinite sum (250) and expressing $%
G_{1}$ from there, one can easily obtain the following quadratic equation
for $G_{1}$%
\begin{equation}
G_{1}^{2}-\frac{\zeta (2)}{16\zeta (4)\widetilde{F}}\left[ 16\zeta (3)%
\widetilde{F}+\frac{\widetilde{c}_{-1}^{(1)}}{\zeta (2)}+16\right] G_{1}+%
\frac{\widetilde{c}_{-1}^{(1)}\zeta (2)\zeta (3)}{16\widetilde{F}\zeta
^{2}(4)}=0\text{ , }  \tag{253}
\end{equation}
where
\begin{equation}
\widetilde{F}\equiv \sum\limits_{m=3}^{\infty }\widetilde{d}_{m}+\frac{19}{%
4\zeta (4)}-\frac{3\zeta (2)}{\zeta ^{2}(4)}\text{ .}  \tag{254}
\end{equation}
Therefore, in the asymptotic limit $\overline{z}\rightarrow 1$ and in the
case of poles at infinity, $G_{1}$is uniquely expressed through the Riemann
zeta-function. As a solution of the quadratic equation, it has two values,
which of course are finite, since the values of the zeta-function are
finite. This again confirms the previously established result (although for
a certain partial case) that $G_{1}$ is a \textit{finite (convergent)
quantity} for the case of the investigated cubic equations.

\section*{ XVI. INFINITE  POINT  OF \ THE \ LINEAR - \\
- FRACTIONAL \ TRANSFORMATION  (31) \\
AND \ THE \ TWO \ COUPLED \ ALGEBRAIC \\
EQUATIONS \ FOR \ THE \ WEIERSTRASS \\
FUNCTION}

\bigskip In this Section again the obtained algebraic equation (53) will be
studied, but for the case of the infinite point $n=dx^{5}=-\frac{d}{c}$ of
the linear-fractional transformation (31), which for convenience shall be
rewritten again
\begin{equation}
dx^{5}\equiv \frac{a\widetilde{dx}^{5}+b\text{ }}{c\widetilde{dx}^{5}\text{ }%
+d\text{ }}  \tag{255}
\end{equation}
. Substituting (255) into equation (53), a cubic algebraic equation is
obtained for the variable $\widetilde{m}=\frac{a}{c}:$%
\begin{equation}
-A\widetilde{m}^{3}+3A\frac{b}{d}\widetilde{m}^{2}+(B\frac{b}{d}-A\frac{b^{2}%
}{d^{2}}-2\frac{d}{c}C)\widetilde{m}+(-G^{(4)}+2\frac{b}{c}C+\frac{b}{d}%
C)\equiv 0\text{ ,}  \tag{256}
\end{equation}
where $A$, $B,C$ and $G^{(4)}$ are given, as previously, by formulaes (37 -
40) in Sect.V.

\bigskip For the purpose of investigating the above equation, it is
appropriate to remind briefly the two approaches, proposed in this paper.
The \textit{first one }was developed in Sect.VI and it contained in itself
the following steps:

1. Transforming the original cubic equation into a cubic two-variable
equation of the type (59), containing cubic terms of a new variable $%
\widetilde{m}=\frac{a}{c}$, related to the linear-fractional transformation,
and also quadratic terms of the original variable $n=dx^{5}$. In a sense,
the appearence of a new variable makes the equation more complicated, but at
the same time this variable turns out to be the parametrizable one (with the
Weierstrass function), and this shall be succesfully exploited also in this
section.

2. ''Fixing up'' the coefficient functions in the quadratic forms $P_{1}(n)$%
, $P_{2}(n)$, $P_{3}(n)$ (60-62), so that the standard parametrizable form
\begin{equation}
\widetilde{n}^{2}=4m^{3}-g_{2}m-g_{3}  \tag{257}
\end{equation}
is obtained with $g_{2},g_{3}-$ definite complex numbers. The consistency of
the parametrization, in the sense of an additionally imposed condition, was
ensured by the obtained quadratic algebraic equation (78) in terms of the
''angular'' type variables $l=(l_{1},l_{2},l_{3})=p\times q$ and $%
f=(f_{1},f_{2},f_{3})=r\times q$%
\begin{equation}
4\overline{q}_{1}l^{1}+g_{2}f^{1}\overline{q_{1}}+l^{1}f^{3}+f^{1}l^{3}=0%
\text{ .}  \tag{258}
\end{equation}
Evidently, there is a certain indeterminacy and freedom in chosing the
elements of the algebraic variety $(l,f)$. However, by means of an
appropriate choice of the ratio of the parameters in the linear-fractional
transformation, equation (258) will acquire a very simple form, and in such
a way this second difficulty can also be overcomed.

\bigskip The \textit{second approach, }outlined in Sect. XII on the base of
the analytical calculations in the preceeding sections,was based on
parametrization with the Weierstrass functiion of the cubic form (257), but
with $g_{2}$ and $g_{3}$ - complex functions. In order to bring the original
cubic equation into the form (257), a linear transformation with an
appropriately chosen parameters was needed.

In the present section, a combination of the two approaches shall be
implemented. First, the original cubic equation shall be ''splitted up''
into the following two algebraic equations \textit{\ }
\begin{equation}
t^{2}=-A\widetilde{m}^{3}-2\frac{d}{c}C\widetilde{m}+(-G^{(4)}+2\frac{b}{c}C)%
\text{ ,}  \tag{259}
\end{equation}
\begin{equation}
t^{2}=-3A\frac{b}{d}\widetilde{m}^{2}-A\frac{b^{2}}{d^{2}}\widetilde{m}+B%
\frac{b}{d}\widetilde{m}\text{ .}  \tag{260}
\end{equation}
The first equation is a cubic one and the second - a quadratic one, and this
fact is important. Moreover, there is no need to perform a linear
transformation, and since it has been proven that an equation of the type
(259) is parametrizable with the Weierstrass function, one can set up:
\begin{equation}
t=\rho ^{^{\prime }}(z)\text{ \ \ \ \ \ \ \ \ \ \ \ \ \ \ \ \ \ \ \ \ \ \ \ }%
\widetilde{m}=\rho (z)\text{ .}  \tag{261}
\end{equation}
Therefore, the second equation (260) can be written in the form
\begin{equation}
\left[ \rho ^{^{\prime }}(z)\right] ^{2}=K\rho (z)[\rho (z)-\alpha ]\text{ ,}
\tag{262}
\end{equation}
where
\begin{equation}
K=-3A\frac{b}{d}\text{ \ \ \ \ \ \ \ \ \ \ \ \ \ \ \ \ \ \ \ \ \ \ }\alpha =%
\frac{1}{3}\frac{b}{d}-\frac{B}{3A}\text{ .}  \tag{263}
\end{equation}
Introducing the notation
\begin{equation}
g(z)\equiv \rho (z)-\alpha  \tag{264}
\end{equation}
and performing the integration in (262), one obtains
\begin{equation}
2\int \frac{dg^{\frac{1}{2}}}{g^{\frac{1}{2}}\sqrt{1+\frac{\alpha }{g}}}%
=\int \sqrt{K}dz\text{ \ \ \ \ .}  \tag{265}
\end{equation}
After two subsequent changes of variables
\begin{equation}
g^{\frac{1}{2}}\equiv y\text{ \ \ \ \ \ \ \ and \ \ \ \ \ \ \ \ \ \ \ }\frac{%
y}{g^{\frac{1}{2}}}\equiv tg\varphi \text{ ,}  \tag{266}
\end{equation}
one derives the following equation
\begin{equation}
\int \sqrt{K}dz=2\int \frac{1}{cos\varphi }d\varphi =2\int \frac{dr}{1-r^{2}}%
=ln\mid \frac{1+r}{1-r}\mid \text{ ,}  \tag{267}
\end{equation}
where
\begin{equation}
r=sin\varphi \text{ .}  \tag{268}
\end{equation}
In terms of the original variables, the solution is easily found to satisfy
the equation
\begin{equation}
\mid \frac{2\rho -\alpha +2\epsilon \rho ^{\frac{1}{2}}(\rho -\alpha )^{%
\frac{1}{2}}}{\alpha }\mid =\lambda _{1}exp(\int \sqrt{K}dz)\text{ .}
\tag{269}
\end{equation}
In (269) $\epsilon =\pm 1$ and $\lambda _{1}$ is the integration constant.
Remembering the initial equation (262) and introducing for convenience the
notations
\begin{equation}
P(z)\equiv \frac{1}{2}\lambda _{1}\sqrt{K}exp(\int \sqrt{K}dz)+\frac{\alpha
\sqrt{K}}{2}\text{ \ \ \ \ \ \ \ \ \ \ \ \ }Q(z)\equiv -\frac{\epsilon }{2}%
\sqrt{K}\text{ ,}  \tag{270}
\end{equation}
the first order differential equation (268) may be rewritten in the simple
form
\begin{equation}
\rho ^{^{\prime }}(z)=P(z)+Q(z)\rho (z)\text{ .}  \tag{271}
\end{equation}
Performing again the integration, one obtains
\begin{equation}
\rho (z)=-\frac{P}{Q}+H\text{ , }  \tag{272}
\end{equation}
where $H$ denotes
\begin{equation}
H\equiv \frac{\lambda _{2}}{Q}exp(\int Qdz)  \tag{273}
\end{equation}
and $\lambda _{2}$ is the second integration constant. From (271) and (272)
one can find
\begin{equation}
\lambda _{2}=\frac{\rho ^{^{\prime }}(z)}{exp(\int Qdz)}\text{ \ \ \ \ \ \ \
\ \ \ \ and \ \ \ \ \ \ \ \ }\rho ^{^{\prime }}(z)=QH\text{\ },  \tag{274}
\end{equation}
from where the integration constant $\lambda _{2}$ can be expressed, if \
one substitutes $\rho ^{^{\prime }}(z)$ with its equivalent and well-known
expression
\begin{equation}
\rho ^{^{\prime }}(z)=-2\sum\limits_{\varpi }\frac{1}{(z-\varpi )^{3}}\text{
.}  \tag{275}
\end{equation}
In order to find the first integration constant, or at least a relation
between the two integration constants, let us substitute $\rho ^{^{\prime
}}(z)$ from (272) and $\rho (z)$ from (271) into the first cubic equation
(259), which shall be written in a more general form
\begin{equation}
\left[ \rho ^{^{\prime }}(z)\right] ^{2}=B_{1}\rho ^{3}(z)+B_{2}\rho
^{2}(z)+B_{3}\rho (z)+B_{4}\text{ .}  \tag{276}
\end{equation}
In the present case
\begin{equation}
B_{1}\equiv -A\text{ \ \ \ \ \ \ }B_{2}\equiv 0\text{ \ \ \ \ \ \ }%
B_{3}\equiv -2\frac{d}{c}C\text{ \ \ \ \ \ \ }B_{4}=-G^{(4)}+2\frac{b}{c}C%
\text{ .}  \tag{277}
\end{equation}
After the substitution, equation (276) acquires the following form in
respect to the variable $X=\frac{P}{Q}$, containing the first integration
constant $\lambda _{1}$(which is to be found)
\begin{equation}
\overline{B}_{1}X^{3}+\overline{B}_{2}X^{2}+\overline{B}_{3}X+\overline{B}%
_{4}\equiv 0\text{ ,}  \tag{278}
\end{equation}
where
\begin{equation}
\overline{B}_{1}\equiv -B_{1}\text{ \ \ \ \ \ \ \ \ \ \ \ \ \ \ \ \ \ \ \ }%
\overline{B}_{2}\equiv 3HB_{1}+B_{2}  \tag{279}
\end{equation}
\begin{equation}
\overline{B}_{3}\equiv -3H^{2}B_{1}-2HB_{2}-B_{3}\text{ \ \ \ \ \ \ \ \ \ \
\ \ \ \ }\overline{B}_{4}\equiv B_{4}+B_{3}H+B_{2}H^{2}+B_{1}H^{3}-Q^{2}H^{2}%
\text{ .}  \tag{280}
\end{equation}
Note that the other integration constant $\lambda _{2}$ through $H$ enters
the coefficient functions $\overline{B}_{1},\overline{B}_{2},\overline{B}%
_{3} $ and

$\overline{B}$\bigskip $_{4}$.

\section*{XVII. FINDING \ THE \ RELATION  BETWEEN \\
THE \ TWO  INTEGRATION \ CONSTANTS}

\bigskip The analyses in the preceeding section concerned the
parametrization of the two coupled equations, and performing \ an
integration only of the second one. Now we shall find the relation between
the two integration constants, inserting the found as a result of the
integration solution into the first equation (276-277).

Let us first observe that if one makes the formal identification
\begin{equation}
A\Leftrightarrow \overline{B}_{1}\text{ \ \ \ \ \ \ \ \ \ \ \ \ }%
B\Leftrightarrow \overline{B}_{2}\text{ \ \ \ \ \ \ \ \ \ }C\Leftrightarrow
\overline{B}_{3}\text{ \ \ \ \ \ \ \ }D\Leftrightarrow G^{(4)}\text{ , }
\tag{281}
\end{equation}
then equation (281) is analogous to equation (36) from Sect. V. Therefore,
one may apply the developed there approach in the same manner.

After performing the \textit{linear-fractional} transformation
\begin{equation}
X=\frac{\overline{a}\widetilde{X}+\overline{b}}{\overline{c}\widetilde{X}+%
\overline{d}}  \tag{282}
\end{equation}
(where naturally $\overline{a},\overline{b},\overline{c},\overline{d}$ are
different from $a,b,c,d$ in the transformation (53 ,255)), and introducing
the notation $\chi =\frac{\overline{a}}{\overline{c}}$, the transformed
equation (278) is obtained in the form
\begin{equation}
\overline{C}_{1}(\widetilde{X})\chi ^{3}+\overline{C}_{2}(\widetilde{X})\chi
^{2}+\overline{C}_{3}(\widetilde{X})\chi +\overline{C}_{4}(\widetilde{X}%
)\equiv 0\text{ .}  \tag{283}
\end{equation}
Since the subsequent step concerns the coefficient functions $\overline{C}%
_{1}(\widetilde{X}),\overline{C}_{2}(\widetilde{X}),\overline{C}_{3}(%
\widetilde{X}),\overline{C}_{4}(\widetilde{X}),$ which are completely
similar to the expressions (60-63) and (54-56), they shall be represented
below in terms of the new notations

\bigskip
\begin{equation}
\overline{C}_{1}(\widetilde{X})\equiv r_{1}\widetilde{X}^{2}+r_{2}\widetilde{%
X}+r_{3}=-3\overline{B}_{1}(\frac{\overline{c}}{\overline{d}})^{2}\widetilde{%
X}^{2}-3\overline{B}_{1}\frac{\overline{c}}{\overline{d}}\widetilde{X}-%
\overline{B}_{1}\text{ ,}  \tag{284}
\end{equation}

\bigskip
\begin{equation}
\overline{C}_{2}(\widetilde{X})\equiv q_{1}\widetilde{X}^{2}+q_{2}\widetilde{%
X}+q_{3}=[3\overline{B}_{1}(\frac{\overline{c}}{\overline{d}})^{2}\frac{%
\overline{b}}{\overline{d}}-2\overline{B}_{2}(\frac{\overline{c}}{\overline{d%
}})^{2}]\widetilde{X}^{2}-3\overline{B}_{2}\frac{\overline{c}}{\overline{d}}%
\widetilde{X}-\overline{B}_{2}\text{ ,}  \tag{285}
\end{equation}
\begin{equation*}
\overline{C}_{3}(\widetilde{X})\equiv p_{1}\widetilde{X}^{2}+p_{2}\widetilde{%
X}+p_{3}=\left[ -6\overline{B}_{3}\frac{\overline{c}}{\overline{d}}+%
\overline{B}_{2}\frac{\overline{b}}{\overline{d}}\frac{\overline{c}}{%
\overline{d}}-\overline{B}_{3}(\frac{\overline{c}}{\overline{d}})^{2}\right]
\widetilde{X}^{2}+
\end{equation*}

\begin{equation}
+\left[ -6\overline{B}_{3}+2\overline{B}_{2}\frac{\overline{c}}{\overline{d}}%
\frac{\overline{b}}{\overline{d}}+3\overline{B}_{1}\frac{\overline{c}}{%
\overline{d}}(\frac{\overline{b}}{\overline{d}})^{2}-\overline{B}_{3}\frac{%
\overline{c}}{\overline{d}}\right] \widetilde{X}-\widetilde{B}_{3}-2\frac{%
\overline{d}}{\overline{c}}\overline{B}_{3}\text{ ,}  \tag{286}
\end{equation}
\begin{equation}
\overline{C}_{4}(\widetilde{X})\equiv \overline{F}\widetilde{X}^{2}+%
\overline{N}\widetilde{X}+\overline{M}\text{ ,}  \tag{287}
\end{equation}
where in the last expression
\begin{equation}
\overline{F}\equiv \overline{B}_{3}\frac{\overline{b}}{\overline{d}}(\frac{%
\overline{c}}{\overline{d}})^{2}+2\overline{B}_{3}\frac{\overline{b}}{%
\overline{d}}\frac{\overline{c}}{\overline{d}}\text{ ,}  \tag{288}
\end{equation}
\begin{equation}
\overline{N}=\overline{B}_{2}\frac{\overline{c}}{\overline{d}}(\frac{%
\overline{b}}{\overline{d}})^{2}+2\overline{B}_{3}\frac{\overline{c}}{%
\overline{d}}\frac{\overline{b}}{\overline{d}}\text{ ,}  \tag{289}
\end{equation}
and
\begin{equation}
\overline{M}=\overline{B}_{1}(\frac{\overline{b}}{\overline{d}})^{3}+%
\overline{B}_{2}(\frac{\overline{b}}{\overline{d}})^{2}+\overline{B}_{3}%
\frac{\overline{b}}{\overline{d}}\text{ .}  \tag{290}
\end{equation}
Now in turn is the folowing important assumption: In order to simplify the
expressions (284-286) and to make the most appropriate choice so that a
relation is imposed only on the coefficients $\overline{a},\overline{b},%
\overline{c},\overline{d}$ of the linear-fractional transformation (and not
on $\overline{B}_{1},\overline{B}_{2},$\ $\overline{B}_{3},$ $\overline{B}%
_{4},$ for example), let us suppose that $\overline{C}_{2}(\widetilde{X}),%
\overline{C}_{3}(\widetilde{X})$ are not quadratic, but \textit{linear} in \
$\widetilde{X}$, i.e. $q_{1=}p_{1}=0$. From (285) for $\overline{C}_{2}(%
\widetilde{X})$ and (286) for $\overline{C}_{3}(\widetilde{X})$ it follows
\begin{equation}
\frac{\overline{b}}{\overline{d}}=\frac{2\overline{B}_{2}}{3\overline{B}_{1}}%
\text{ ,}  \tag{291}
\end{equation}
\begin{equation}
\frac{\overline{c}}{\overline{d}}=-6+\frac{2\overline{B}_{2}^{2}}{3\overline{%
B}_{1}\overline{B}_{3}}\text{ .}  \tag{292}
\end{equation}
After performing the same change of variables as in (65), but in terms of
our notations

\begin{equation}
\overline{X}=\sqrt{\overline{F}}\left( \widetilde{X}+\frac{\overline{N}}{2%
\overline{F}}\right) \text{ \ \ \ \ \ }\Rightarrow \text{ \ \ \ \ \ }%
\widetilde{X}=\frac{\overline{X}}{\sqrt{\overline{F}}}-\frac{\overline{N}}{2%
\overline{F}}\text{ ,}  \tag{293}
\end{equation}
the transformed equation (283) will be
\begin{equation}
\overline{X}^{2}=\overline{D}_{1}(\overline{X})\chi ^{3}+\overline{D}_{2}(%
\overline{X})\chi ^{2}+\overline{D}_{3}(\overline{X})\chi +\overline{D}_{4}(%
\overline{X})\text{ .}  \tag{294}
\end{equation}
Let us find also the quadratic algebraic equation (258) for $%
(l_{1},l_{2},l_{3})$ and $(f_{1},f_{2},f_{3})$ for the case $p_{1}=q_{1}=0$.
We have
\begin{equation}
l=(l_{1},l_{2},l_{3})=(0,\overline{p}_{2}\overline{q}_{3},0)\text{ \ \ \ \ \
\ \ \ \ \ \ \ \ \ \ \ \ \ \ \ \ }f=(f_{1},f_{2},f_{3})=(\overline{r}_{1}%
\overline{q}_{2},\overline{r}_{2}\overline{q}_{3},-\overline{r}_{1}\overline{%
q}_{3})\text{\ \ \ .\ \ \ \ }  \tag{295}
\end{equation}
It can easily be checked that eq.(258) is \textit{identically satisfied!} In
other words, we have succeeded to find such a (trivial!) algebraic variety $%
(p,q,r)$, so that the quadratic algebraic equation (221) holds and
therefore, the parametrization with the Weierstrass function can be applied:
\begin{equation}
\overline{X}\equiv \rho ^{^{\prime }}(v)\text{ \ \ \ \ \ \ \ }\chi =\frac{%
\overline{a}}{\overline{c}}=\rho (v)\text{ .}  \tag{296}
\end{equation}
In principle, $v$ should be another complex variable, but here for
simplicity it shall be assumed that $v=z$. Now it is interesting to note
that relations (291) for $\frac{\overline{b}}{\overline{d}},$(292) for $%
\frac{\overline{c}}{\overline{d}}$ and (296) for $\frac{\overline{a}}{%
\overline{c}}$ allow us to determine all the ratios between $\overline{a},%
\overline{b},\overline{c},\overline{d}.$ Consequently, $\overline{F}$ and $%
\overline{N}$ in (288 - 289) are also determined, and taking into account
transformation (293) and expresion (274) for $\lambda _{2}$ (expressed
through $\rho ^{^{\prime }}(z)$)$,$ the linear-fractional transformation
(282) can be represented as
\begin{equation}
X=\frac{\frac{\overline{a}}{\overline{c}}\widetilde{X}+\frac{\overline{b}}{%
\overline{c}}}{\widetilde{X}+\frac{\overline{d}}{\overline{c}}}=\frac{\rho
(z)[\overline{F}^{-\frac{1}{2}}\lambda _{2}exp(\int Qdz)-\frac{\overline{N}}{%
2\overline{F}}]+\frac{\overline{b}}{\overline{c}}}{\overline{F}^{-\frac{1}{2}%
}\lambda _{2}exp(\int Qdz)+\frac{\overline{d}}{\overline{c}}}\text{ .}
\tag{297}
\end{equation}
As already mentioned, the ratios $\overline{b}:\overline{c},\overline{d}:%
\overline{c}$ can easily be found from (291-292)
\begin{equation}
\frac{\overline{b}}{\overline{c}}=\frac{\overline{B}_{2}\overline{B}_{3}}{%
\overline{B}_{2}^{2}-9\overline{B}_{1}\overline{B}_{2}}\text{ \ \ \ \ \ \ \
\ \ \ }\frac{\overline{d}}{\overline{c}}=\frac{3\overline{B}_{1}\overline{B}%
_{2}}{2(\overline{B}_{2}^{2}-9\overline{B}_{1}\overline{B}_{2})}\text{ .}
\tag{298}
\end{equation}
On the other hand, $X=\frac{P}{Q}$ can be determined also from (270)
\begin{equation}
X=\frac{P}{Q}=-\epsilon \lambda _{1}exp(\int \sqrt{K}dz)-\epsilon \alpha
\text{ .}  \tag{299}
\end{equation}
From the two equations (297) and (299), the first integration constant $%
\lambda _{1}$ can be expressed through the other constant $\lambda _{2}$.
Moreover, (297) can be substituted into (292) and after some
transformations, the following quadratic equation can be found for $\lambda
_{2}$
\begin{equation}
\lambda _{2}^{2}+K_{1}\lambda _{2}+K_{2}(\rho (z)-\frac{\overline{b}}{%
\overline{d}})=0\text{ ,}  \tag{300}
\end{equation}
where
\begin{equation}
K_{1}\equiv \frac{1}{2\sqrt{\overline{F}}}\frac{\overline{d}}{\overline{c}}%
exp(\frac{\varepsilon }{2}\int \sqrt{K}dz)[2\overline{F}-\overline{N}\frac{%
\overline{c}}{d}]\text{ ,}  \tag{301}
\end{equation}
\begin{equation}
K_{2}\equiv \frac{\varepsilon }{2}\sqrt{\overline{F}}\frac{\overline{d}}{%
\overline{c}}\sqrt{K}exp(\varepsilon \int \sqrt{K}dz)\text{ ,}  \tag{302}
\end{equation}
The two roots of the quadratic equation (300) can be expressed as
\begin{equation}
\lambda _{2}=\frac{-K_{1}+\varepsilon \sqrt{K_{1}^{2}-4(\rho (z)-\frac{%
\overline{b}}{\overline{d}})K_{2}}}{2}\text{ .}  \tag{303}
\end{equation}
But $\lambda _{2}$ can also be expressed from (274) and (262) in the
following way
\begin{equation}
\lambda _{2}^{2}=\frac{\left[ \rho ^{^{\prime }}(z)\right] ^{2}}{%
exp[-\varepsilon \int \sqrt{K}dz]}=\frac{K\rho ^{2}(z)-K\alpha \rho (z)}{%
exp(-\varepsilon \int \sqrt{K}dz)}\text{ .}  \tag{304}
\end{equation}
Now taking the square of $\lambda _{2}$ from (303) and expressing the
equality of (303) and (304), one can obtain
\begin{equation*}
2K\rho ^{2}(z)+2(K_{2}-\alpha K)\rho (z)-\overline{L}=
\end{equation*}
\begin{equation}
=-\varepsilon K_{1}exp(-\varepsilon \int \sqrt{K}dz)\sqrt{K_{1}^{2}-4(\rho
(z)-\frac{\overline{b}}{\overline{d}})K_{2}}\text{ ,}  \tag{305}
\end{equation}
where $\overline{L}$ denotes
\begin{equation}
\overline{L}\equiv 2K_{2}\frac{\overline{b}}{\overline{d}}+exp(-\varepsilon
\int \sqrt{K}dz)K_{1}^{2}\text{ .}  \tag{306}
\end{equation}
In order to get rid of the square root in (305) and thus to obtain an
algebraic equation, let us take the square of both sides of (305). The
result is a fourth-order algebraic equation in respect to the Weierstrass
function $\rho (z)$%
\begin{equation}
E_{1}\rho ^{4}(z)+E_{2}\rho ^{3}(z)+E_{3}\rho ^{2}(z)+E_{4}\rho (z)+E_{5}=0%
\text{ ,}  \tag{307}
\end{equation}
where
\begin{equation}
E_{1}\equiv 4K^{2}\text{ \ \ \ \ \ \ \ \ \ \ }E_{2}\equiv 8K(K_{2}-\alpha K)%
\text{ \ \ \ \ \ \ \ \ \ \ \ }E_{3}\equiv 4[(K_{2}-\alpha K)^{2}-\overline{L}%
K\text{ ,}  \tag{308}
\end{equation}
\begin{equation}
E_{4}\equiv 4[K_{1}^{2}K_{2}exp(-2\varepsilon \int \sqrt{K}dz)-\overline{L}%
(K_{2}-\alpha K)]\text{ ,}  \tag{309}
\end{equation}
\begin{equation}
E_{5}\equiv -\left[ \overline{L}^{2}+K_{1}^{4}exp(-2\varepsilon \int \sqrt{K}%
dz)+4\frac{\overline{b}}{\overline{d}}K_{2}K_{1}^{2}exp(-2\varepsilon \int
\sqrt{K}dz)\right] \text{ .}  \tag{310}
\end{equation}
In a sense, it is surprising that (307) is a \textit{fourth-rank algebraic
equation}, while after parametrizing the ''original'' system of two coupled
equations (259 - 260), one is left with a \textit{third-rank algebraic
equation}. It can be rewritten as
\begin{equation}
\rho ^{3}(z)=3\frac{b}{d}\rho ^{2}(z)+(\frac{B}{A}\frac{b}{d}-\frac{b^{2}}{%
d^{2}}-2\frac{C}{A}\frac{d}{c})\rho (z)+(-\frac{G^{(4)}}{A}+2\frac{b}{c}%
\frac{C}{A}+\frac{b}{d}\frac{C}{A})\text{ .}  \tag{311}
\end{equation}
It should be remembered, however that the \textit{fourth-rank algebraic
equation }(307) has been obtained after the differential equation (262) has
been solved, and through the integration constant this amounts to a
redefinition of the complex variable and the functions. Also, a more
''specific'' form of the linear-fractional transformation (282) has been
performed, which in a way 'adjusts'' the parametrizable variable in (294) $%
\overline{X}=\rho ^{^{\prime }}(v)$, and it was assumed that $v=z$. What is
perhaps interesting, from a purely mathematical poini of view, is that this
''redefinition'' and transformation resulted in a higher-rank algebraic
equation.

It is important to note that in writing down the system of the two coupled
equations (259-260), only the proven in the previous sections fact about the
parametrization of a cubic equation of the kind $%
t^{2}=4m^{3}-g_{2}(z)m-g_{3}(z)$ has been used. Since $(t,m)$ turn out to be
parametrized with $(\rho ^{^{\prime }}(z),\rho (z))$, and $(t,m)$ enter also
the second equation (260), this explains why the Weierstrass function
''parametrizes' the second equation. But this does not mean yet that the
more general cubic equation (256) can be parametrized with the Weierstrass
function. As it has been already mentioned in Section VI, this fact needs to
be proved, if this is possible at all. In the next Section, some necessary
conditions will be found for parametrizing the cubic equation \ (256) of a
more general kind (with all non-zero coefficient functions) . Although for a
particular equation, the investigation in the next section might provide a
partial answer to the raised in Sect. VIII problem whether it is possible to
parametrize a cubic equation in its general form.

\section*{ XVIII. FINDING \ THE \ NECESSARY \\
CONDITIONS  FOR \ PARAMETRIZATION \\
OF \ THE \ CUBIC \ EQUATION (256)}

\bigskip The analyses in this section concern equations (307) and (311). If $%
\rho ^{3}(z)$ is substituted into equation (307), then the following
equation is obtained in respect to $\rho (z)$%
\begin{equation}
S_{1}\rho ^{4}(z)+S_{2}\rho ^{2}(z)+S_{3}\rho (z)+S_{4}=0\text{ ,}  \tag{312}
\end{equation}
where
\begin{equation}
S_{1}\equiv 4K^{2}\text{ \ \ \ \ \ \ \ \ \ }S_{2}\equiv 4\left[
(K_{2}-\alpha K)^{2}+6\frac{b}{d}K(K_{2}-\alpha K)\right] \text{ ,}
\tag{313}
\end{equation}
\begin{equation*}
S_{3}\equiv 4[K_{1}^{2}K_{2}exp(-2\varepsilon \int Kdz)-\overline{L}%
(K_{2}-\alpha K)+
\end{equation*}
\begin{equation}
+2K(K_{2}-\alpha K)(\frac{B}{A}\frac{b}{d}-\frac{b^{2}}{d^{2}}-2\frac{C}{A}%
\frac{d}{c})\text{ ,}  \tag{314}
\end{equation}
\begin{equation*}
S_{4}\equiv 8K(K_{2}-\alpha K)(2\frac{b}{c}\frac{C}{A}+\frac{b}{d}\frac{C}{A}%
-\frac{G^{(4)}}{A})-\overline{L}^{2}-
\end{equation*}
\begin{equation}
-K_{1}^{4}exp(-2\varepsilon \int \sqrt{K}dz)-4\frac{\overline{b}}{\overline{d%
}}K_{2}K_{1}^{2}exp(-2\varepsilon \int \sqrt{K}dz)\text{ . }  \tag{315}
\end{equation}
Two possibilities result from the above equation (312), and they shall be
investigated here.

The \textit{first possibility} is when (312) holds for arbitrary values of $%
z $ and $\rho (z)$, and therefore the coefficient functions $%
S_{1},S_{2},S_{3},S_{4}$ must be identically zero. This means that
\begin{equation}
S_{1}=K\equiv 0\text{ .}  \tag{316}
\end{equation}
But if $K=0$, then from the defining equality (302) for $K_{2}$ it follows
\begin{equation}
K_{2}=0\text{ .}  \tag{317}
\end{equation}
. If (316-317) are satisfied, then it will follow also
\begin{equation}
S_{2}=S_{3}=0\text{ ,}  \tag{318}
\end{equation}
and this is exactly what is needed. Now it remains only to impose the
condition
\begin{equation}
0\equiv S_{4(K=0)}=-\overline{L}^{2}-K_{1}^{4}=-2K_{1}^{4}\text{ .}
\tag{319}
\end{equation}
Using expressions (301) for $K_{1}$ and (288-289) for $\overline{F}$ and $%
\overline{N}$, it can be shown that the above equation is equivalent to the
following equation
\begin{equation}
\frac{\overline{b}}{\overline{d}}\left[ 4\overline{B}_{3}-\overline{B}_{2}%
\frac{\overline{c}}{\overline{d}}\frac{\overline{b}}{\overline{d}}\right] =0%
\text{ . }  \tag{320}
\end{equation}
The first opportunity for this equation to be satisfied is when
\begin{equation}
\frac{\overline{b}}{\overline{d}}=\frac{2\overline{B}_{2}}{3\overline{B}_{1}}%
=\frac{2HB_{1}}{3\overline{B}_{1}}\equiv 0\text{ .}  \tag{321}
\end{equation}
Since we do not want to impose the rather restrictive condition $B_{1}=-A=0$%
, it remains (with account also of (273)) that
\begin{equation}
0\equiv H\equiv -\frac{2\lambda _{2}}{\varepsilon \sqrt{K}}exp(-\frac{%
\varepsilon }{2}\int \sqrt{K}dz)\text{ \ \ \ \ \ \ \ \ }\Rightarrow \text{ \
\ \ \ \ \ }\lambda _{2}=0\text{ .}  \tag{322}
\end{equation}
If $\ \lambda _{2}=0,$ then from (297) and (299) it can be found that
\begin{equation}
\lambda _{1}=-\alpha +\varepsilon \rho (z)\text{ .}  \tag{323}
\end{equation}
But perhaps it is more interesting to see from (270)) and (271) that for the
particular case it will follow
\begin{equation}
P=Q=0\text{ \ \ \ \ \ \ \ \ \ \ }\rho ^{^{\prime }}(z)=0\text{ \ \ \ \ \ \ \
\ \ \ \ \ }\Rightarrow \rho (z)=const.\text{ .}  \tag{324}
\end{equation}
This represents the \textit{first found} very nice and simple condition
under which a parametrization of (256) with the Weierstrass function is
possible. The \textit{second possibility} for equation (320) to hold is when
\begin{equation}
4\overline{B}_{3}-\overline{B}_{2}\frac{\overline{c}}{\overline{d}}\frac{%
\overline{b}}{\overline{d}}=\frac{[9\overline{B}_{3}^{2}\overline{B}%
_{1}^{2}+9\overline{B}_{1}\overline{B}_{3}\overline{B}_{2}^{2}-\overline{B}%
_{2}^{4}]}{9\overline{B}_{1}^{2}\overline{B_{3}}}=0\text{ .}  \tag{325}
\end{equation}
Additionally, if equations (277), (279-280) and (273) for $H$ are taken into
account, the above equation can be represented in the following form in
respect to the integration constant $\lambda _{2}$%
\begin{equation}
-16.3^{3}A^{2}\lambda _{2}^{2}=0\text{ .}  \tag{326}
\end{equation}
Again, since $A\neq 0$, we come to the case $\lambda _{2}=0$.

\section*{ XIX. ANOTHER \ NECESSARY \ CONDITION \\
FOR \ PARAMETRIZATION \ OF \ (256) \\
AND APPLICATION  OF  THE \ THEORY \\
OF \ RIEMANN  SURFACES}

\bigskip Now the \textit{second possibility}$\,$\ when equation (312) holds
shall be investigated. If one denotes $\rho (z)\equiv w$, then the equation
\begin{equation}
S_{1}w^{4}+S_{2}w^{2}+S_{3}w+S_{4}=0  \tag{327}
\end{equation}
and the couple of \textit{complex variables }$(w,z)$ determine a Riemann
surface for the equation
\begin{equation}
w=F(z,S_{1}(z),S_{2}(z),S_{3}(z),S_{4}(z))\text{ ,}  \tag{328}
\end{equation}
which represents a solution of the complex algebraic equation (327). Since
the equation always has complex solutions and the Weierstrass function in
principle is also a complex function, let us write $w$ as
\begin{equation}
\rho (z)\equiv w\equiv w_{1}(z)+iw_{2}(z)\text{ \ .}  \tag{329}
\end{equation}
Remember also that no matter that from the original definition of the
Weierstrass function in (35) it is not evident that the ''imaginary'' part $%
w_{2}(z)$ is present, this can be guessed from the presence of complex
poles, which after decomposing into real and imaginary numbers will give the
$"w_{1}(z)"$ and $"w_{2}(z)"$ terms. If one substitutes (329) into (327),
then equation (327) for the Riemann surface ''splits up'' into two equations
for a couple of Riemann surfaces, determined by the complex pairs $(w_{1},z)$
and $(w_{2},z)$. The obtained equations for the \textit{real }and \textit{%
imaginary} parts of the complex algebraic equation (327) are respectively
\begin{equation}
S_{1}(w_{1}^{2}-w_{2}^{2})^{2}-4S_{1}w_{1}^{2}w_{2}^{2}+S_{2}(w_{1}^{2}-w_{2}^{2})+S_{3}w_{1}+S_{4}=0%
\text{ ,}  \tag{330}
\end{equation}
\begin{equation}
w_{2}\left[ 4S_{1}w_{1}(w_{1}^{2}-w_{2}^{2})+2S_{2}w_{1}+S_{3}\right] =0%
\text{ .}  \tag{331}
\end{equation}
The last equation for the imaginary part is identically satisfied when $%
w_{2}=0$. Then the corresponding $w_{1}$ has to be determined from (330),
which becomes a quartic equation
\begin{equation}
S_{1}w_{1}^{4}+S_{2}w_{1}^{2}+S_{3}w_{1}+S_{4}=0  \tag{332}
\end{equation}
with roots $x_{1},x_{2},x_{3},x_{4}$, satisfying the relation $%
-x_{3}-x_{4}=x_{1}+x_{2}=u$, and according to the general theory$^{13}$ the
roots can be found after solving the equation
\begin{equation}
u^{6}+2\frac{S_{2}}{S_{1}}u^{4}+[(\frac{S_{2}}{S_{1}})^{2}-4\frac{S_{4}}{%
S_{1}}]u^{2}-(\frac{S_{3}}{S_{1}})^{2}=0\text{.}  \tag{333}
\end{equation}
The other case when (331) is identically zero is when the expression in the
square brackets is zero, and from there
\begin{equation}
w_{2}^{2}=\frac{4S_{1}w_{1}^{3}+2S_{2}w_{1}+S_{3}}{4S_{1}w_{1}}\text{ .}
\tag{334}
\end{equation}
Again, substituting (334) into (330) for the real part and setting up $%
w_{1}^{2}\equiv \widetilde{w}_{1},$ one can obtain the following third-rank
equation for $\widetilde{w}_{1}$%
\begin{equation}
64S_{1}^{3}\widetilde{w}_{1}^{3}+32S_{1}^{2}S_{2}\widetilde{w}%
_{1}+(4S_{1}S_{2}^{2}-16S_{1}^{2}S_{4})\widetilde{w}_{1}-S_{1}S_{3}^{2}=0%
\text{ .}  \tag{335}
\end{equation}
This equation has three complex roots for $\widetilde{w}_{1}$ and therefore
- six roots for $w_{1}.$ Note also that each time when $\varepsilon $
appears in the coefficient functions, then for each function $%
S_{1},S_{2},S_{3},S_{4}$ and for every root, the corresponding expressions
for the roots have to be taken once with $\varepsilon =1$ and once with $%
\varepsilon =-1.$ So there will be a finite number of combinations, but more
than six.

\section*{XX. NECESSARY \ AND  SUFFICIENT \\
CONDITIONS  FOR \ \ PARAMETRIZATION  \\
WITH  A\ CONSTANT  WEIERSTRASS \\
FUNCTION \ AND \ THE \ RESULTING \\
NONLINEAR \ EQUATIONS}

\bigskip In this section it shall be assumed that a constant Weierstrass
function $\rho (z)\equiv const\equiv w\equiv w_{1}(z)+iw_{2}(z)$
parametrizes the algebraic equation (327) , defined over the Riemann surface
$(w,z)$. If the Weierstrass function is constant then it follows that
\begin{equation}
\frac{\partial w}{\partial z}=\frac{\partial \overline{w}}{\partial z}\text{
,}  \tag{336}
\end{equation}
and therefore
\begin{equation}
\frac{\partial w_{1}}{\partial z}+i\frac{\partial w_{2}}{\partial z}=\frac{%
\partial w_{1}}{\partial z}-i\frac{\partial w_{2}}{\partial z}=0\text{ .}
\tag{337}
\end{equation}
Consequently,
\begin{equation}
\frac{\partial w_{1}}{\partial z}=\frac{\partial w_{2}}{\partial z}=0\text{ .%
}  \tag{338}
\end{equation}

Taking this into account and differentiating by $z$ equation (334) for $%
w_{2} $, one can obtain
\begin{equation}
w_{1}=-\frac{X^{^{\prime }}}{Y^{^{\prime }}}\text{ ,}  \tag{339}
\end{equation}
where $X$ and $Y$ denote
\begin{equation}
X=\frac{S_{3}}{4S_{1}}\text{ \ \ \ \ \ \ \ \ \ \ \ \ \ \ \ \ \ \ }Y=\frac{%
S_{2}}{2S_{1}}\text{ ,}  \tag{340}
\end{equation}
and $^{\prime }$ means a derivative in respect to the complex variable $z.$
Denoting $\beta =w_{2}^{2}=const$ and inserting back $w_{1}$ from (339) into
(334), one can obtain the following \textit{nonlinear equation}
\begin{equation}
\frac{(X^{^{\prime }})^{2}}{(Y^{^{\prime }})^{2}}-X\frac{Y^{^{\prime }}}{%
X^{^{\prime }}}+Y-\beta =0\text{ .}  \tag{341}
\end{equation}
Introducing the notation $Z=\frac{X^{^{\prime }}}{Y^{^{\prime }}}$ allows us
to rewrite the equation as a \textit{cubic algebraic equation} in respect to
$Z$
\begin{equation}
Z^{3}+(Y-\beta )Z-X=0\text{ .}  \tag{342}
\end{equation}
Finding the roots of this equation, one can easily express $X^{^{\prime }}$
as
\begin{equation}
X^{^{\prime }}=\widetilde{F}(X,Y)Y^{^{\prime }}\text{ ,}  \tag{343}
\end{equation}
where $\widetilde{F}_{1}(X,Y)$ is the expression for the first root of the
cubic equation
\begin{equation}
\widetilde{F}_{1}(X,Y)=\sqrt[3]{\frac{X}{2}+\sqrt{\frac{X^{2}}{4}+\frac{%
(Y-\beta )^{3}}{27}}}+\sqrt[3]{\frac{X}{2}-\sqrt{\frac{X^{2}}{4}+\frac{%
(Y-\beta )^{3}}{27}}}\text{ .}  \tag{344}
\end{equation}
The other roots are correspondingly
\begin{equation}
\widetilde{F}_{2}(X,Y)=-\sqrt[3]{-\frac{X}{2}+\sqrt{\frac{X^{2}}{4}-\frac{%
(Y-\beta )^{3}}{27}}}-\sqrt[3]{-\frac{X}{2}-\sqrt{\frac{X^{2}}{4}-\frac{%
(Y-\beta )^{3}}{27}}}\text{ ,}  \tag{345}
\end{equation}
\begin{equation}
\widetilde{F}_{3}(X,Y)=\sqrt[3]{\frac{X}{2}+\sqrt{\frac{X^{2}}{4}+\frac{%
(Y-\beta )^{3}}{27}}}-\sqrt[3]{\frac{X}{2}-\sqrt{\frac{X^{2}}{4}+\frac{%
(Y-\beta )^{3}}{27}}}\text{ .}  \tag{346}
\end{equation}
Further the calculations will be done only for the first root $\widetilde{F}%
_{1}(X,Y).$

\bigskip Now let us perform a differentiation by $z$ of equation (330) for
the real part, taking into account also the defining expressions (340),
(339) and introducing also the notation $T\equiv \frac{S_{4}}{S_{1}}.$ The
result be another nonlinear differential equation
\begin{equation}
2YY^{\prime }-4\frac{(X^{^{\prime }})^{2}}{Y^{^{\prime }}}+T^{^{\prime }}=0%
\text{ .}  \tag{347}
\end{equation}
Equations (341) and (347) represent a \textit{system of coupled nonlinear
differential equations for }$X$\textit{\ and }$Y.$ Now we are going to
establish an interesting mathematical property of this system : due to the
peculiar structure of equation (341) as an algebraic equation with respect
to the ratio of the derivatives of $X$ and $Y$ and subsequently expressing \
$X^{^{\prime }}$ as linearly proportional to $Y^{^{\prime }}$ (with a
coefficient function $\widetilde{F}$ depending only on $X$ and $Y$)$,$ this
system will turn out to be an integrable one! So, we shall treat equations
(343) and (347) instead of (341) and (347) . Substituting (343) into (347),
we receive
\begin{equation}
2YY^{^{\prime }}-4\widetilde{F}^{2}Y^{^{\prime }}+T^{^{\prime }}=0\text{ .}
\tag{348}
\end{equation}
Fortunately, this equation can be integrated to give
\begin{equation}
T+f(X,....)=-Y^{2}+\frac{4}{27}i^{\frac{2}{3}}(Y-\beta )^{2}+4I_{1}+4I_{2}
\tag{349}
\end{equation}
and the integrals can be written as
\begin{equation}
I_{1},I_{2}\equiv \int \left( \frac{X}{2}\mp \sqrt{\frac{X^{2}}{4}+\frac{%
(Y-\beta )^{3}}{27}}\right) ^{\frac{2}{3}}dY\text{ .}  \tag{350}
\end{equation}
The integration function $f(X,...)$ depends on $X$ and other constants or
functions, different from $Y$. Now it remains to calculate (analytically)
the remaining two integrals. For the purpose, let us perform the following
three consequent variable changes
\begin{equation}
\frac{Y-\beta }{3}\equiv t\text{ ,}  \tag{351}
\end{equation}
\begin{equation}
\widetilde{t}\equiv (\frac{4}{X^{2}})^{\frac{1}{3}}t  \tag{352}
\end{equation}
and
\begin{equation}
\overline{t}\equiv \sqrt{1+\widetilde{t}^{3}}=\sqrt{1+\frac{4}{X^{2}}\frac{%
(Y-\beta )^{3}}{27}}\text{ .}  \tag{353}
\end{equation}
It is straightforward to check that the integrals can be brought to the
following form
\begin{equation}
I_{1}=\int (1+\overline{t})^{\frac{2}{3}}\overline{t}d\overline{t}=\frac{3}{5%
}\overline{t}(1+\overline{t})^{\frac{5}{3}}\mid _{z=z_{1}}^{z=z_{2}}-\frac{3%
}{5}\int (1+\overline{t})^{\frac{5}{3}}d\overline{t}\text{ ,}  \tag{354}
\end{equation}
\begin{equation}
I_{2}=\int (1-\overline{t})^{\frac{2}{3}}\overline{t}d\overline{t}=-\frac{3}{%
5}\overline{t}(1-\overline{t})^{\frac{5}{3}}\mid _{z=z_{1}}^{z=z_{2}}-\frac{3%
}{5}\int (1-\overline{t})^{\frac{5}{3}}d\overline{t}\text{ .}  \tag{355}
\end{equation}
Note that if a closed contour is chosen, then the first terms in the above
two integrals will be zero. If the integrals are calculated and expression
(353) is used, then the following expression is obtained
\begin{equation}
I_{1}+I_{2}=-\frac{9}{40}\left\{ \left( 1+\sqrt{1+\frac{4}{X^{2}}\frac{%
(Y-\beta )^{3}}{27}}\right) ^{\frac{8}{3}}+\left( 1-\sqrt{1+\frac{4}{X^{2}}%
\frac{(Y-\beta )^{3}}{27}}\right) ^{\frac{8}{3}}\right\} \text{ .}  \tag{356}
\end{equation}
Substituting back into expression (349), one obtains a solution of the
nonlinear differential equation (348), which depends on terms in powers of
fractional numbers.

\section*{XXI. CONCLUSION\protect\bigskip}

Let us summarize the obtained results.

In this paper a cubic algebraic equation has been obtained in respect to the
differentials $dX^{i}$ of some generalized coordinates $X^{i}.$ The
derivation of the equation was possible due to the representation of the
contravariant metric tensor in terms of differential quantities. Also, in
Sect. III the equation was derived upon assuming that $dX^{i}$ is either an
exact differential, or that $dX^{i}$ are \textbf{zero-helicity vector field
components}.

The derived equation (20) clearly reflects the structure of the
gravitational Lagrangian, and can be regarded as an equation for its all
possible coordinate transformations (admissable parametrizations), provided
the Christoffell's connection $\Gamma _{ij}^{k}$ and the Ricci tensor $R_{ij%
\text{ }}$ are given.

The main problems, which one encounters when investigating such algebraic
equations are several, and in this paper only one of them is resolved in
more details.

The \textbf{first }and most serious problem is that the equation is defined
on an algebraic variety of several variables, since in gravity theory one
usualle deals with at least four-dimensional (and higher-dimensional also)
manifolds. At the same time, the standard and known methods from algebraic
geometry for parametrizing algebraic curves by means of the Weierstrass
function concern only algebraic curves of two variables. That is why in the
paper one of the variables - $dx^{5}$ has been singled out ot the base of
the physical consideration of the Randall-Sundrum models, and the other
variable for convenience is chosen to be the ratio $\frac{a}{c}$ of the
functions $a(z)$ and $c(z)$, which enter in the \textbf{linear-fractional
transformation \ }of $dx^{5}.$ The rest of the variables $%
dx^{1},dx^{2},dx^{3},dx^{4}$ enter the cubic equation in scalar quantities
(functions). Of course, if a two-dimensional manifold is considered, then
the variables might be related to the two variables of the manifold. Such an
analyses of a two dimensional algebraic equation and its parametrization may
find application in string and brane theory (also in gravity theory) .

For the purpose of higher-dimensional \ algebraic varieties and equations,
probably methods from the theory of abelian varieties and hyper-elliptical \
(Weierstrass) functions have to be applied. However, these methods are
developed at an abstract mathematical level,$^{42}$ far from being adjusted
to any concrete application.

The \textbf{second problem} concerns the methods for bringing the algebraic
equation to a parametrizable form.The standard approach of applying a
\textbf{linear-fractional transformation} has been chosen for the purpose,
but in Sect. XII \ it was demonstrated also how a \textbf{linear
transformation} can also be applied. In the last case, a couple of cubic
equations is investigated, and what is interesting is that by means of
parametrization of the first equation with the Weierstrass function $\rho
(z) $ the second equation is obtained in a parametrizable form, but in the
general form of a cubic equation in respect to $\rho (z).$ This justifies
the performed investigation in Sect. VIII, concerning the (eventual)
possibility for the Weierstrass function to satisfy a cubic equation of the
general kind (108). Sect. XVI - XIX demonstrate how a differential equation
with the Weierstrass function can be integrated, and after performing an
additional linear-fractional transformation and finding the relation between
the two integration constants in Sect. XVII, one comes to a purely algebraic
(and not differential) equation. Interestingly, in some special cases, for
example the case of parametrization with a constant in Weierstrass function
in Sect. XX, one can effectively work with the \textit{approach of Riemann
surfaces} and even to obtain an integrable equation at the end. It should be
noted here that this is possible in some cases only. For example, if one
changes the formulation of the problem and would like to find the conditions
for parametrization with a real-valued Weierstrass function, then it may be
shown that the equation will not be an integrable one. This is not performed
in this paper.

As for the \textbf{linear-fractional transformation}, its advantages are the
following: 1. It contains more parameters (in the case - functions $a,b,c,d$
of a complex variable) and it makes possible to take account of the \textbf{%
point at infinity}. In principle, in complex analyses and projective
geometry this is a well-developed procedure, but as far as physical
applications in gravity theory and in relativistic hydrodynamics are
concerned - this is still un unexplored area. 2. In Sect. VI it was proved
that by means of a suitable change of variables it is possible to derive a
\textbf{second-order (quadratic) algebraic equation in }terms of
''angular''- type\textbf{\ \ }variables from the initial cubic algebraic
equation. Since a quadratic equation is easier to deal with, this simplifies
the analyses and moreover, a transition to the original variables can also
be performed.

The \textbf{third problem, }investigated in more details in the present
paper, is the \textbf{form of the parametrizable cubic curve}. This was
discussed in the Introduction, and evidently a concrete physical problem
from gravity theory \ has shown the necessity to investigate the case when $%
g_{2}$ and $g_{3}$ are complex functions and not complex numbers, as it is
in the standard theory of elliptic functions. In regard to this, in Sect.
VIII and \ Sect. IX two mutually related with each other problems for
resolving are being stated: 1. Can the Weierstrass function $\rho (z)$
parametrize an \textbf{arbitrary cubic curve} with coefficient functions of
a \textbf{complex variable }? 2. Can the Weierstrass function parametrize
the well-known parametrizable form of the cubic equation, but again with
coefficient functions depending on a complex variable? Although the explicit
form of the equations for the Loran coefficient functions are presented in
Sect. VIII, the first question still remains unanswered, and perhaps
computer simulations only can help for its resolution. As for the second
question, the answer is affirmative, and after solving a system of algebraic
equations for various values of $\ m$ and $n$, the explicit form of all the
Loran coefficient functions $c_{m}^{(1)}$ , $c_{m}^{(0)}$ was found, both
from the negative- power and the positive- power expansion. A confirmation \
of \ the consistency of the derived equations is equation (153) for a value
of $m=-6$, which is being satisfied by the previously derived equations.
However, the values of the coefficient functions are perhaps not so
important as the result, which follows from this calculation, namely:
\textbf{the infinite sums }$G_{1}$\textbf{and }$G_{2}$\textbf{, which in the
general case might be divergent, in the particular case of the
''parametrizable'' form of the cubic algebraic equation with }$%
g_{2}=g_{2}(z) $\textbf{\ and }$g_{3}=g_{3}(z)$\textbf{, should be convergent%
}! This fact, although of pure mathematical nature, probably deserves more
attention and further elaboration from another point of view and by applying
different mathematical approaches. The finiteness of $G_{1}$ and $G_{2}$ is
not imposed ''by hand'', but is obtained as a consequence of the fulfillment
of the above mentioned equation with the Weierstrass function. \textbf{\ }

It is to be noted also that if some other assumptions are made - for example
in Sect. XI about poles at infinity in the positive-power decomposition of $%
g_{2}$\bigskip $(z)$ and $g_{3}(z)$, then within the large $n$ asymptotic
approximation, a method for \textbf{comparing \ the convergency radius} of
two infinite sums may be used for the determination of the coefficient
functions $c_{-k}^{(1)}$. Comparing the expressions for $c_{-k}^{(1)}$ in
the negative- power Loran decomposition and in the other case of positive- \
power Loran decomposition with poles at infinity, it is seen that in the two
cases $c_{-k}^{(1)}$ are expressed in different ways. For example, in the
first case from (143)\ $c_{-k}^{(1)}$ is expressed through the sums $G_{k}$,
while in the second (asymptotic) case $G_{n}$ do not appear (185), and
instead of them the Gamma function $\Gamma (l)$ appears and some finite and
infinite summation formulaes of the kind $\sum%
\limits_{k=0}^{l-1}l(l-1)...(l-k)$ and $\sum\limits_{n=0}^{\infty }\frac{%
(-1)^{n}}{n!(l+1+n)}$. Therefore, the assumption about \textbf{poles at
infinity} and the consequent appearence of the zeta-function substantially
changes the calculations. However, while the infinite point of the
linear-fractional transformation may have some physical justification, the
''poles at infinity'' case for the moment does not have a definite physical
meaning.

\bigskip

\section*{ACKNOWLEDGMENTS}

\bigskip

\bigskip The author is greateful to Dr. L. Alexandrov and Dr. D. M. Mladenov
(BLTP, Joint Institute for Nuclear Research, Dubna) for useful comments and
their interest towards this work, to Prof. N.A. Chernikov (BLTP, Joint
Institute for Nuclear Research, Dubna) for his support and encouragement and
to Dr.V.I. Zhuravliev and the Directorate of the Bogoliubov Laboratory for
Theoretical Physics for their support during his stay in Dubna (Russia).

The work on this paper was supported also by a Shoumen University grant
under contract  005/2002. The author is greateful also to Dr. P.L. Bozhilov
(Shoumen University, Bulgaria) for the possibility to participate in this
grant.

\section*{APPENDIX A: ADDITIONAL SYSTEM OF EQUATIONS FOR \ m= -3, -1}

\bigskip

In this Appendix the system of equations for $m=-3,-1$ will be presented,
which were not investigated in Sect. IX. However, the method for their
derivation is completely the same, but some new interesting consequences
will appear.

\bigskip For the purpose, let us first rewrite the two sums in expression
(111) for $\left[ \rho ^{^{\prime }}(z)\right] ^{2}$, putting in the first
sum $2(n-1)=m$ and in the second term $n-4=m$. Then expression (111)
acquires the form
\begin{equation*}
\left[ \rho ^{^{\prime }}(z)\right] ^{2}=\frac{4}{z^{6}}+\frac{1}{16}%
\sum\limits_{m=0,2,4,...}^{\infty }G_{m+6}(m+2)^{2}(m+4)^{2}z^{m}-
\end{equation*}
\begin{equation}
-4\sum\limits_{m=-3,-2,-1,0,1,..}G_{m+6}(m+4)(m+5)z^{m}\text{ .}  \tag{A1}
\end{equation}
For $m=-3$ only the term from the second sum in (A1) will contribute.
Putting also $m=-3$ in the R.H.S. of expression (118) for $M(z)\rho
^{3}+N(z)\rho ^{2}+P(z)\rho +E(z)$ \ for the case of $M=4,$ $N=0,$ $%
P(z)\equiv -g_{2}(z)$ and $E(z)\equiv -g_{3}(z)$ (i.e. $c_{m}^{(2)}=0$ for
all $m$ , $c_{0}^{(3)}\equiv 4$ and $c_{m}^{(3)}\equiv 0$ for $m\neq 0$)$,$%
one has to take into account that terms with a ''negative-valued'' indice
like $c_{-1-2n}^{(3)},c_{-1-n}^{(3)}$, $c_{-3-3n}^{(3)}$ ($n=1,2....)$ are
zero.

The obtained equation for $m=-3$ is
\begin{equation}
-8G_{3}=2(n+1)G_{n}c_{1-n}^{(3)}+c_{-1}^{(1)}+c_{-3-n}^{(1)}G_{n}+c_{-3}^{(0)}%
\text{ .}  \tag{A2}
\end{equation}
For $n=1$, when the first term on the R.H.S. is non-zero, the equation is
\begin{equation}
-8G_{3}=16G_{1}+c_{-1}^{(1)}+c_{-4}^{(1)}G_{1}+c_{-3}^{(0)}\text{ .}
\tag{A3}
\end{equation}
Since from expressions (142) and (136) from Section 7 $c_{-3}^{(1)}$ and $%
c_{-4}^{(1)}$ can be found, from the above equation (A3) $c_{-3}^{(0)}$ can
be expressed.

For $n=p>1$ equation (A2) is
\begin{equation}
-8G_{3}=c_{-1}^{(1)}+c_{-3-p}^{(1)}G_{p}+c_{-3}^{(0)}\text{ .}  \tag{A4}
\end{equation}
Substracting the two equations (A3-A4), one can find the following
expression for $c_{-3-p}^{(1)}$:
\begin{equation}
c_{-3-p}^{(1)}=\frac{G_{1}}{G_{p}}(16+c_{-4}^{(1)})=16\frac{G_{1}}{G_{p}}%
\frac{\left[ \frac{G_{2}}{G_{4}}G_{p}+3(\frac{G_{2}}{G_{4}})^{2}G_{p}-(3%
\frac{G_{2}}{G_{4}}+21)G_{p-2}\right] }{\frac{G_{2}}{G_{4}}G_{p}-G_{p-2}}%
\text{ .}  \tag{A5}
\end{equation}
An expression for $c_{-3-p}^{(1)}$ can also be found from formulae (143) for
$k=p+3$
\begin{equation}
c_{-3-p}^{(1)}=-\frac{16G_{4}}{G_{p+3}-\frac{G_{4}}{G_{3}}G_{p+1}}\text{ .}
\tag{A6}
\end{equation}
From the equality of the two expressions, $G_{1}$ can again be expressed as
a convergent expression. Note also that the formulaes $G_{2p}=G_{3}(\frac{%
G_{4}}{G_{2}})^{\frac{2p-3}{2}}$ and $G_{2p+1}=G_{3}(\frac{G_{4}}{G_{2}}%
)^{p-1}$from (151) satisfy the equality expression since then the
denominators in (A5) and (A6) will be zero. This precludes the investigation
of the system of equations for $m=-3.$

For $m=-1$, the general equation can be written as
\begin{equation}
--48G_{5}=2(n+1)G_{n}c_{3-n}^{(3)}+(n+1)G_{n}c_{1-n}^{(3)}+G_{n}c_{-1-n}^{(1)}+c_{1}^{(1)}+c_{-1}^{(0)}%
\text{ ,}  \tag{A7}
\end{equation}
and for $n=1$ and $n=2$ the corresponding equations are
\begin{equation}
-48G_{5}=8G_{1}+c_{-2}^{(1)}G_{1}+c_{1}^{(1)}+c_{-1}^{(0)}\text{ ,}  \tag{A8}
\end{equation}
\begin{equation}
-48G_{5}=G_{2}c_{-3}^{(1)}+c_{1}^{(1)}+c_{-1}^{(0)}\text{ .}  \tag{A9}
\end{equation}
The coefficient $c_{-3}^{(1)}$ can also be found from (143) for $k=3$
\begin{equation}
c_{-3}^{(1)}=-\frac{16G_{4}}{G_{3}-\frac{G_{4}}{G_{2}}G_{1}}  \tag{A10}
\end{equation}
. Substituting (A10) into (A9) gives an opportunity to express $%
c_{1}^{(1)}+c_{-1}^{(0)}$ as
\begin{equation}
c_{1}^{(1)}+c_{-1}^{(0)}=-48G_{5}+\frac{16G_{2}G_{4}}{G_{3}-\frac{G_{4}}{%
G_{2}}G_{1}}\text{ .}  \tag{A11}
\end{equation}
This expression, together with formulae (137) for $c_{-2}^{(1)}$ and (143)
for $c_{-1}^{(1)}$, represented as $c_{-1}^{(1)}=\frac{F}{G_{1}},$can be
substituted into the first equation (A8) to obtain the following quadratic
equation for $G_{1}$
\begin{equation}
(G_{3}-\frac{G_{4}}{G_{2}}G_{1})(\frac{G_{1}}{G_{2}}F-24G_{1})+16G_{2}G_{4}=0%
\text{ .}  \tag{A12}
\end{equation}
In a similar way, one can write down the equation for $n=3$
\begin{equation}
-48G_{5}=32G_{3}+c_{-4}^{(1)}G_{3}+c_{1}^{(1)}+c_{-1}^{(0)}\text{ .}
\tag{A13}
\end{equation}
Substituting $c_{-4}^{(1)}$ and $c_{1}^{(1)}+c_{-1}^{(0)}$ from (136) and
(A11), one can derive
\begin{equation}
2L=\frac{G_{2}}{G_{3}G_{4}}\frac{G_{k}}{G_{k-2}}-\frac{1}{G_{3}}  \tag{A14}
\end{equation}
where
\begin{equation}
L=\frac{G_{3}-\frac{G_{4}}{G_{2}}G_{1}}{7G_{3}(G_{3}-\frac{G_{4}}{G_{2}}%
)+2G_{2}G_{4}}\text{ .}  \tag{A15}
\end{equation}
From the above two equations a relation, similar to (147) can be obtained
\begin{equation*}
G_{k}=\beta G_{k-2}=\beta ^{\frac{2p-1}{2}}G_{1}\text{ for }k=2p
\end{equation*}
\begin{equation}
=\beta ^{p}G_{1}\text{ for }k=2p+1\text{ ,}  \tag{A16}
\end{equation}
where
\begin{equation}
\beta =\frac{G_{3}G_{4}}{G_{2}}(2L+\frac{1}{G_{3}})\text{ .}  \tag{A17}
\end{equation}
Of course, in order to have an unique determination of $G_{k},$one has to
require $\beta =\gamma $, where from (148) $\beta =2\frac{G_{4}}{G_{2}}-%
\frac{G_{3}}{G_{1}}$. This will result again in a quadratic equation for $%
G_{1}.$

Much more important and informative in the investigated case $m=-1$ turns
out to be the equation for a general $n>3$
\begin{equation}
-48G_{5}=c_{-1-n}^{(1)}G_{n}+c_{1}^{(1)}+c_{-1}^{(0)}\text{ .}  \tag{A18}
\end{equation}
Let us remind that an expression for $c_{-1-n}^{(1)}$ can be written from
(143)
\begin{equation}
c_{-n-1}^{(1)}=-\frac{16}{G_{n+1}-\frac{G_{4}}{G_{2}}G_{n-1}}\text{ .}
\tag{A19}
\end{equation}
Also, from (A11) one has an expression for $c_{1}^{(1)}+c_{-1}^{(0)}.$ These
two expressions can be substituted into equation (A18), which acquires the
form
\begin{equation}
G_{n}(G_{3}-\frac{G_{4}}{G_{2}}G_{1})=G_{2}G_{4}(G_{n+1}-\frac{G_{2}}{G_{4}}%
G_{n-1})\text{ .}  \tag{A20}
\end{equation}
Now it is interesting to note that using the formulaes
\begin{equation}
G_{n+1}=G_{2p+1}=G_{3}(\frac{G_{4}}{G_{2}})^{p-1};\text{ \ \ \ \ }%
G_{n-1}=G_{2p-1}=G_{3}(\frac{G_{4}}{G_{2}})^{p-2}\text{\ \ \ \ \ \ \ \ \ \ \
}  \tag{A21}
\end{equation}
from (151), it can easily be checked that the R.H. S. of (A20) is equal to
zero for the case $n=2p$. The other case $n=2p+1$ gives the same result.
Therefore, from (A20) the following concise relation is obtained, expressing
the proportionality of $G_{1}$ and $G_{2}$ with a coefficient of
proportionality the ratio $\frac{G_{3}}{G_{4}}$
\begin{equation}
G_{1}=\frac{G_{3}}{G_{4}}G_{2}\text{ .}  \tag{A22}
\end{equation}

\bigskip

\section*{APPENDIX B: ADDITIONAL SYSTEM OF EQUATIONS \ FOR \ m=2k}

\bigskip

This Appendix will preclude the proof, started in Sect. IX that all the
coefficient functions in the Loran function decomposition of the equation $%
\left[ \rho ^{^{\prime }}(z)\right] ^{2}=4\rho ^{3}-g_{2}(z)\rho -g_{3}(z)$
can be uniquely expressed, and especially those from the positive-power
decomposition.

For $m=2k>0$, the corresponding equation is
\begin{equation*}
(k+2[(k+1)^{2}(k+2)-8(2k+5)]G_{2k+6}=2(n+1)G_{n}c_{2(k+2)-n}^{(3)}+(n+1)^{2}G_{2n}c_{2(k+1)-2n}^{(3)}+
\end{equation*}
\begin{equation}
+(n+1)G_{n}c_{2(k+1)-n}^{(3)}+2(n+1)^{2}G_{2n}c_{2(k+1)-2n}^{(3)}+(n+1)^{3}G_{3n}c_{2k-3n}^{(3)}+c_{2(k+1)}^{(1)}+c_{2k-n}^{(1)}G_{n}+c_{2k}^{(0)}%
\text{ .}  \tag{B1}
\end{equation}
Additionally fixing the value of $n=k+1$, one can obtain from (B1)

\begin{equation}
(k+2)[(k+1)^{2}(k+2)-8(2k+5)]G_{2k+6}=12(k+2)^{2}G_{2(k+1)}+c_{2(k+1)}^{(1)}+G_{k+1}c_{k-1}^{(1)}+c_{2k}^{(0)}%
\text{ .}  \tag{B2}
\end{equation}
For $n=2(k+2)$ the equation is
\begin{equation}
(k+2)[(k+1)^{2}(k+2)-8(2k+5)]G_{2k+6}=8(2k+5)G_{2(k+2)}+c_{2(k+1)}^{(1)}+c_{-4}^{(1)}G_{2(k+2)}+c_{2k}^{(0)}%
\text{ .}  \tag{B3}
\end{equation}
Substracting the two equations, one can express $c_{k-1}^{(1)}$ as

\begin{equation}
c_{k-1}^{(1)}=\frac{1}{G_{k-1}}[%
-12(k+2)^{2}G_{2(k+1)}+8(2k+5)G_{2(k+2)}+c_{-4}^{(1)}G_{2(k+2)}]\text{ .}
\tag{B4}
\end{equation}
Since $k\geq 1$, from this formulae it is clear that all the Loran
coefficients $c_{m}^{(1)}$ in the positive-power decomposition can be
expressed, including the coefficient $c_{0}^{(1)}$, through which the
coefficients from the negative-power decomposition in Sect. IX were
expressed. Also, from (B4) $c_{2(k+1)}^{(1)}$ can be expressed (by
performing the indice change $k-1\rightarrow 2(k+1)$. If $c_{2(k+1)}^{(1)}$
is substituted back into equation (B3), one can express also the \textit{%
even positive-power coefficients} $c_{2k}^{(0)}$ as
\begin{equation*}
c_{2k}^{(0)}=(k+2)[(k+1)^{2}(k+2)-16(k+3)]G_{2k+6}-[c_{-4}^{(1)}+8(2k+5)]G_{2(k+2)}+
\end{equation*}
\begin{equation}
+\frac{1}{G_{2(k+2)}}[12(2k+5)^{2}G_{4(k+2)}-(32k+88+c_{-4}^{(1)})G_{2(2k+5)}%
]\text{ .}  \tag{B5}
\end{equation}
Now let us write down equation (B1) for another possible value of $n=2(k+1)$
\begin{equation}
(k+2)[(k+1)^{2}(k+2)-8(2k+5)]G_{2k+6}=4(2k+3)G_{2(k+1)}+c_{2(k+1)}^{(1)}+c_{2k}^{(0)}+c_{-2}^{(1)}G_{2(k+1)}%
\text{ .}  \tag{B6}
\end{equation}
Combining (B3) and (B6), $c_{-2}^{(1)}$ can be expressed as
\begin{equation}
c_{-2}^{(1)}=\frac{1}{G_{2(k+1)}}[%
-4(2k+3)G_{2(k+1)}+8(2k+5)G_{2(k+2)}+c_{-4}^{(1)}G_{2(k+2)}]\text{ .}
\tag{B7}
\end{equation}
The coefficient $c_{-4}^{(1)}$ can easily be calculated from (136) and (142)
to be
\begin{equation}
c_{-4}^{(1)}=\frac{16\left[ 3\frac{G_{2}}{G_{4}}%
(G_{2}G_{k}-G_{4}G_{k-2})-20G_{4}G_{k-2}\right] }{G_{2}G_{k}-G_{4}G_{k-2}}%
\text{ .}  \tag{B8}
\end{equation}
On the other hand, it is important to observe that $c_{-2}^{(1)}$ can be
calculated independently also from equations (100) and (142)
\begin{equation}
c_{-2}^{(1)}=\frac{12(G_{2}G_{k}-G_{4}G_{k-2})-16G_{4}G_{k}}{%
G_{2}G_{k}-G_{4}G_{k-2}}\text{ .}  \tag{B9}
\end{equation}
Note also that from relations (151) for $G_{2p}$ and $G_{2p+1}$ it follows
\begin{equation}
\frac{G_{k-2}}{G_{k}}=\frac{G_{2}}{G_{4}}\text{ for }k=2p\text{ \ and \ }%
k=2p+1  \tag{B10}
\end{equation}
or written in another way - $G_{2}G_{k}-G_{4}G_{k-2}=0$.

Setting up equal the two expressions (B7) and (B9) for $c_{-2}^{(1)}$,
cancelling the equal denominators and \textit{subsequently} taking into
account (B10), one can obtain the following concise recurrent relation
\begin{equation}
G_{k}=\alpha (k)G_{k-2}\text{ ,}  \tag{B11}
\end{equation}
where $\alpha (k)$ denotes
\begin{equation}
\alpha (k)=20\frac{G_{2(k+2)}}{G_{2(k+1)}}\text{ .}  \tag{B12}
\end{equation}
Continuing further the recurrent relation (B11), one can derive
\begin{equation*}
G_{k}=\alpha (k)\alpha (k-2).....\alpha (k-(k-3))G_{k-(k-1)}=
\end{equation*}
\begin{equation}
=coeff.\frac{G_{2(k+2)}}{G_{2(k+1)}}\frac{G_{2k}}{G_{2(k-1)}}\frac{G_{2(k-2)}%
}{G_{2(k-3)}}....\frac{G_{2.5}}{G_{2.4}}G_{1}\text{ .}  \tag{B13}
\end{equation}
If $k=2p$, the numerical coefficient in (B13) will be $20^{p}$.

A similar relation can be obtained by fixing $n=2(k+1)$. Then the
corresponding equation is
\begin{equation*}
(k+2)[(k+1)^{2}(k+2)-8(2k+5)]G_{2k+6}=
\end{equation*}
\begin{equation}
=4(2k+3)G_{2(k+1)}+c_{-1}^{(1)}G_{2(k+1)}+c_{2(k+1)}^{(1)}+c_{2k}^{(0)}\text{
.}  \tag{B14}
\end{equation}
Substracting this equation from (B16) for $n=2(k+1)$ and taking into account
expression (B4) for $c_{k-1}^{(1)}$, one can obtain
\begin{equation*}
\frac{G_{k+1}}{G_{k-1}}[c_{-4}^{(1)}+8(2k+5)]G_{2(k+2)}=
\end{equation*}
\begin{equation}
=G_{2(k+1)}[c_{-2}^{(1)}+4(2k+3)+12(k+2)^{2}\frac{(G_{k+1}-G_{k-1})}{G_{k-1}}%
\text{ ,}  \tag{B15}
\end{equation}
where $c_{-4}^{(1)}$ is given by (B8) and $c_{-1}^{(1)}$ by (B9).
Substituting the above expressions into (B15) and again taking into account
that $G_{2}G_{k}-G_{4}G_{k-2}=0$, one derives the following recurrent
relation
\begin{equation}
G_{2(k+1)}=20\frac{G_{k-1}}{G_{k+1}}\text{ .}  \tag{B16}
\end{equation}
If this relation is substituted into (B11), \ then it can be derived that
\begin{equation}
G_{k}^{2}G_{k+1}=20G_{k-1}G_{k-2}G_{k+2}\text{ .}  \tag{B17}
\end{equation}
This equality is valid for $k\geq 3.$ For $k=3,5,7,9$ the above relation may
be written as
\begin{equation}
G_{3}^{2}G_{8}=20G_{1}G_{5}\text{ \ \ \ \ \ \ \ \ \ \ \ \ \ \ \ \ \ \ \ \ \
\ \ \ \ \ }G_{5}^{2}G_{12}=G_{3}G_{7}\text{\ \ \ \ ,\ \ \ \ }  \tag{B18}
\end{equation}
\begin{equation}
G_{7}^{2}G_{16}=G_{5}G_{9}\text{ \ \ \ \ \ \ \ \ \ \ \ \ \ \ \ \ \ \ \ \ \ \
\ \ \ }G_{9}^{2}G_{20}=G_{7}G_{11}\text{\ \ \ \ \ \ \ .\ \ \ \ \ \ }
\tag{B19}
\end{equation}
Since on the L.H.S. of (B18) and (B19) $G_{8},G_{12},G_{16}$ and $G_{20}$
are zero, the R. H. S. should also be zero. If $G_{1}\neq 0,$ $G_{3}\neq 0$,
the R. H. S. of the first pair of equations (B18) equals to zero if $%
G_{5}=G_{7}=0$. But since $G_{5}$ and $G_{7}$ appear also in the R.H.S. of
the second pair of equations (B19), the R.H.S. will be zero and therefore $%
G_{9}$ and $G_{11}$ may be different from zero. The treatment of the
subsequent equations is analogous. That is why one may conclude that a pair
of even sums $G_{2l+1}$, $G_{2l+3}$ $(l\geq 2)$ is zero, but the next pair $%
G_{2l+5}$, $G_{2l+7}$ may be different from zero.

The last fixing of the value of $n=\frac{2k}{3}$\bigskip\ for the case $m=2k$
gives the equation

\begin{equation}
(k+2)[(k+1)^{2}(k+2)-8(2k+5)]G_{2k+6}=4(1+\frac{2k}{3}%
)^{3}G_{2k}+c_{2(k+1)}^{(1)}+c_{\frac{4k}{3}}^{(1)}G_{\frac{2k}{3}%
}+c_{2k}^{(0)}\text{ .}  \tag{B20}
\end{equation}
Substracting from (B20) equation \ (B6) for $n=2(k+1)$ and setting up $\frac{%
2k}{3}=p$, one can derive
\begin{equation}
\left[ 12(p+1)+c_{-2}^{(1)}\right]
G_{3p+2}-4(1+p)^{3}G_{3p}-c_{2p}^{(1)}G_{p}=0\text{ .}  \tag{B21}
\end{equation}
Similarly, substracting from (B20) equation (B3) for $n=2(k+2),$ one obtains
\begin{equation}
4(1+p)^{3}G_{3p}+c_{2p}^{(1)}G_{p}-8(3p+5)G_{3p+4}-c_{-4}^{(1)}G_{3p+4}=0%
\text{ .}  \tag{B22}
\end{equation}
From the two equations it follows
\begin{equation}
\left[ 12(p+1)+c_{-2}^{(1)}\right] G_{3p+2}=\left[ 8(3p+5)+c_{-4}^{(1)}%
\right] G_{3p+4}\text{ .}  \tag{B23}
\end{equation}
Again taking into account (B8) for $c_{-2}^{(1)}$ and (B8) for $c_{-4}^{(1)}$
for value of $k=3p$, one can obtain
\begin{equation}
G_{3p+4}=20\frac{G_{3p+2}}{G_{3p-2}}G_{3p}\text{ .}  \tag{B24}
\end{equation}
However, in view of the relations (B18-19) and the consequences from them,
the last relation will make sense only when \textit{each} of the indices $%
3p+4,3p+2,$ $3p-2,3p$ equals one of the indices $2l+5,2l+7$ ($l\geq 2$) and
then the relation (B24) will be nonzero.

\section*{APPENDIX C: ADDITIONAL SYSTEM OF EQUATIONS FOR \ m=2k+1 AND m=-k}

\bigskip

For $m=2k+1$ the corresponding equation is
\begin{equation*}
-8(2k+5)(k+3)G_{2k+7}=c_{2k+5}^{(3)}+2(n+1)G_{n}c_{2k+5-n}^{(3)}+(n+1)^{2}G_{2n}c_{2k+3-2n}^{(3)}+
\end{equation*}
\begin{equation*}
+(n+1)G_{n}c_{2k+3-n}^{(3)}+2(n+1)^{2}G_{2n}c_{2k+3-2n}^{(3)}+(n+1)^{3}G_{3n}c_{2k+1-3n}^{(3)}+
\end{equation*}
\begin{equation}
+c_{2k+3}^{(1)}+c_{2k+1-n}^{(1)}G_{n}+c_{2k+1}^{(0)}\text{ .}  \tag{C1}
\end{equation}
The important conclusion, which can be made from this equation is the
following: \textit{if }$c_{2k+3}^{(1)}$\textit{\ is calculated from (B4) for
value of }$k^{^{\prime }}-1=2k+3$\textit{, then the odd number coefficients }%
$c_{2k+1}^{(0)}$\textit{\ can also be found!} Remember also that in Appendix
B only the even number coefficients $c_{2k}^{(0)}$ were found (form. B5). In
order to express $c_{2k+1}^{(0)}$, it is enough to set up $n=2k+3$ in (C1),
when from all the coefficients $c_{m}^{(3)}$ only the second term on the
R.H.S. will be non-zero. Then
\begin{equation}
c_{2k+1}^{(0)}=-c_{2k+3}^{(1)}-c_{-4}^{(1)}G_{2k+5}-8(k+3)(2k+5)G_{2k+7}-16(k+3)G_{2k+5}%
\text{ .}  \tag{C2}
\end{equation}
For another value of $n=\frac{2k+3}{2}$, equation (C1) acquires the
following form
\begin{equation}
-8((k+3)(2k+5)G_{2k+7}=3(2k+5)^{2}G_{2k+3}+c_{2k+3}^{(1)}+c_{\frac{2k-1}{2}%
}^{(1)}G_{\frac{2k+3}{2}}+c_{2k+1}^{(0)}\text{ .}  \tag{C3}
\end{equation}
But since $c_{\frac{2k-1}{2}}^{(1)}$ and $G_{\frac{2k+3}{2}}$ have to be
integer numbers, this will be possible if for example $2k-1=2p$. For this
value of $k$, one can express $c_{2p+2}^{(0)}$ from (C3)
\begin{equation}
c_{2p+2}^{(0)}=-c_{2p+4}^{(1)}-c_{p}^{(1)}G_{p+2}-12(p+3)^{2}G_{2p+4}-8(p+3)(2p+7)G_{2p+8}%
\text{ .}  \tag{C4}
\end{equation}
However, $c_{2p+4}^{(0)}$ can be expressed also from equation (B2) for
values of $m=2k$, $n=k+1$ and $k=p+2$
\begin{equation*}
c_{2p+2}^{(0)}=-c_{2p+4}^{(1)}-c_{p}^{(1)}G_{p+2}-12(p+3)^{2}G_{2p+4}+
\end{equation*}
\begin{equation}
+(p+3)\left[ (p+3)(p+2)^{2}-8(2p+7)\right] G_{2p+8}\text{ .}  \tag{C5}
\end{equation}
From the two equations (C4-C5), one easily obtains
\begin{equation}
16(p+3)(2p+7)G_{2p+8}=0\text{ .}  \tag{C6}
\end{equation}
Since the coefficient in front of $G_{2p+8}$ is a positive one, (C6) will be
fulfilled if
\begin{equation}
G_{2p+8}=G_{2k+6}=0\text{ .}  \tag{C7}
\end{equation}
The last means that \textit{the even-number sums }$%
G_{6},G_{8},G_{10},....... $\textit{\ are zero! }

Again fixing the value of $n=2k+3$, one derives from (C1) the equation
\begin{equation}
-8(k+3)(2k+5)G_{2k+7}=[8(k+2)+c_{-2}^{(1)}]G_{2k+3}+c_{2k+3}^{(1)}+c_{2k+1}^{(0)}%
\text{ .}  \tag{C8}
\end{equation}
Combining this equation with (C3), setting up $2k-1=2p$, one can express $%
c_{p}^{(1)}$%
\begin{equation}
c_{p}^{(1)}=-\frac{(A(p)-c_{-2}^{(1)})G_{2p+4}}{G_{p+2}}\text{ ,}  \tag{C9}
\end{equation}
\qquad \qquad \qquad \qquad\ where
\begin{equation}
A(p)=3(2p+1)^{2}+26(2p+1)+59\text{ .}  \tag{C10}
\end{equation}
Comparing this expression with formulae (B4) for $c_{p}^{(1)}$ and taking
into account (B8) and (B9), one derives the following relation
\begin{equation}
G_{2(p+3)}=\frac{1}{20}\frac{G_{2(p+2)}}{G_{2(p-2)}}\text{ .}  \tag{C11}
\end{equation}
Expressing by means of (B15) $G_{2(p+2)}$ and $G_{2(p-2)}$ and substituting
into (C11), one can obtain the relation also in another form
\begin{equation}
G_{2(p+3)}=\frac{1}{20}\frac{G_{p+2}}{G_{p}}\frac{G_{p-4}}{G_{p-2}}\text{ ,}
\tag{C12}
\end{equation}
and from the two expressions one can obtain also the ratio $\frac{G_{2(p+2)}%
}{G_{2(p-2)}}$ without any numerical coefficients. Note also that (C11-C12)
refer to non-zero even numbers of $G_{m}$ since we have $2p=2k-1$, and the
relations should be written in respect to $k$ and not $p$.

The corresponding equation is
\begin{equation}
-8(k+3)(2k+5)G_{2k+7}=\frac{4}{27}(2k+4)^{3}G_{2k+1}+c_{2k+3}^{(1)}+G_{\frac{%
2k+1}{3}}c_{\frac{4k+2}{3}}^{(1)}+c_{2k+1}^{(0)}\text{ .}  \tag{C13}
\end{equation}
Setting up $2k+1=3p$ and keeping in mind that $c_{2p}^{(1)}$ and $%
c_{3p+2}^{(1)}$ can be found from (B4), one can express $c_{3p}^{(0)}$ as
\begin{equation}
c_{3p}^{(0)}=-\frac{4}{3}%
(3p+4)(3p+5)G_{3p+6}-4(p+1)^{3}G_{3p}-c_{3p+2}^{(1)}-G_{p}c_{2p}^{(1)}\text{
.}  \tag{C14}
\end{equation}
The equation for the last case of $m=-k$ ($k>3,$ $k\neq 6$) is
\begin{equation*}
c_{4-k}^{(3)}+2(n+1)G_{n}c_{-k+4-n}^{(3)}+(n+1)^{2}G_{2n}c_{-k+2-2n}^{(3)}+(n+1)G_{n}c_{-k+2-n}^{(3)}+
\end{equation*}
\begin{equation}
+2(n+1)^{2}G_{2n}c_{-k+2-2n}^{(3)}+(n+1)^{3}G_{3n}c_{-k-3n}^{(3)}+c_{-k+2}^{(1)}+c_{-k-n}^{(1)}G_{n}+c_{-k}^{(0)}=0%
\text{ .}  \tag{C15}
\end{equation}
Since it has been shown already how all the Loran coefficient functions can
be expressed and the treatment of this equation is completely analogous to
the preceeding ones, equation (C15) shall not be considered.

\bigskip

\section{APPENDIX\ D:\ COEFFICIENT \ FUNCTIONS \\
$N_{1},N_{2},N_{3}$ and $\ N_{4}$ \ DEPENDING \\
ON \ THE \ ''BAR''\ VARIABLES}

\bigskip The coefficient functions $N_{1},N_{2},N_{3}$ and $N_{4}$ in the
cubic algebraic equation (106) for $T$ in Section VII are the following
\begin{equation*}
N_{1}\equiv 2\overline{p}_{1}^{2}Z^{2}-2Z^{4}\overline{p}_{1}(\frac{O^{2}%
\overline{p}_{1}}{4}+\frac{O}{2}\overline{p}_{2}+\overline{p}_{3})-Z^{4}%
\overline{p}_{1}(O^{2}\overline{p}_{1}+O\overline{p}_{2}+2\overline{p}_{3})+
\end{equation*}
\begin{equation}
+Z^{6}(O^{2}\overline{p}_{1}+O\overline{p}_{2}+2\overline{p}_{3})(\frac{O^{2}%
\overline{p}_{1}}{4}+\frac{O\overline{p}_{2}}{2}+\overline{p}_{3})  \tag{D1}
\end{equation}
\begin{equation*}
N_{2}\equiv 8Z\overline{p}_{1}^{2}-Z^{2}\overline{p}_{1}(\overline{p}_{2}+O%
\overline{p}_{1})+\left[ Z^{4}(\overline{p}_{2}+O\overline{p}_{1})-16Z^{3}%
\overline{p}_{1}\right] (\frac{O^{2}}{4}\overline{p}_{1}+\frac{O}{2}%
\overline{p}_{2}+\overline{p}_{3})+
\end{equation*}
\begin{equation}
+(O^{2}\overline{p}_{1}+O\overline{p}_{2}+2\overline{p}_{3})\left[ -\frac{15%
}{2}Z^{3}\overline{p}_{1}+\frac{23}{2}Z^{5}(\frac{O^{2}}{4}\overline{p}_{1}+%
\frac{O}{2}\overline{p}_{2}+\overline{p}_{3})\right]  \tag{D2}
\end{equation}
\begin{equation*}
N_{3}\equiv (\frac{O^{2}}{4}\overline{p}_{1}+\frac{O}{2}\overline{p}_{2}+%
\overline{p}_{3})\left[ 30Z^{4}(O^{2}\overline{p}_{1}+O\overline{p}_{2}+2%
\overline{p}_{3})+8Z^{3}(\overline{p}_{2}+O\overline{p}_{1})-24\overline{p}%
_{1}Z^{2}\right] +
\end{equation*}
\begin{equation}
+8\overline{p}_{1}^{2}-4Z\overline{p}_{1}(\overline{p}_{2}+O\overline{p}%
_{1})-8Z^{2}\overline{p}_{1}(O^{2}\overline{p}_{1}+O\overline{p}_{2}+2%
\overline{p}_{3})  \tag{D3}
\end{equation}

\begin{equation*}
N_{4}\equiv -4\overline{p}_{1}(\overline{p}_{2}+O\overline{p}_{1})+6Z%
\overline{p}_{1}(O^{2}\overline{p}_{1}+O\overline{p}_{2}+2\overline{p}_{3})+
\end{equation*}
\bigskip
\begin{equation}
+\left[ -18Z^{3}(O^{2}\overline{p}_{1}+O\overline{p}_{2}+2\overline{p}%
_{3})+12Z^{2}(\overline{p}_{2}+O\overline{p}_{1})\right] (\frac{O^{2}}{4}%
\overline{p}_{1}+\frac{O}{2}\overline{p}_{2}+\overline{p}_{3})\text{ \ \ .}
\tag{D4}
\end{equation}

\bigskip

\bigskip

\bigskip $^{1}$S.S. Roan, \textit{Proceedings of the Third Asian Conference
2000, }edited by T.\ Sunada, P. W.Sy, and Y. Lo (World Scientific,
Singapore, 2002) [arXiv:math-ph/0011038].

$^{2}$M. Kaku, \textit{Introduction to Superstrings} (Springer-Verlag,
Berlin-Heidelberg, 1988).

$^{3}$M. Green, J. Schwartz, E. Witten, \textit{Superstring Theory}, vol. 1
and 2 \ (Cambridge University Press, Cambridge, 1987).

$^{4}$G. V. Kraniotis, S. B. Whitehouse, Class. Quant. Grav. \textbf{19, }%
5073 (2002) \ [arXiv:gr-qc/0105022].

$^{5}$A. I. Markushevich, \textit{Theory of Analytical Functions } (State
Publ.House, Moscow, 1950).

$^{6}$N. I. Ahiezer, \textit{Elements of the Elliptic Functions Theory} \
(Nauka Publish. House, Moscow, 1979).

$^{7}$I. I. Privalov, \textit{Introduction to the Theory of Complex Variable
Functions (}Higher School. Publish. House, Moscow, 1999).

$^{8}$E. I. Whittaker, \textit{A Treatise on the Analytical Dynamics of
Particles and Rigid Bodies} (Cambridge University Press, Cambridge, 1927).

$^{9}$V. V. Golubev, \textit{Lectures on the Integration of the Equations of
Motions of a Heavy Rigid Body Around a Fixed Point \ (Moscow}, 1953).

$^{10}$S. Lang, \textit{Elliptic Functions} \ (Addison-Wesley Publishing
Company, Inc., London-Amsterdam, 1973).

$^{11}$N. Koblitz, \textit{Introduction to Elliptic Curves and Modular Forms}
(Springer-Verlag, New-York - Berlin, 1984).

$^{12}$S. Lang, \textit{Introduction to Modular Forms} \ (Springer- Verlag,
Berlin-Heidelberg, 1976).

$^{13}$V. V. Prasolov, Y. P. Soloviev, \textit{Elliptic Functions and
Algebraic Equations} \ (Factorial Publishing House, Moscow, 1997).

$^{14}$S. Manoff, Part. Nucl. \textbf{30}, 517 (1999) [Rus. Edit.
Fiz. Elem. Chast.Atomn.Yadra. \textbf{30} (5), 1211 (1999)
[arXiv:gr-qc/0006024].

$^{15}$C. Johnson, \textit{D-Brane Primer. Lectures, given at ICTP,
TASI and SUSSTEPP} \ [arXiv:hep-th/0007170].

$^{16}$P. Di Vecchia, A. Liccardo, \textit{D-Branes in String Theory I.
Lectures, presented at the 1999 NATO-ASI on Quantum Geometry \ in Akureyri}
[arXiv:hep-th/9912161].

$^{17}$P. Di Vecchia, A. Liccardo, \textit{D-Branes in String Theories II.
Lectures, presented at the YITP\ Workshop on Developments in Superstring and
M-Theory, Kyoto,\ Japan, October 1999 \ [}arXiv\textit{:}hep-th/9912275].

$^{18}$B. G. Dimitrov, in\textit{\ ''Perspectives of Complex Analyses,
Differential Geometry and Mathematical Physics. Proceedings of the 5th
International Workshop on Complex Structures and Vector Fields'', }edited by
S. Dimiev, and K. Sekigawa (World Scientific, Singapore, \textit{2001) [}%
arXiv:gr-qc/0107089\textit{].}

$^{19}$Y. I. Manin, \textit{Cubic Forms: Algebra, Geometry, Arithmetic} \
(North Holland, Amsterdam, 1974).

$^{20}$W. Fulton, \textit{Algebraic Curves. An Introduction to
Algebraic Geometry} \ (W.A. Benjamin, Inc., New York, Amsterdam, 1969).

$^{21}$R. J. Walker, \textit{Algebraic Curves} (Princeton, New Jersey, 1950).

$^{22}$A. I. Markushevich, \textit{Introduction to the Classical Theory of
Abelian Functions} (Nauka Publish. House, Moscow, 1979).

$^{23}$M. Schlichenmaier, \textit{An Introduction to Riemann Surfaces,
Algebraic Curves and Moduli Spaces}, Lecture Notes in Physics 322
(Springer-Verlag, Berlin-Heidelberg, 1989).

$^{24}$A. Hurwitz, R. Courant, \textit{Allgemeine Funktionentheorie und
Elliptische Funktionen} \ (Springer-Verlag, Berlin-Heidelberg, 1964).

$^{25}$S. Manoff, Class. Quant. Grav. \textbf{18, }1111 (2001)
[arXiv:gr-qc/9908061].

$^{26}$N. Efimov, E.P. Rosendorn, \textit{Linear Algebra and
Multidimensional Geometry} \ (Nauka Publishing House, Moscow, 1974).

$^{27}$M. Reid, \textit{Undergraduate} \textit{Algebraic Geometry, }London
Math. Soc. Student Texts 12 (Cambridge University Press, Cambridge).

$^{28}$D. Mumford, \textit{Algebraic Geometry. Complex Projective Varieties}
\ (Springer-Verlag, New York, Berlin, Heidelberg, 1976).

$^{29}$L Randall, R. Sundrum, Phys. Rev. Lett. \textbf{83, }3370 (1999)
[arXiv: hep-th/9905221].

$^{30}$L Randall, R. Sundrum, Phys. Rev. Lett. \textbf{83, }4690 (1999)
[arXiv:hep-th/9906064].

$^{31}$K. Kuchar, Journ.Math.Phys. \textbf{17,} 777 (1976); \textbf{17, }792
(1976); \textbf{17,} 801 (1976); \textbf{18} 1589 (1977).

$^{32}$E. Witten, Nucl. Phys. B\textbf{\ 311, }46 (1988).

$^{33}$O. Aharony, S. Gubser, J. Maldacena, H. Ooguri and Y. Oz, Phys.
Reports \textbf{323, }183 (2000) [arXiv:hep-th/9905111].

$^{34}$J. Maldacena and H. Ooguri, J. Math. Phys. \textbf{42, }2929 (2001)
[arXiv:hep-th/0001053].

$^{35}$J. Maldacena, H. Ooguri and J. Son, Journ. Math. Phys.\textbf{42, }%
2961 (2001) [arXiv:hep-th/0005183].

$^{36}$A. Giveon, D. Kutasov, N. Seiberg, Adv. Theor. Math. Phys. \textbf{2}%
, 733 (1998) [arXiv:hep-th/9806194].

$^{37}$M. Spradlin, A. Strominger, A. Volovich, Lectures at the LXXVI\
Houches School \textit{''Unity from Duality: \ Gravity, Gauge Theory and
Strings''}, August 2001 [arXiv:hep-th/0110007] .

$^{38}$N.V. Efimov, \textit{A Higher Geometry} \ (State Publish. House,
Moscow, 1961).

$^{39}$D. Cox, J. Little, D. O'Shea, \textit{Ideals, Varieties, and
Algorithms. An Introduction to Computational Algebraic Geometry and
Commutative Algebra } (Springer-Verlag, New York, 1998).

$^{40}$\textit{From Number Theory to Physics}, edited by M. Waldschmidt, P.
Moussa, J.-M. Luck, and C. Itzykson \ \ (Springer-Verlag, Berlin,
Heidelberg, 1992).

$^{41}$E. C. Titchmarsh, \textit{Theory of Functions}, Oxford, 1932.

$^{42}$D. Mumford, \textit{Tata Lectures on Theta} (Birkhauser,
Boston-Basel-Stuttgart, vol. 1, 1983; vol.2, 1984).

\end{document}